\documentclass[journal]{new-aiaa}
\usepackage[utf8]{inputenc}
\usepackage{graphicx}
\usepackage{amsmath}
\usepackage[version=4]{mhchem}
\usepackage{siunitx}
\usepackage{longtable,tabularx}
\usepackage{epstopdf}
\usepackage{enumerate}
\usepackage{psfrag}
\usepackage{amsbsy}
\usepackage{amsfonts}
\usepackage{setspace}
\usepackage{float}
\usepackage{subfig}
\usepackage{tabularx}
\usepackage{epsfig}
\usepackage{booktabs,array,dcolumn}
\usepackage{arydshln}
\usepackage{dashrule}
\usepackage{wasysym}
\usepackage{multicol}
\usepackage{multirow}
\usepackage[normalem]{ulem}
\usepackage{framed} 
\usepackage{nomencl}
\usepackage{xpatch}
\usepackage{fancyhdr}
\usepackage{color}\definecolor{RED}{rgb}{1,0,0}\definecolor{BLUE}{rgb}{0,0,1}
\newcommand{\reffig}[1]{figure \ref{#1}}
\newcommand{\reffigs}[1]{figures \ref{#1}}
\newcommand{\refFig}[1]{Figure \ref{#1}}
\newcommand{\refFigs}[1]{Figures \ref{#1}}
\newcommand{\reftab}[1]{table \ref{#1}}

\newcommand{\refeq}[1]{Eq. \eqref{#1}}
\newcommand{\refse}[1]{Section \ref{#1}}

\title{Unsteady aeroelastic characterization and scaling relations of flexible membrane wings}

\author{Guojun Li \footnote{Corresponding Author: guojunli@xjtu.edu.cn. Assistant Professor, State Key Laboratory for Manufacturing Systems Engineering, Shaanxi China 710049}}
\affil{State Key Laboratory for Manufacturing Systems Engineering, Xi'an Jiaotong University, Xi'an, Shaanxi, China}
\affil{School of Mechanical Engineering, Xi'an Jiaotong University, Xi'an, Shaanxi, China}
\author{Rajeev Kumar Jaiman\footnote{Associate Professor, Department of Mechanical Engineering, BC Canada V6T 1Z4, and AIAA Senior Member.}}
\affil{Department of Mechanical Engineering, The University of British Columbia, Vancouver, British Columbia, Canada}

\begin{document}

\maketitle

\begin{abstract}
We present a numerical study to characterize nonlinear unsteady aeroelastic interactions of two-dimensional flexible wings at high angles of attack. The coupled fluid-flexible wing system is solved by a body-fitted variational aeroelastic solver based on the fully-coupled Navier-Stokes and nonlinear structural equations. Using the coupled fluid-structure analysis, this study is aimed to provide physical insight and correlations for the aeroelastic behavior of flexible wings in the parameter space of the angle of attack and the aeroelastic number. The phase diagrams of the aerodynamic performance are established to obtain the envelope curves of the optimal performance and determine the transition line of the drag variation. The effects of the angle of attack and the aeroelastic number on the aeroelastic behaviors are systematically examined. The time-averaged membrane deformation is positively correlated with a non-dimensional number, the so-called Weber number. A new scaling relation is proposed based on the dynamic equilibrium between the aerodynamic force fluctuation and the combined inertia-elastic fluctuation. The unsteady aerodynamic force can be adjusted by manipulating the membrane vibration, the mass ratio, the Strouhal number and the aeroelastic number. The numerical investigations provide design guidelines and have the potential to enhance the maneuverability and flight agility of micro air vehicles with flexible wing structures.
\end{abstract}

\section{Introduction}
\lettrine{F}{exible} morphing wing structures have received much attention from the aerospace engineering community in the context of bio-inspired swimming/flying vehicles \cite{shyy2010recent,jiakun2021review,hefler2021distinct,tiomkin2021review}. The flexible wings of natural fliers and swimmers with membrane components and appendages have light-weight structures, shape reconfigurable capability and possibly perception ability to surrounding environments \cite{hubel2010time,wang2015lift}. These favorable features enable the fliers/swimmers to adapt to complex environments with optimal performance and high maneuverability for locomotion purposes \cite{wu2011fish,alexander2015wing,hedenstrom2015bat}. Given the superior aerodynamic performance of flexible wings, there has been an increasing interest to incorporate the aeroelastic effect of flexible wings into the next-generation intelligent and efficient morphing air vehicles. However, the complex fluid-structure coupling phenomena and the wide range of aeroelastic parameters pose a serious challenge when applying flexible wings to flying vehicles. There is an urgent need to grasp the underlying correlation between the aerodynamic loads and the membrane kinematics from a system design point of view. The physical understanding of the coupled fluid-structure dynamics can provide design guidance and active control strategies for morphing air vehicles with flexible smart skin.

During the past decades, several experimental, numerical and theoretical investigations have been carried out to explore the aerodynamic characteristics of the flexible wings with membrane structures \cite{rojratsirikul2009unsteady,rojratsirikul2010effect,tregidgo2013unsteady,tzezana2019thrust,bleischwitz2018near,serrano2018fluid,mavroyiakoumou2021dynamics}. The effects of structural material properties, freestream inflow conditions and wing geometries on the unsteady membrane dynamics and the aerodynamic performance were widely explored \cite{song2008aeromechanics,gordnier2014impact,mavroyiakoumou2021dynamics,mavroyiakoumou2020large,li2020flow_accept}. Flexible membrane wings can outperform their rigid counterparts by generating greater lift, delaying stall, suppressing drag and reducing noise. Based on the previous studies \cite{li2022aeroelastic,gordnier2014impact}, the membrane dynamics and the aerodynamic performance are primarily governed by the camber effect and the flow-induced vibration. The flexible membrane wings deform passively under aerodynamic loads. Thus, the local angle of attack near the leading edge is reduced, resulting in suppressed separated flows. When the flow-excited instability occurs, the vortex shedding process is synchronized with the membrane vibration, which leads to the well-known frequency lock-in phenomenon \cite{gordnier2009high,rojratsirikul2011flow,li2020flow_accept}. By proper harnessing of such flow-excited instability, one can regulate the membrane kinematics and the vortex-shedding features through the fluid-membrane coupling effect. Furthermore, through the wing morphing and continuous alternation of the surrounding flows, there is a potential to achieve superior flight efficiency and robust performance in gusty wind environments and confined spaces.

Bats exhibit agile flying skills when turning directions, tumbling and hanging upside down on a tree to rest and chasing prey in confined spaces \cite{swartz2015advances}. To achieve the flight task, bats usually rotate their wings and stretch or release the membrane by skeletons. These abundant flying behaviors result in a wide range of angles of attack from zero to extreme vertical state. Natural fliers can still maintain strong maneuverability and certain aerodynamic forces at high angles of attack with the aid of their flexible morphing wings. Compared with the biological flight with strong environmental adaptability, human-made flying vehicles perform poorer flight efficiency at high angles of attack. Plenty of studies have been performed to investigate the aeroelasticity of flexible membrane wings ranging from low to relatively large angles of attack \cite{song2008aeromechanics,rojratsirikul2010effect,bleischwitz2015aspect}. A few of the works extended the angle of attack to an extreme value up to $90^\circ$ \cite{newman1987aerodynamic,fan2014simulation,das2020compliant}. The flow features and the membrane responses at high angles of attack exhibit obvious differences from those at low angles. The flexible membrane was found to lose its stability when the angle of attack exceeded a critical value \cite{sun2017nonlinear,tiomkin2019membrane}. Complex synchronization phenomena between the vortex shedding frequency and the membrane vibration frequency are established at certain ranges of angles of attack \cite{rojratsirikul2011flow,li2022aeroelastic}. The results indicate that the aerodynamic performance is closely related to the membrane deformation and the flow-induced vibration. Fully understanding the physics of the aerodynamic force adjustment mechanism over a wide range of angles of attack is the key to designing efficient flexible morphing wings. 

As mentioned above, bats change their wing shapes by stretching or releasing the membrane structures during maneuvering. The balance between the aerodynamic loads and the membrane tension can be described by a non-dimensional number the so-called aeroelastic number, which is defined as $Ae=\frac{E^s h}{\frac{1}{2} \rho^f U_{\infty}^2 c}$. Here, $E^s$ denotes Young's modulus, $h$ is the wing thickness, $c$ represents the characteristic length of the wing, $\rho^f$ is the air density and $U_{\infty}$ is the freestream velocity. The membrane rigidity governed by $E^s h$ \cite{sun2017nonlinear,gordnier2009high} and the aeroelastic number \cite{li2020flow_accept} are found to reduce lift forces by suppressing the membrane deformation. The effect of aeroelastic numbers on the membrane dynamics was mainly explored at a fixed angle of attack. The agile flight of bats is a combined action of the changes in the angle of attack and the aeroelastic number. In other words, the optimal combination of these two parameters corresponding to desirable aerodynamic performance exists and follows certain dynamical relations, which are not systematically studied in the literature. The present study aims to investigate the coupled fluid-membrane dynamics in the parameter space of the angle of attack and the aeroelastic number. The optimal parameter combination is determined based on our high-fidelity numerical simulations.

Previous studies found that the membrane deformation and the aerodynamic characteristics were governed by various aeroelastic parameters (e.g., angle of attack, Young's modulus, material density and free-stream velocity). To characterize the membrane dynamics, three typical non-dimensional parameters are usually employed, namely the Reynolds number, the mass ratio and the aeroelastic number. For a flexible membrane with uniform aerodynamic loads and no pretension, a non-dimensional parameter, the so-called Weber number, is introduced to characterize the maximum membrane displacement of the coupled fluid-membrane system \cite{song2008aeromechanics,waldman2017camber}. The Weber number can be defined as the ratio of the aerodynamic force on the membrane surface and the membrane rigidity, which is given as
\begin{equation}
We = \frac{L}{E^s h} = C_L \frac{\frac{1}{2}\rho^s U_{\infty}^2 c}{E^s h} = \frac{C_L}{Ae}
\end{equation}
where $L$ is the lift force and $C_L$ denotes the lift coefficient. Waldman et al. \cite{waldman2017camber} examined the relationship between the maximum membrane camber and the Weber number by employing the Young-Laplace equation for nonlinear deformation and a potential flow model at low angles of attack. A good agreement was found between the predicted data from the established models and the experimental data with different angles of attack, aeroelastic numbers and pretension values. The one-third power relation between the membrane camber and the Weber number was suggested in their studies. Das et al. \cite{das2020compliant} introduced a force-kinematics scaling relation between unsteady force fluctuations and membrane vibrations for a bluff membrane-like structure vertically placed in a uniform fluid flow. This simple scaling relation suggests a positive correlation between the aerodynamic force fluctuations and the inertial force fluctuations caused by the flow-induced membrane vibration. The scaling relations discussed above show some potential in explaining the coupled fluid-membrane mechanism while devising simple design guidelines.

The present study aims to investigate the aeroelastic performance of a two-dimensional flexible membrane wing over a wide range of parameter space. Of particular interest is to explore the optimal parameter combination corresponding to desirable aerodynamic performance and determine the scaling relations between the aerodynamic loads and the membrane deformation. To fully characterize the membrane aeroelasticity, we establish the aeroelastic performance phase diagrams in the $Ae$-$\alpha$ space through a series of coupled numerical simulations. The angle of attack is chosen from zero to extreme $90^\circ$. The empirical solutions of the contour line envelopes including the optimal aerodynamic performance regions are determined by exponential curve fitting functions. The effects of the angle of attack and the aeroelastic number are further investigated by comparing the mean membrane responses and the instantaneous flow features. The role of flexibility is examined by comparing the flexible membrane wing and its rigid flat and rigid cambered counterparts over a wide range of angles of attack. Two scaling relations are derived based on the membrane governing equation. These new scaling relations for the aerodynamic loads and the membrane deformation are suggested for the characterization of the coupled fluid-membrane dynamics. 

This paper is organized as follows. In \refse{sec:section2}, the governing equations of the coupled fluid-flexible structure system are described. \refse{sec:section3} presents the problem setup and the mesh convergence study of the two-dimensional flexible wing. We discuss the aeroelastic characteristics of flexible wings at various angles of attack and the effect of aeroelastic numbers in \refse{sec:section4}. Two scaling relations for the mean membrane responses and the unsteady membrane dynamics are proposed. Finally, the key findings are summarized in \refse{sec:section5}.

\section{Computational Aeroelasticity Framework} \label{sec:section2}
To simulate the coupled fluid-flexible structure system, the incompressible Navier-Stokes equations are coupled with the nonlinear structure equations via a partitioned iterative scheme. The Navier-Stokes equations are discretized via a stabilized Petrov-Galerkin finite element method in an arbitrary Lagrangian-Eulerian (ALE) reference frame. The structure motion equations are solved via nonlinear co-rotational finite element method in a Lagrangian coordinate \cite{jaiman2016stable,li2018novel}. The body-fitted moving boundary is applied between the interface of the fluid domain and the structure domain. The large eddy simulation (LES) model is employed to simulate the unsteady viscous flows around the flexible membrane. A partitioned iterative coupling algorithm is adopted to couple the fluid and structural motion equations. A predictor-corrector approach is utilized to solve the coupled framework advanced in time. The compactly-supported radius basis function (RBF) is employed to transfer the fluid loads and the structural displacements along the non-matching fluid-solid interface, which naturally ensures energy conservation. The body-fitted spatial fluid meshes are updated based on the efficient RBF-based remeshing method to preserve the high mesh quality. The nonlinear interface force correction (NIFC) scheme \cite{jaiman2016stable} is implemented in the coupled framework to correct the fluid forces at each iterative step to avoid the numerical instability caused by the significant added mass effect. This high-fidelity fluid-structural interaction solver has been applied and extensively validated on flexible flapping wings \cite{li2021high} and fluid-membrane interaction \cite{li2020flow_accept}.

\section{Problem Description and Mesh Convergence Study} \label{sec:section3}
\subsection{Problem Description}
\begin{figure}[H]
	\centering
	\includegraphics[width=0.75\textwidth]{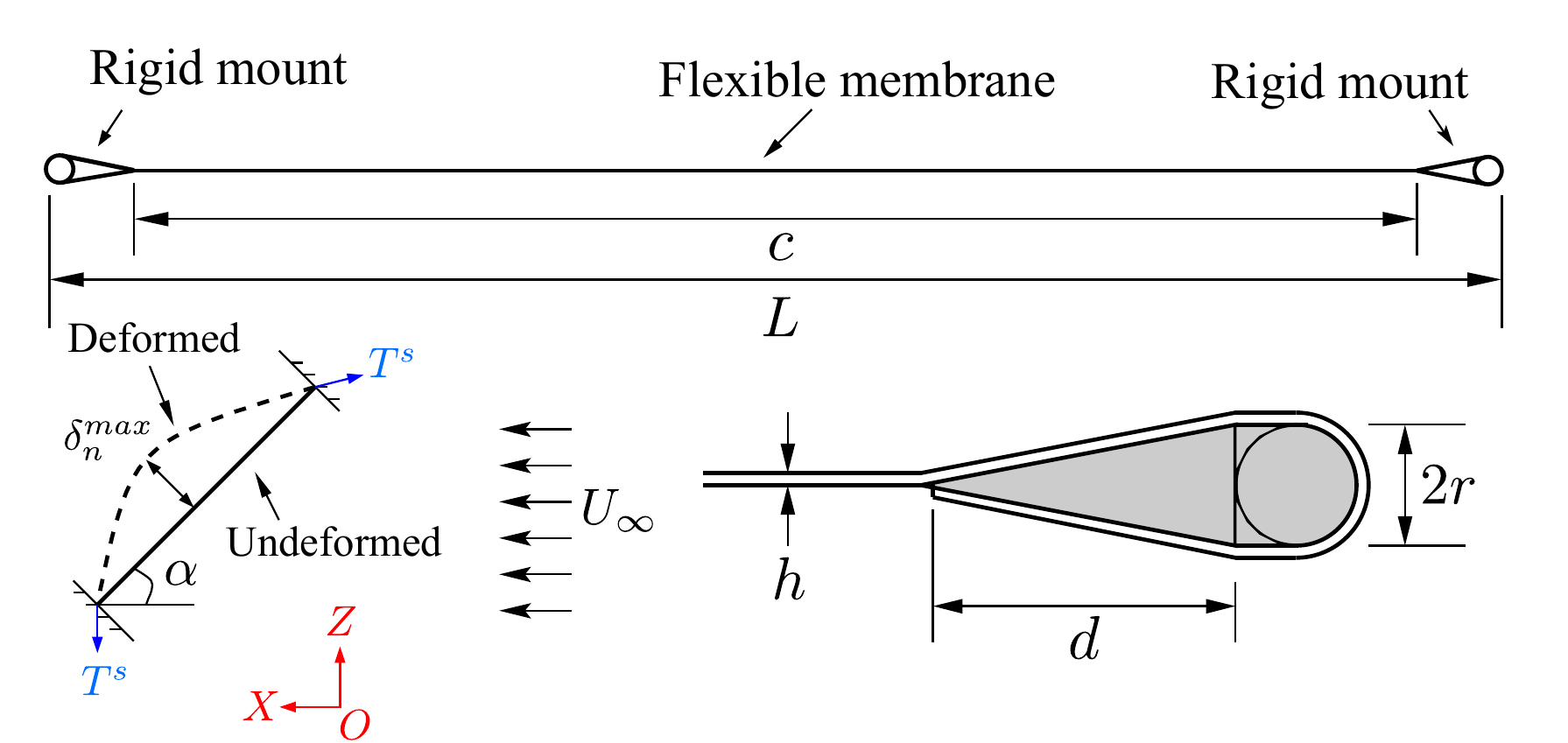}
	\caption{Schematic of the 2D flexible membrane wing geometry and the details of leading and trailing edge mounts.}
	\label{2dmembranegeo}
\end{figure}

A two-dimensional flexible membrane wing model employed in the experimental study \cite{rojratsirikul2009unsteady} is considered as the numerical simulation model to investigate the aerodynamic characteristics and the membrane dynamics at different angles of attack. In the experiment, this flexible membrane wing made of a black latex rubber sheet with a thickness of $h=0.2$ mm is attached to the rigid airfoil-shaped mount at the leading and trailing edges. The schematic of this membrane wing geometry and the fixed mounts is presented in \reffig{2dmembranegeo}. The chord length of the membrane wing is $c=136.5$ mm with a span length of $b=450$ mm. The rigid mount consists of a triangle part with a length of $d=5$ mm and circle support with a diameter of $2r=1.5$ mm, which results in a total length of $L=150$ mm for the wing.

\begin{figure}[H]
	\centering
	\includegraphics[width=0.95\textwidth]{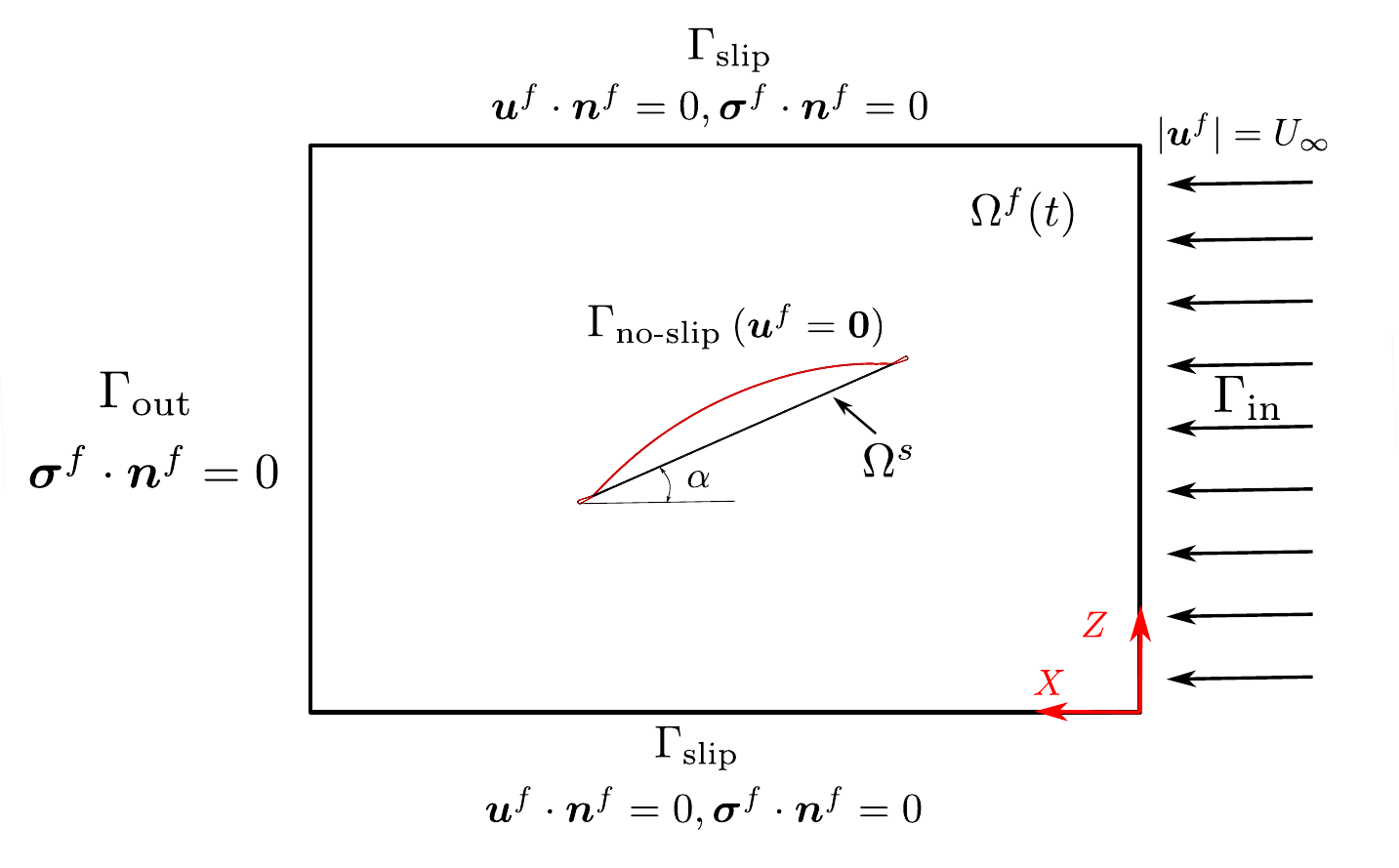}
	\caption{Schematic of the computational domain for a uniform flow past a 2D flexible membrane wing. }
	\label{2dcomputational}
\end{figure}

The schematic of the 2D computational fluid domain is displayed in \reffig{2dcomputational}. The flexible membrane wing is clamped at the leading and trailing edges to form an angle of attack of $\alpha$ between the uniform freestream and the initial chord directions. The distance between the inlet ($\Gamma_{\rm{in}}$) and outlet ($\Gamma_{\rm{out}}$) boundaries is 20$L$. The same size is set for the top ($\Gamma_{\rm{top}}$) and bottom ($\Gamma_{\rm{bottom}}$) sides of the computational domain. A uniform freestream velocity $U_{\infty}$ is applied along the inlet boundary and the traction-free boundary condition is employed at the outlet boundary. The slip-wall boundary condition is implemented at both sides of the computational domain. The no-slip wall condition is applied at the surface of the membrane wing. The LES model is employed here to capture the unsteady flow features around the flexible membrane.

\subsection{Mesh Convergence Study}
\begin{figure}[H]
	\centering 
	\subfloat[][]{\includegraphics[width=0.8\textwidth]{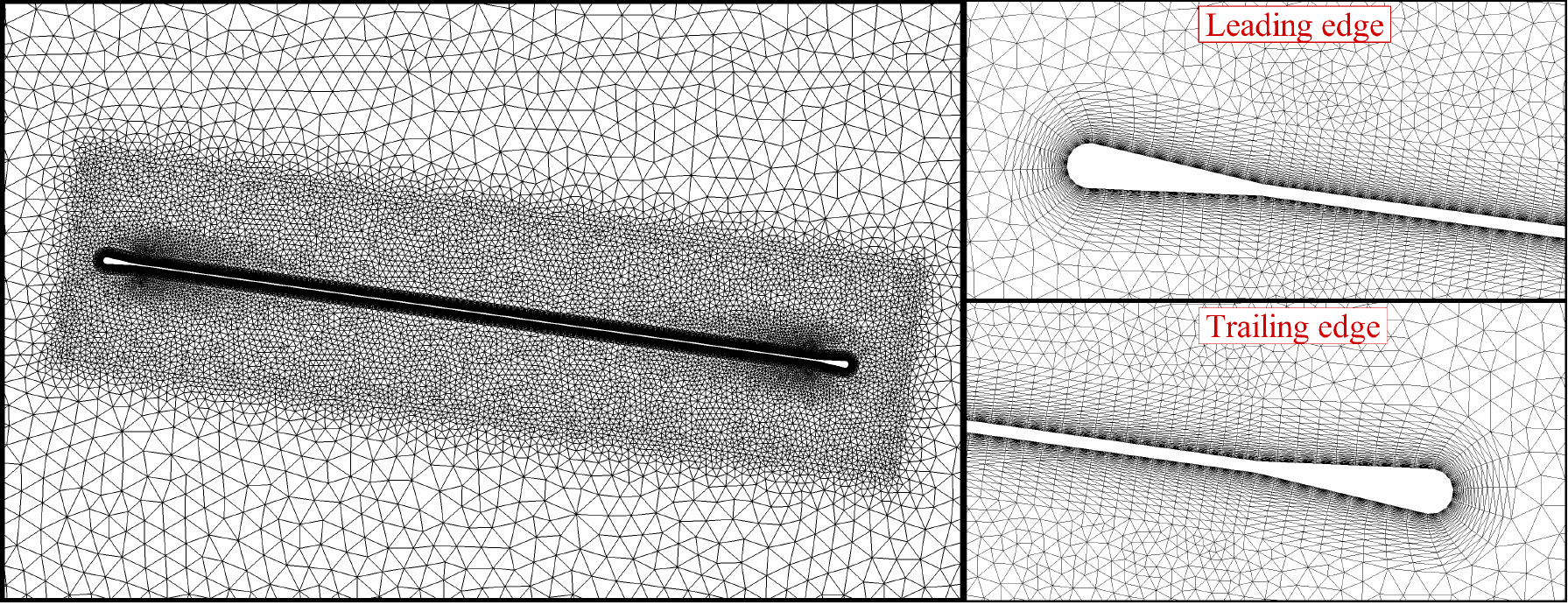}\label{mesh1}}
    \\
	\subfloat[][]{\includegraphics[width=0.8\textwidth]{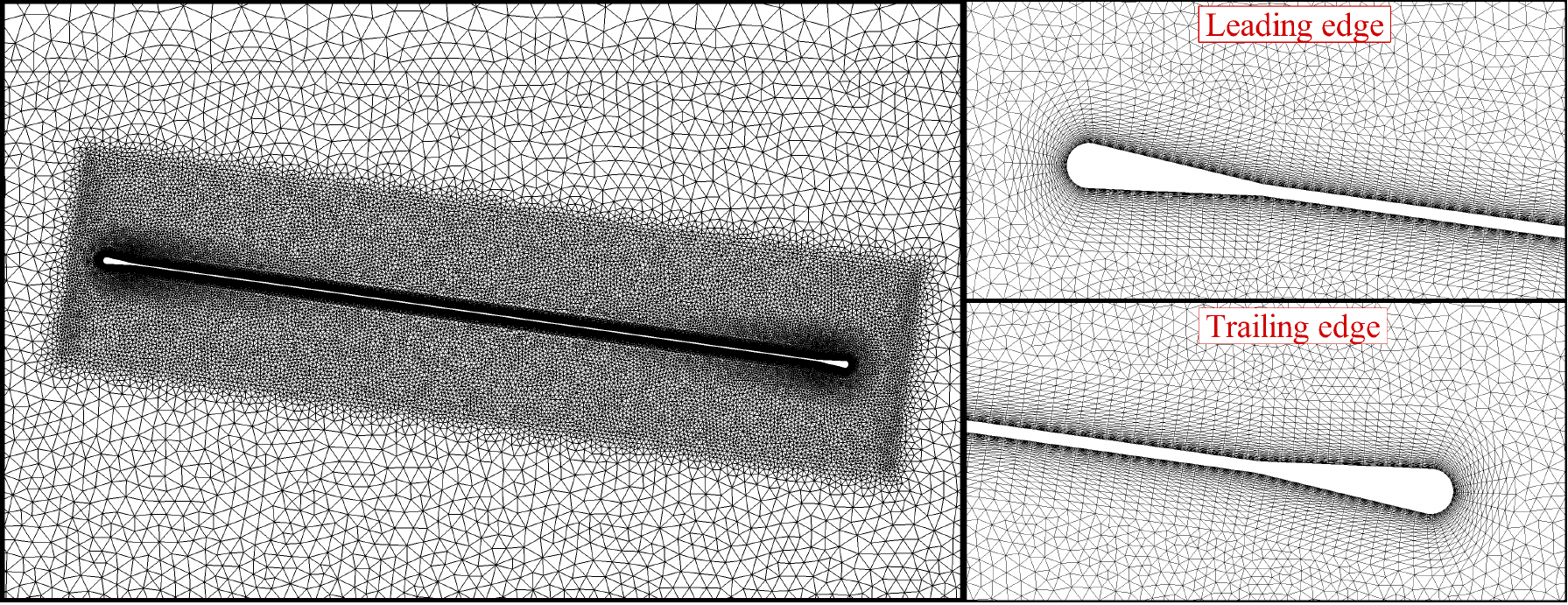}\label{mesh2}}
    \\
    \subfloat[][]{\includegraphics[width=0.8\textwidth]{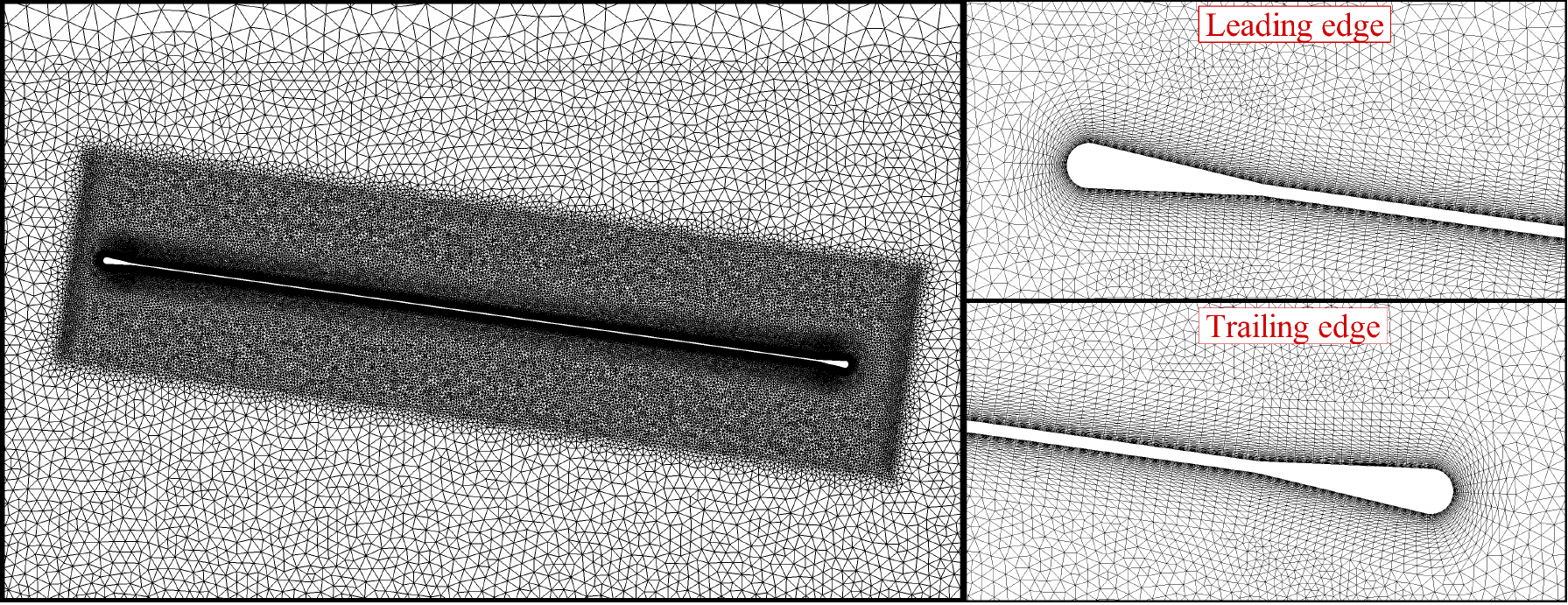}\label{mesh3}}
	\caption{ Comparison of three mesh discretizations in the vicinity of the membrane surface (a) M1, (b) M2 and (c) M3. Close-up view of meshes at leading and trailing edges are shown.}
	\label{mesh_plot}
\end{figure}

The mesh convergence study is performed to choose an appropriate mesh resolution for the numerical study of the coupled fluid-membrane dynamics. Three different meshes M1, M2 and M3 are designed to discretize the fluid domain by 50 616, 87 278 and 135 624 triangular elements, respectively. The boundary layer mesh is built to maintain $y^+<1$ with a stretching ratio of 1.15 for these three meshes. A comparison of the three meshes in the vicinity of the membrane is shown in \reffig{mesh_plot}. The structural model of the flexible membrane is discretized by 30, 50 and 70 structured four-node rectangular finite elements along the chord direction and one element in the span direction. This multibody structural model consists of the rigid mounts at the leading as well as the trailing edges and the flexible membrane component, which are simulated by the rigid body elements and the geometrically exact co-rotational shell elements, respectively. A clamped condition is applied to the rigid mounts and the flexible membrane is attached to the mounts. There is no pretension applied to this 2D flexible membrane. In the mesh convergence study, the freestream velocity is set to $U_{\infty}=0.2886$ m/s and the air density is chosen as $\rho^f=1.1767$ kg/m$^3$, leading to a Reynolds number of 2500. The flexible membrane has Young's modulus of $E^s=3346$ Pa and a structure density of $\rho^s=473$ kg/m$^3$. Thus, the structure-to-fluid mass ratio is $m^*=\rho^s h/\rho^f c$=0.589 and the aeroelastic number is $Ae$=100.04. The flexible membrane is placed in the fluid flow with an angle of attack of $8^\circ$. The non-dimensional time step size, $\Delta t U_{\infty}/c$ is set as 0.00423 in the simulation.

The mean membrane deformation, the forces acting on the membrane wing and the pressure distribution on the wing surface are evaluated from the simulation data. The air load is calculated by integrating the surface traction taking into account the first layer of elements on the membrane surface. The instantaneous lift, the drag and the normal force coefficients are defined as
\begin{equation}
	C_L = \frac{1}{\frac{1}{2} \rho^f U_{\infty}^2 S} \int_{\Gamma} (\boldsymbol{\hat{\sigma}}^f \cdot \boldsymbol{e}) \cdot \boldsymbol{e}_z \rm{d \Gamma}
	\label{CL_2d} 
\end{equation}
\begin{equation}
	C_D = \frac{1}{\frac{1}{2} \rho^f U_{\infty}^2 S} \int_{\Gamma} (\boldsymbol{\hat{\sigma}}^f \cdot \boldsymbol{e}) \cdot \boldsymbol{e}_x \rm{d \Gamma}
	\label{CD_2d} 
\end{equation}
\begin{equation}
	C_n = \frac{1}{\frac{1}{2} \rho^f U_{\infty}^2 S} \int_{\Gamma} (\boldsymbol{\hat{\sigma}}^f \cdot \boldsymbol{e}) \cdot \boldsymbol{e}_c \rm{d \Gamma}
	\label{CN_2d} 
\end{equation}
where $\boldsymbol{e}_x$ and $\boldsymbol{e}_z$ are the Cartesian components of the unit normal $\boldsymbol{e}$ to the membrane surface and $\boldsymbol{e}_c$ is the unit normal to the chord line. $\boldsymbol{\hat{\sigma}}^f$ is the fluid stress tensor with $\Gamma$ being the surface boundary of the membrane. The pressure coefficient is defined as
\begin{equation}
	C_p = \frac{p-p_\infty}{\frac{1}{2} \rho^f U_{\infty}^2}
	\label{Cp_2d} 
\end{equation}
where $p$ and $p_\infty$ are the pressure at the concerned point and the pressure at the far-field, respectively. 

\begin{table}[H]
	\vspace{1.2em}
	\centering
	\caption{Mesh convergence of a 2D flexible membrane wing at $Re$=2500 and $\alpha=8^\circ$. The percentage differences are calculated by using M3 results as the reference.}
	\begin{tabular}{ccccccc}
		\toprule  
		Mesh & $\overline{C}_L$ & $\overline{C}_D$ & $\overline{{C}_L/{C}_D}$ \\
		\midrule  
		Present (M1) & 0.9793 ($-1.54 \%$)  & 0.1114 ($1.09 \%$) & 8.7895 ($-2.62 \%$) \\
		Present (M2) & 0.9948 ($0.02 \%$) & 0.1097 ($0.45 \%$) & 9.0684 ($0.47 \%$) \\
		Present (M3) & 0.9946 & 0.1102 & 9.0257 \\
		\bottomrule 
	\end{tabular}
	\label{2dmeshforce}
	\vspace{-1.5em}
\end{table}

\begin{figure}[H]
	\centering 
	\subfloat[][Time-averaged normalized membrane surface displacement]{\includegraphics[width=0.5\textwidth]{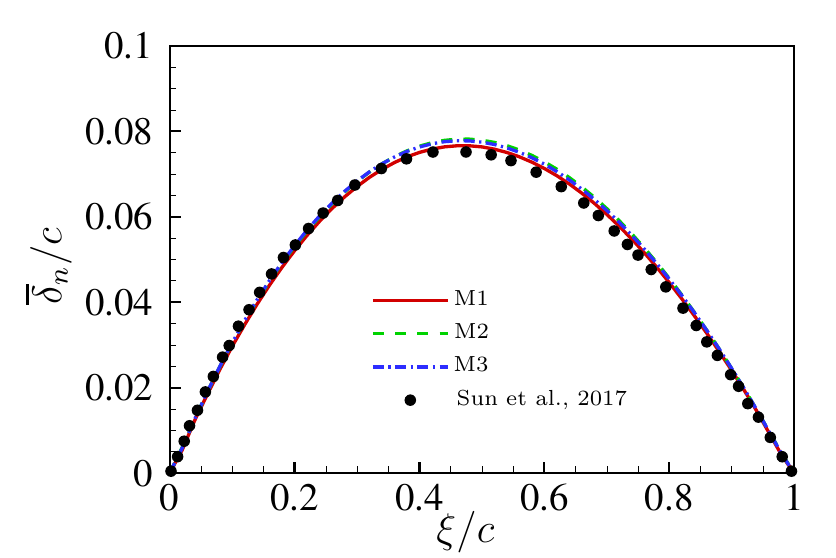}\label{dis_cp_2da}}
	\subfloat[][Time-averaged pressure coefficient distribution]{\includegraphics[width=0.5\textwidth]{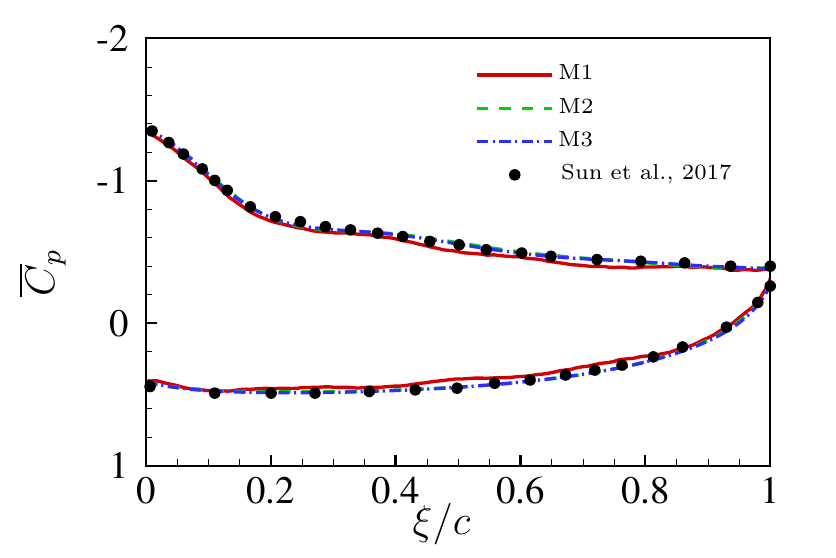}\label{dis_cp_2db}}
	\caption{ Comparison of membrane dynamics along membrane chord between different meshes at $Re$=2500 and $\alpha=8^\circ$.}
	\label{dis_cp_2d}
\end{figure}

A comparison of the mean lift coefficient, the mean drag coefficient and the mean lift-to-drag ratio for three different meshes is summarized in \reftab{2dmeshforce}. The percentage differences for M1 and M2 are calculated with respect to M3. It can be observed that the absolute differences of the time-averaged drag coefficient and the lift-to-drag ratio between M1 and M3 are greater than 2$\%$. The differences for all values between M2 and M3 are less than 1$\%$. For further comparison, the time-averaged membrane displacement (normal to the chord) and the mean pressure coefficient distribution on the surface are evaluated, which are shown in \reffig{dis_cp_2d}. The mean membrane displacement and the mean pressure coefficient are converged when the mesh is refined to M2. A good agreement is observed between the present data and the simulation results in Sun et al. \cite{sun2017effect}. A further comparative study with the numerical simulations from Sun et al. \cite{sun2017effect} can be found in the appendix. Hence, the mesh M2 is adequate and will be considered for further numerical study.

\section{Results and Discussion} \label{sec:section4}
The current study aims to investigate the aerodynamic performance behaviors and the membrane dynamics of two-dimensional flexible membrane wings from low to very high angles of attack. We further extend the investigations to different aeroelastic numbers and characterize the membrane dynamics. The parameter combinations of the angle of attack and the aeroelastic number for the optimal lift coefficient and the optimal lift-to-drag ratio regions are determined from the time-averaged aerodynamic performance phase diagram. A comparison of the aerodynamic performance between the flexible membrane wings and their rigid counterparts is conducted to explore the role of flexibility. Scaling relations for the mean membrane responses and the unsteady aeroelastic dynamics are explored.

\subsection{Coupled Fluid-Membrane Dynamics} \label{sec:dynamics}
\begin{figure}[H]
	\centering
	\subfloat[][Time-averaged lift coefficient]{\includegraphics[width=0.5\textwidth]{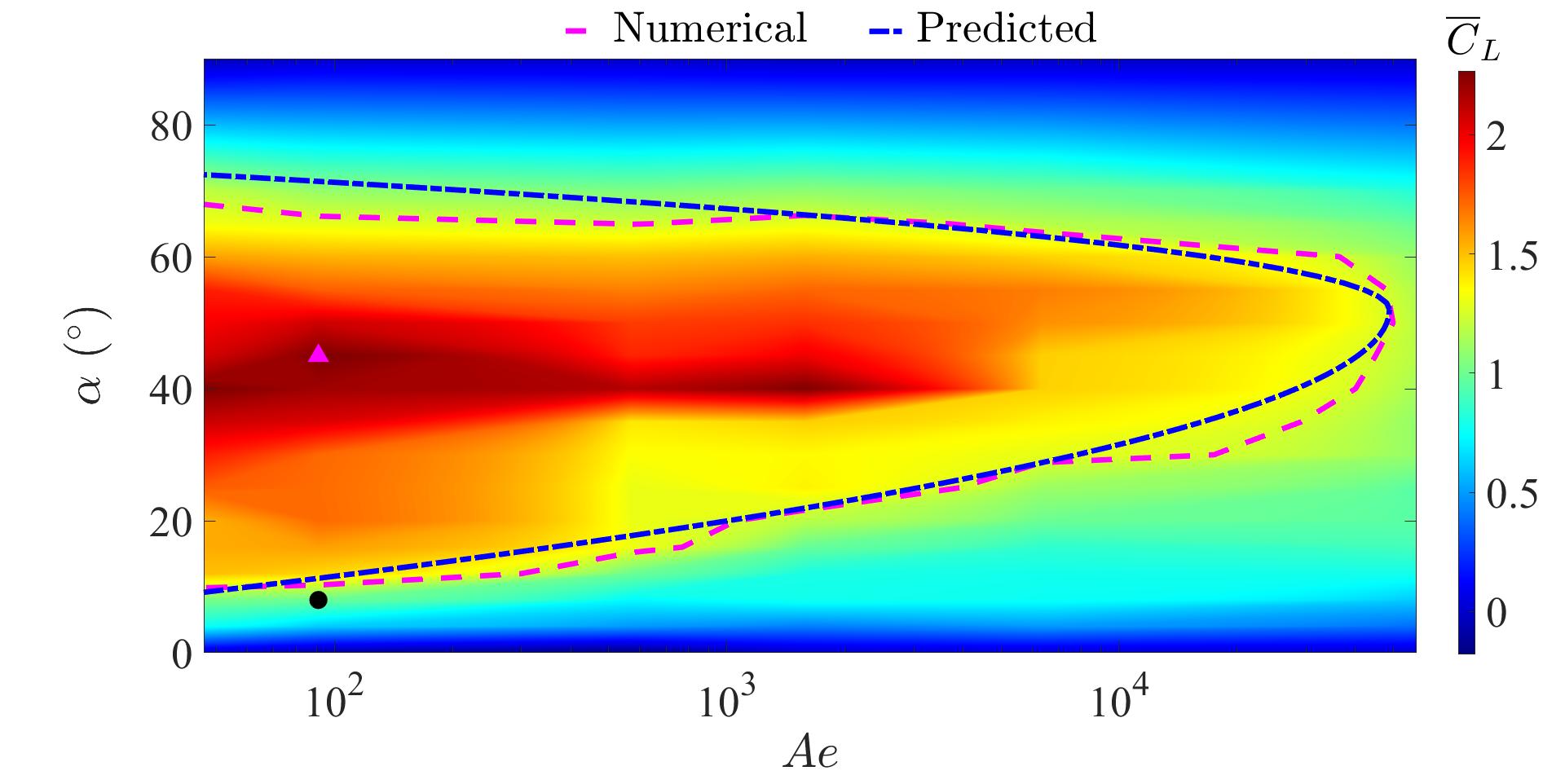}\label{fig:Ae_effecta}}
	\subfloat[][Time-averaged drag coefficient]{\includegraphics[width=0.5\textwidth]{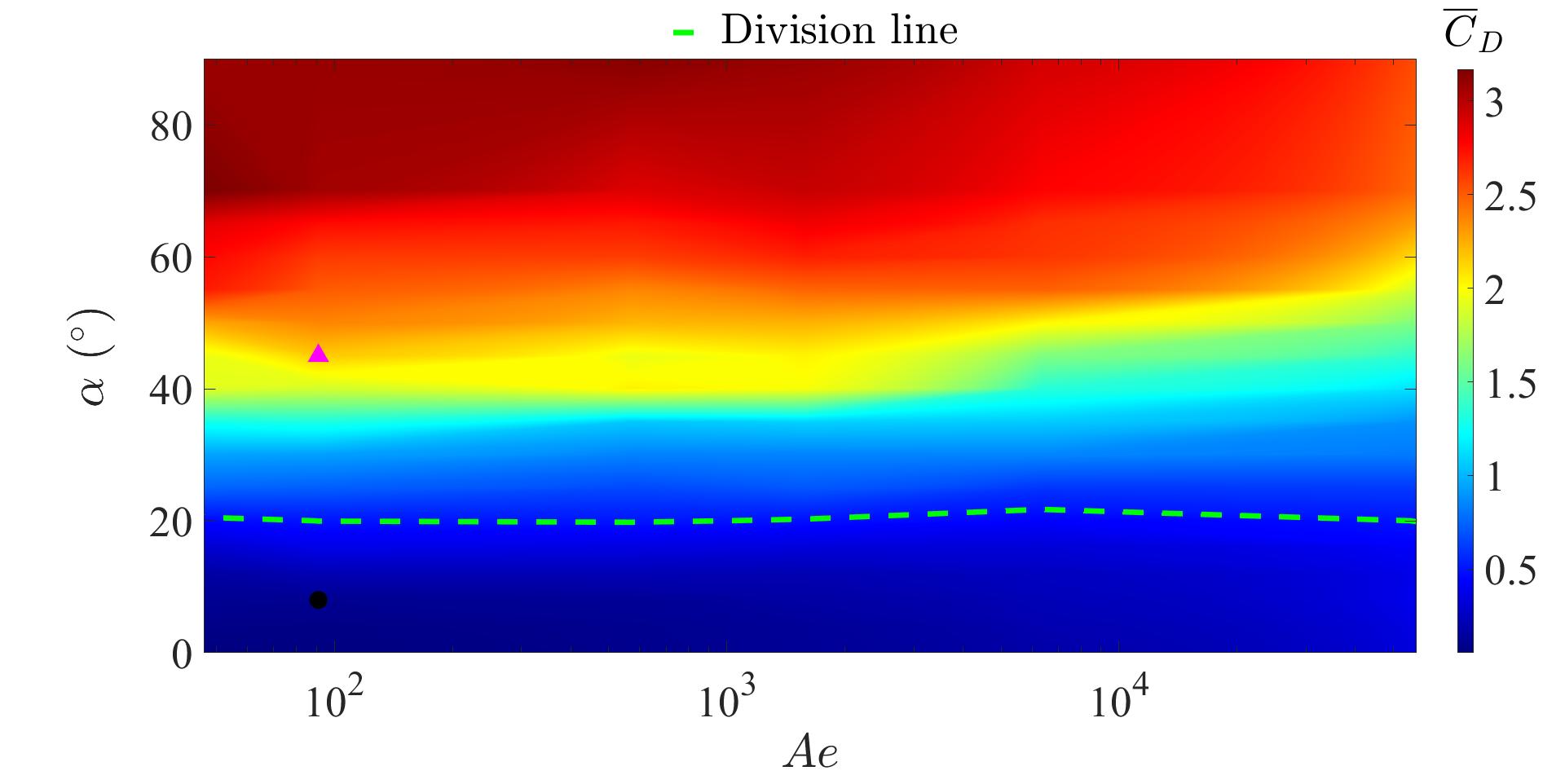}\label{fig:Ae_effectb}}
	\\
	\subfloat[][Time-averaged lift-to-drag ratio]{\includegraphics[width=0.5\textwidth]{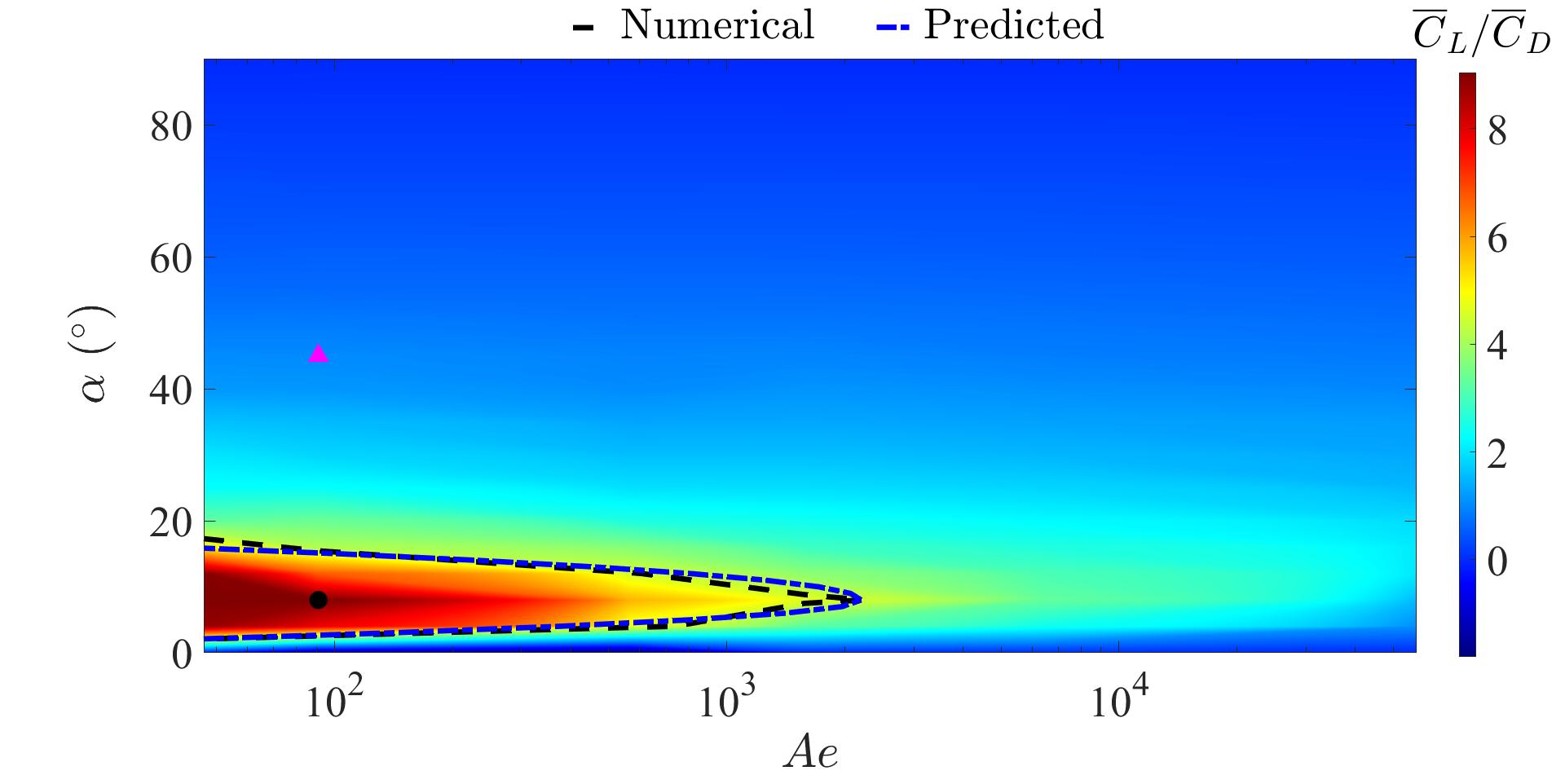}\label{fig:Ae_effectc}}
	\subfloat[][Time-averaged maximum displacement normal to membrane chord]{\includegraphics[width=0.5\textwidth]{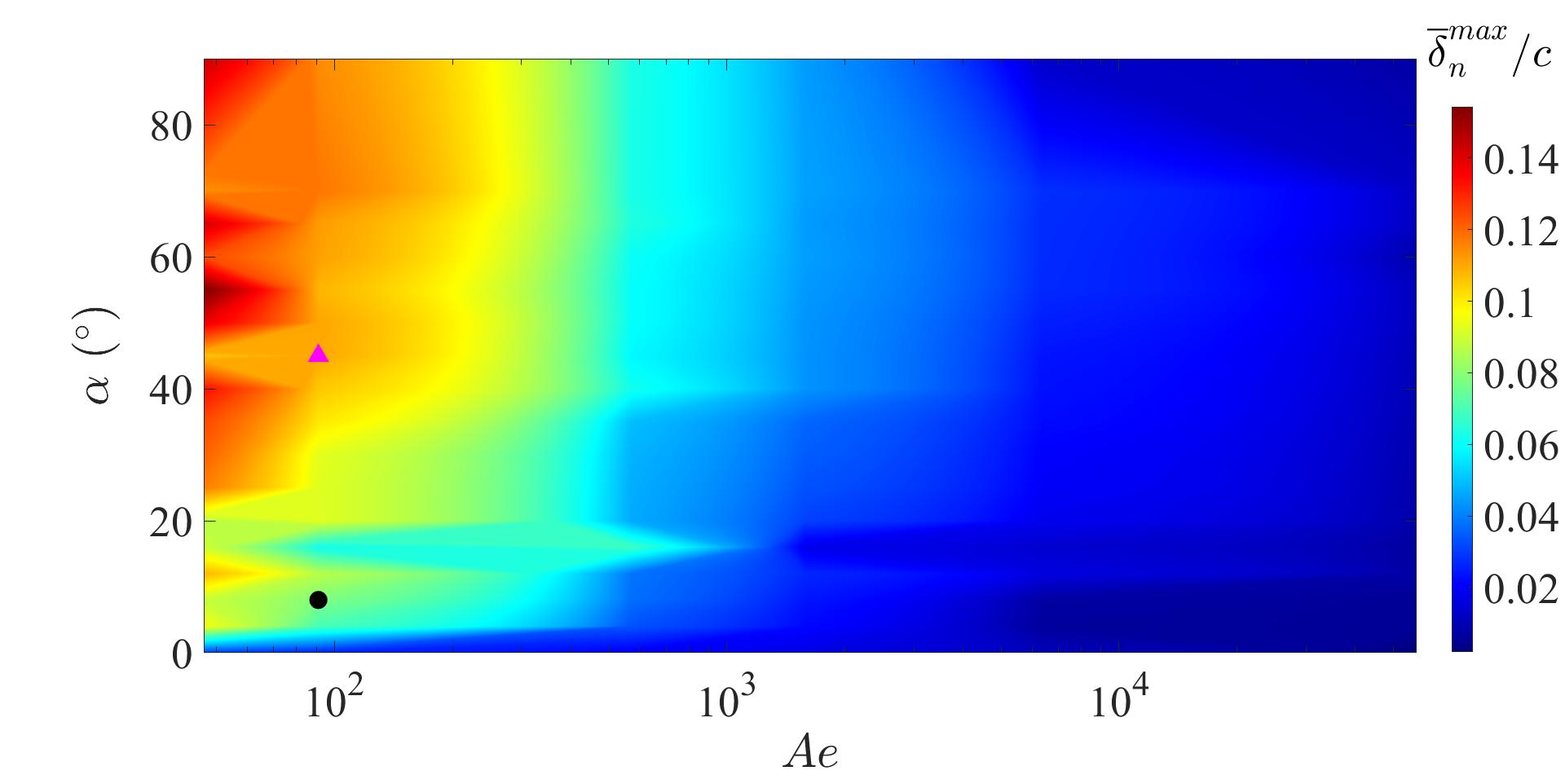}\label{fig:Ae_effectd}}
	\caption{\label{fig:Ae_effect}  Comparison of time-averaged flexible membrane dynamics in the parameter space of $Ae$-$\alpha$.}
\end{figure}

We perform a series of numerical simulations in the parameter space of $Ae$-$\alpha$ to investigate the coupled membrane dynamics over a wide range of parameter space. Six groups of aeroelastic numbers within [46.46, 57336.77] and seventeen angles of attack ranging from $0^\circ$ to $90^\circ$ are chosen to form the parameter space. \refFig{fig:Ae_effect} summarizes the time-averaged aerodynamic forces and membrane displacements in the studied parameter space. The triangle label in magenta color represents the parameter combination with the largest lift coefficient in the selected parameter space. The circle in black color denotes the parameter combination with the highest lift-to-drag ratio in the parameter space.

\refFig{fig:Ae_effect} \subref{fig:Ae_effecta} shows the mean lift coefficient distributions of the two-dimensional flexible membrane wings in the studied parameter space. The stall angle with the maximum lift coefficient is observed near 50$^\circ$. For the flexible membrane with a fixed aeroelastic number, the mean lift coefficient increases to its maximum value at the critical stall angle and then decreases rapidly in the post-stall condition. The mean lift coefficient shows an overall downward trend when the aeroelastic number becomes larger. In other words, the lift performance of the membrane wing degrades when the wing becomes more rigid. The overall optimal lift coefficient region is observed at moderate angles of attack and low aeroelastic numbers. The dashed line in magenta color represents a contour line with the same lift coefficient value in the phase diagram based on numerical simulations. From an engineering design point of view, the flexible membrane wing can achieve better lift performance when the parameter combination falls into this envelope.  One can select a specific lift coefficient value to meet the design requirement and obtain the corresponding envelope. We can notice that the top and bottom halves of the envelope are asymmetrical along the critical stall angles. This is mainly caused by the rapid degradation of the lift performance after the stall occurs. We use piecewise functions to fit the envelope. The predicted empirical solution of the envelope indicated by the blue dashed dot line that connects the angle of attack with the aeroelastic number is given as
\begin{equation}
	Ae=\left\{
	\begin{aligned}
		a_0  \mathrm{exp}^{-\left(\frac{\alpha-a_1}{a_2}\right)^2} \quad \alpha < \alpha_{\rm{stall}}\\
		a_0  \mathrm{exp}^{-\left(\frac{\alpha-a_1}{a_3}\right)^2} \quad \alpha \ge \alpha_{\rm{stall}}\\
	\end{aligned}
	\right
	.
\end{equation}
where $a_0$, $a_1$, $a_2$ and $a_3$ are the coefficients of the piecewise functions and $\alpha_{\rm{stall}}$ denotes the stall angle. It can be seen from the expressions of the empirical solutions that the only difference is the coefficients $a_2$ and $a_3$ in the denominator of the exponent. In \reffig{fig:Ae_effect} \subref{fig:Ae_effecta}, the value of the contour line in magenta color is selected as 1.25. Thus, the coefficients are determined as $a_0=48880$, $a_1=52$, $a_2=14.24$ and $a_3=7.75$ by using the curve fitting.

\refFig{fig:Ae_effect} \subref{fig:Ae_effectb} exhibits the mean drag coefficient distributions in the parameter space. The membrane wing achieves more drag penalty at higher angles of attack. The green dashed line is nearly a straight line around $20^\circ$, which means the mean drag coefficient is roughly the same at $20^\circ$ for all aeroelastic numbers. This line divides the drag coefficient phase diagram into two parts. In the lower part within $\alpha \in [0^\circ, 20^\circ)$, the mean drag coefficient becomes larger when the membrane wing becomes more rigid. In the upper part within $\alpha \in [20^\circ, 90^\circ]$, flexible wings can achieve drag enhancement than rigid wings.

\refFig{fig:Ae_effect} \subref{fig:Ae_effectc} is the mean lift-to-drag ratio distribution in the parameter space. The maximum mean lift-to-drag ratio is observed at $8^\circ$, which is indicated by a black-filled circle. The mean lift-to-drag ratio shows a rapid upward trend and then quickly drops at higher angles of attack. The membrane wing exhibits overall better flight efficiency for flexible conditions. Similar to the plot in \reffig{fig:Ae_effect} \subref{fig:Ae_effecta}, a dashed line in black color represents a contour line with a specific lift-to-drag ratio value. One can choose the parameter combination of the angle of attack and the aeroelastic number within this envelope. It can ensure that the flight performance is better than the desired value when designing flexible membrane wings. The black dashed dot line denotes the empirical solutions of the envelope by using exponent curve fitting. The expressions of the empirical solutions are given as
\begin{equation}
	Ae=\left\{
	\begin{aligned}
		b_0 \mathrm{exp}^{-\left(\frac{\alpha-b_1}{b_2}\right)^2} \quad \alpha < \alpha_{\rm{optimal}}\\
		b_0 \mathrm{exp}^{-\left(\frac{\alpha-b_1}{b_3}\right)^2} \quad \alpha \ge \alpha_{\rm{optimal}}\\
	\end{aligned}
	\right
	.
\end{equation}
where $b_0$, $b_1$, $b_2$ and $b_3$ are the coefficients of the piecewise functions. $\alpha_{\rm{optimal}}$ represents the angle of attack with the maximum mean lift-to-drag ratio. The value of the contour line shown in \reffig{fig:Ae_effect} \subref{fig:Ae_effectc} is selected as 4.5. Thus, the coefficients are given as $b_0=2200$, $b_1=8$, $b_2=3$ and $b_3=4$, respectively.

In \reffig{fig:Ae_effect} \subref{fig:Ae_effectd}, the nondimensional maximum mean membrane displacements normal to the chord are displayed in the parameter space of $Ae$-$\alpha$. We can observe that the membrane shows larger displacements at higher angles of attack. Increasing rigidity can suppress membrane deformation. 

By comparing the lift coefficient in \reffig{fig:Ae_effect} \subref{fig:Ae_effecta} and the membrane deformation in \reffig{fig:Ae_effect} \subref{fig:Ae_effectd}, we observe that the improvement of the lift coefficient is positively correlated with larger membrane displacements within the envelope. When the angle of attack exceeds $20^\circ$, larger membrane displacements lead to more drag penalty. Thus, there must be some underlying physical connections between the membrane deformation and the aerodynamic performance under different parameter combinations. We will further explore the connections based on the proposed scaling relations in \refse{sec:scaling1}.

\subsection{Effect of Angle of Attack} \label{sec:angle}
As discussed in \refse{sec:dynamics}, the aerodynamic performance and the membrane dynamics are strongly affected by the angle of attack. In this section, the time-averaged membrane dynamics and the unsteady membrane responses in the range of $\alpha \in [4^\circ, 90^\circ]$ at a fixed aeroelastic number of $Ae=100.04$ are investigated. The membrane responses at the selected aeroelastic number include the global optimal lift and lift-to-drag ratio performance conditions. To examine how the flexible membrane responds as the angle of attack changes, we also simulate a rigid flat wing and a rigid cambered wing for comparison purposes. The rigid flat wing has an identical geometry to the undeformed membrane wing. The rigid cambered wing model is constructed based on the mean shape of the flexible membrane immersed in unsteady flows.

\begin{figure}[H]
	\centering
	\subfloat[][Time-averaged lift coefficient]{\includegraphics[width=0.5\textwidth]{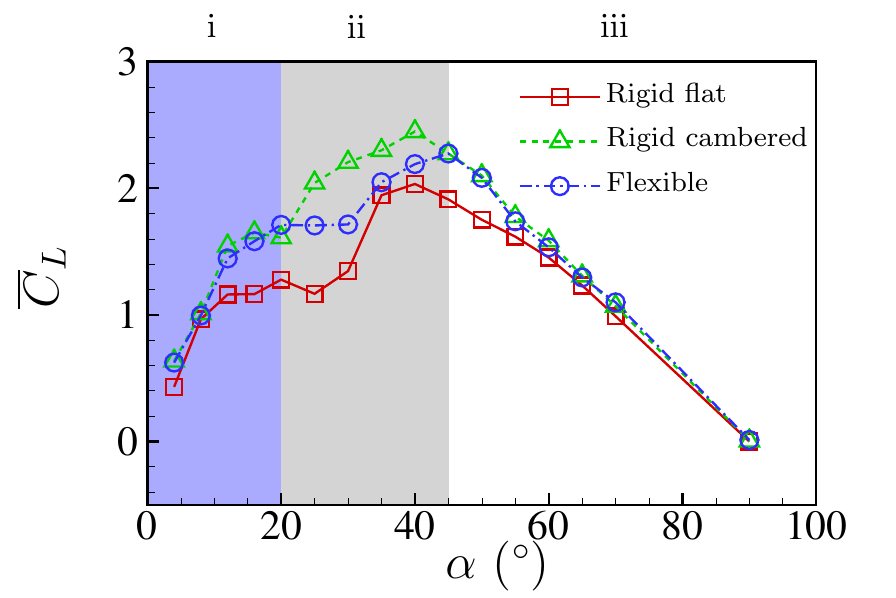}\label{fig:aoa_effecta}}
	\subfloat[][Time-averaged drag coefficient]{\includegraphics[width=0.5\textwidth]{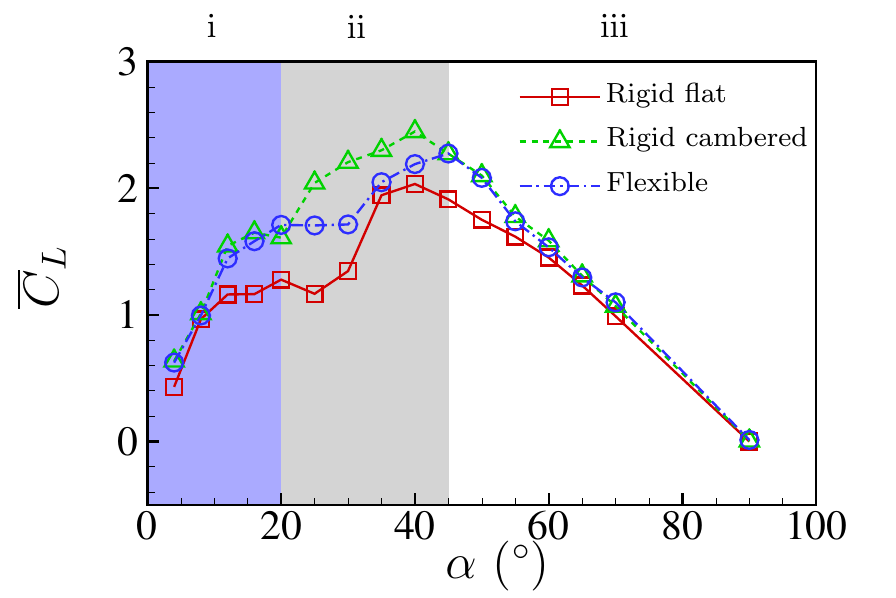}\label{fig:aoa_effectb}}
	\\
	\subfloat[][Time-averaged lift-to-drag ratio]{\includegraphics[width=0.5\textwidth]{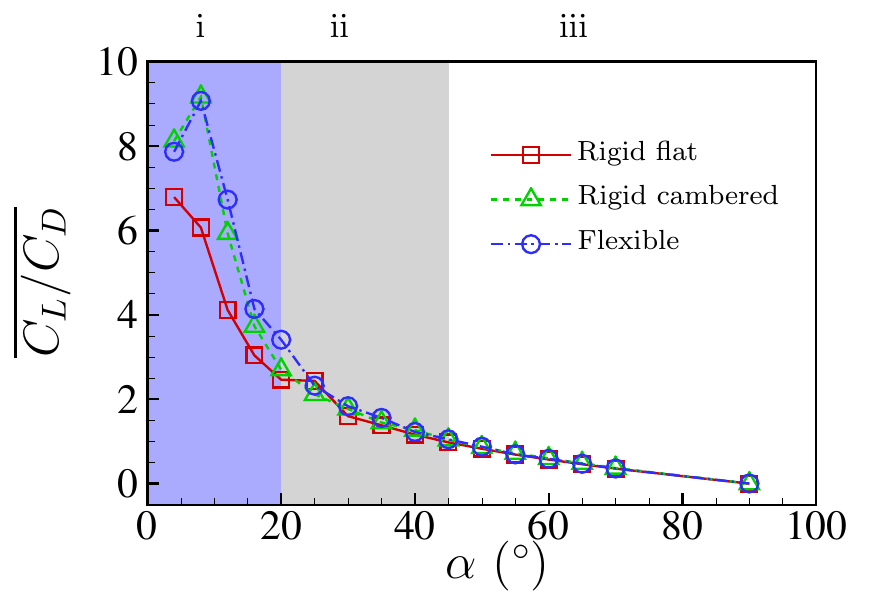}\label{fig:aoa_effectd}}
	\caption{\label{fig:aoa_effect}  Comparison among a rigid flat wing, a rigid cambered wing and a flexible membrane at various angles of attack.}
\end{figure}

\refFig{fig:aoa_effect} presents the comparison of the time-averaged lift coefficient, the time-averaged drag coefficient and the time-averaged lift-to-drag ratio of a rigid flat wing, a rigid cambered wing and a flexible membrane at different angles of attack. Based on the flexible membrane responses, the variation of the aerodynamic characteristics can be categorized into three regimes: (i) pre-stall ($4^\circ \leq \alpha < 20^\circ $), (ii) transitional stall ($20^\circ \leq \alpha < 45^\circ $) and (iii) deep stall ($45^\circ \leq \alpha \leq 90^\circ $). 

It can be seen from \reffig{fig:aoa_effect} \subref{fig:aoa_effecta} that both the rigid cambered wing and the flexible membrane produce overall larger mean lift forces than the rigid flat wing at the examined angle of attack range. The deep stall angle is delayed from $40^\circ$ to $45^\circ$ for the flexible membrane. The flexible membrane produces mean lift forces approaching to those of the rigid cambered wing except for the range of $\alpha \in [20^\circ,45^\circ]$. When the flow-induced vibration is taken into account, the flexible membrane has smaller mean lift forces than the rigid cambered wing in this range.

As presented in \reffig{fig:aoa_effect} \subref{fig:aoa_effectb}, both the rigid cambered wings and the flexible membrane wings generate smaller drag forces than the rigid flat wing at small angles of attack. The mean drag produced by the rigid cambered wing becomes larger than that of the rigid flat wing when the angle of attack is larger than $16^\circ$. The mean drag of the flexible membrane exceeds that of the rigid flat wing when the flexible membrane enters into the deep stall conditions at $\alpha=45^\circ$. The rigid cambered wing produces more mean drag forces than the flexible membrane within the transitional range.

It can be observed from \reffig{fig:aoa_effect} \subref{fig:aoa_effectd} that the rigid flat wing shows the lowest mean lift-to-drag ratio. Both the rigid cambered wing and the flexible membrane achieve the high mean lift-to-drag ratio at $\alpha=8^\circ$. Before the deep stall condition, the flexible membrane shows better performance than the cambered wing. The mean lift-to-drag ratio of the flexible membrane becomes slightly smaller than that of the rigid cambered wing when the angle of attack is larger than $45^\circ$.

\begin{figure}[H]
	\centering
	\subfloat[][]{\includegraphics[width=0.5\textwidth]{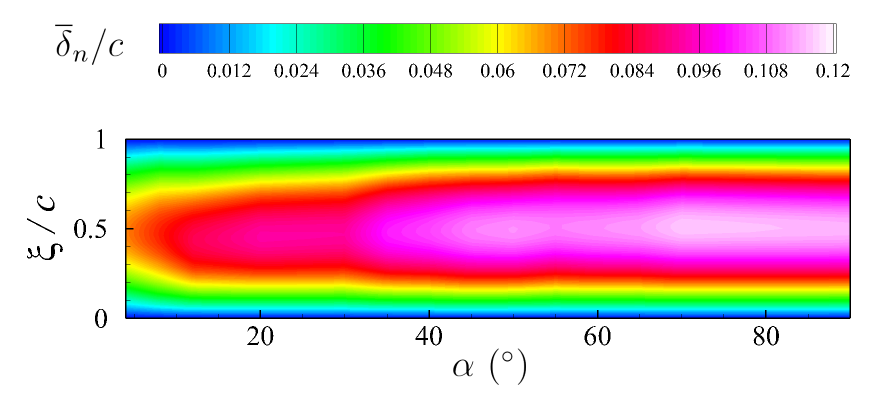}\label{fig:aoa_effect_dis_cpa}}
	\subfloat[][]{\includegraphics[width=0.5\textwidth]{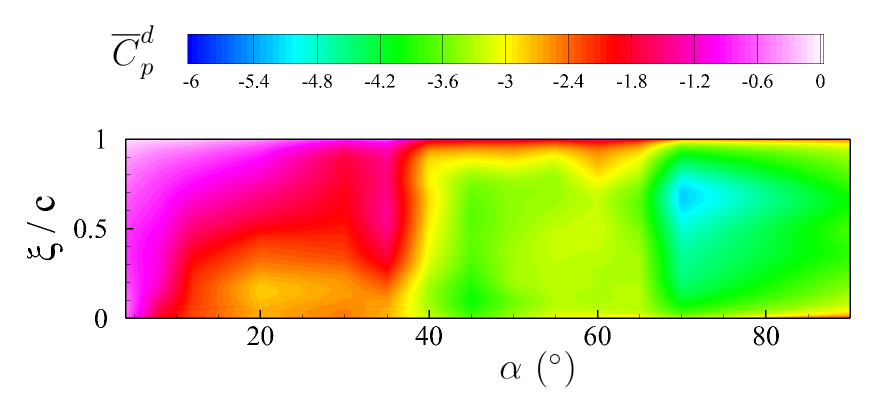}\label{fig:aoa_effect_dis_cpb}}
    \\
    \subfloat[][]{\includegraphics[width=0.5\textwidth]{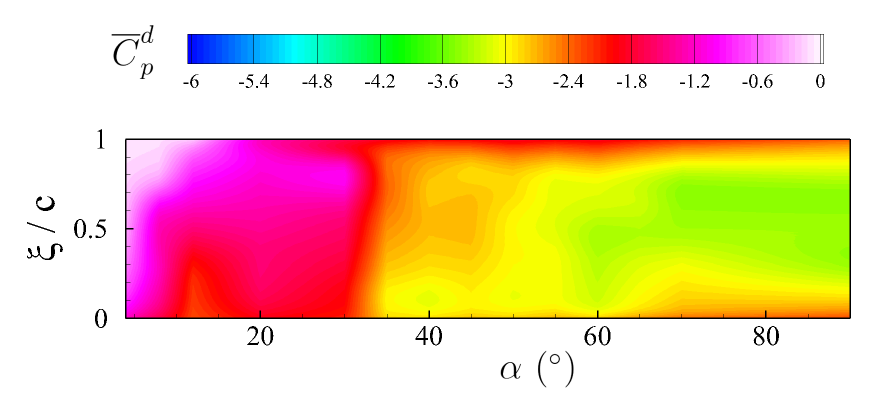}\label{fig:aoa_effect_dis_cpc}}
	\subfloat[][]{\includegraphics[width=0.5\textwidth]{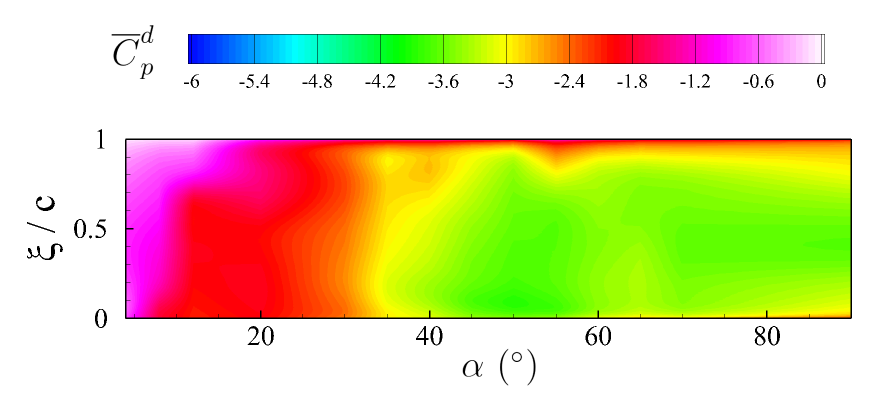}\label{fig:aoa_effect_dis_cpd}}
	\caption{\label{fig:aoa_effect_dis_cp}  Contour of time-averaged  (a) membrane displacement normal to the chord of the flexible membrane; pressure coefficient difference between the upper and lower surfaces for (b)  flexible membrane, (c)  rigid flat wing and (d) rigid cambered wing.}
\end{figure}

We further examine the mean membrane responses for all angles of attack. \refFigs{fig:aoa_effect_dis_cp} \subref{fig:aoa_effect_dis_cpa} and \subref{fig:aoa_effect_dis_cpb} plot the contour of the time-averaged membrane displacement normal to the chord and the time-averaged pressure coefficient difference between the upper and lower surfaces along the membrane chord. It can be seen that the maximum membrane displacement increases quickly when the angle of attack changes from $4^\circ$ to $45^\circ$. As the angle of attack further increases, the variation of the mean membrane displacement becomes somewhat saturated. The pressure coefficient difference is directly related to the normal force. The negative pressure coefficient difference means a suction force. We notice that the large suction area expands from the trailing edge to the leading edge significantly as the angle of attack increases from $4^\circ$ to $45^\circ$. Then, the pressure coefficient difference distribution exhibits a slight change as the angle of attack further increases to the vertical state.

\begin{figure}[H]
	\centering
	\subfloat[][Lift coefficient]{\includegraphics[width=0.5\textwidth]{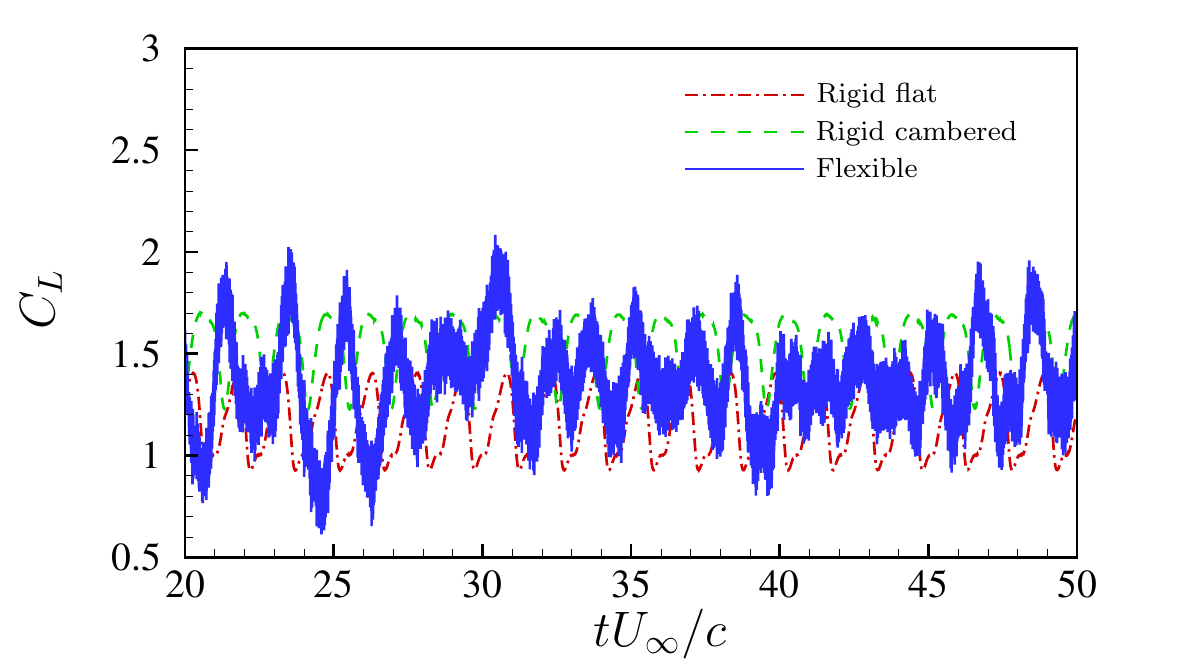}\label{fig:force_12a}}
	\subfloat[][Drag coefficient]{\includegraphics[width=0.5\textwidth]{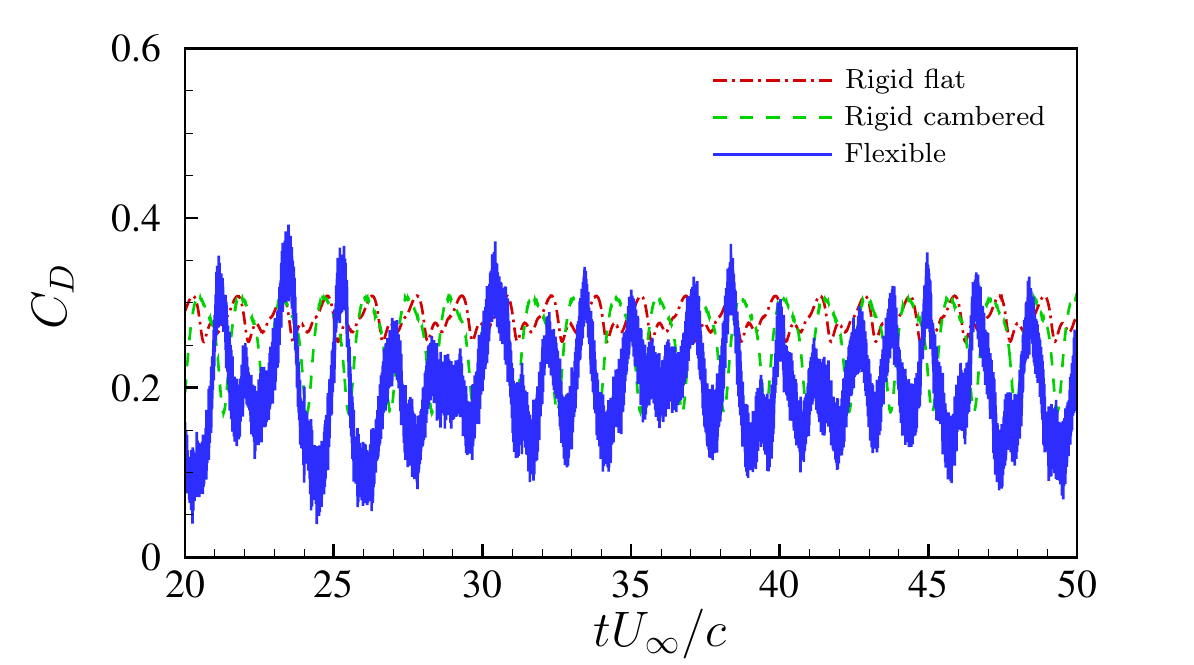}\label{fig:force_12b}} 
    \\
    \subfloat[][Maximum displacement]{\includegraphics[width=0.5\textwidth]{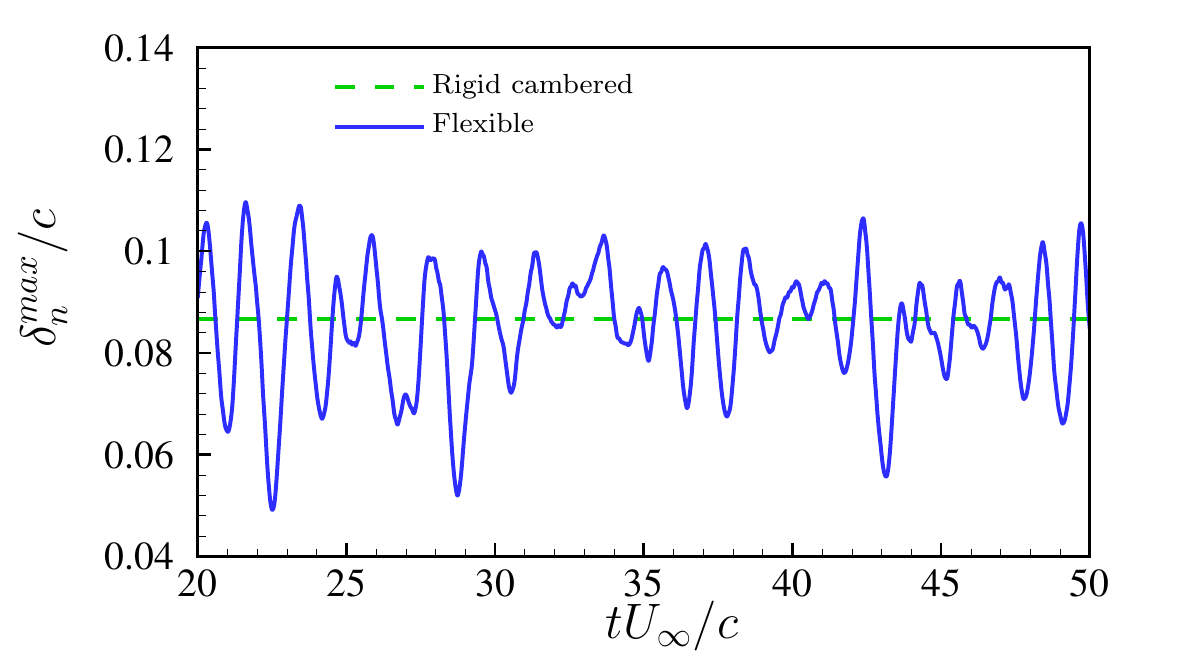}\label{fig:force_12c}}
    \quad \quad
	\subfloat[][Deflection envelope]{\includegraphics[width=0.44\textwidth]{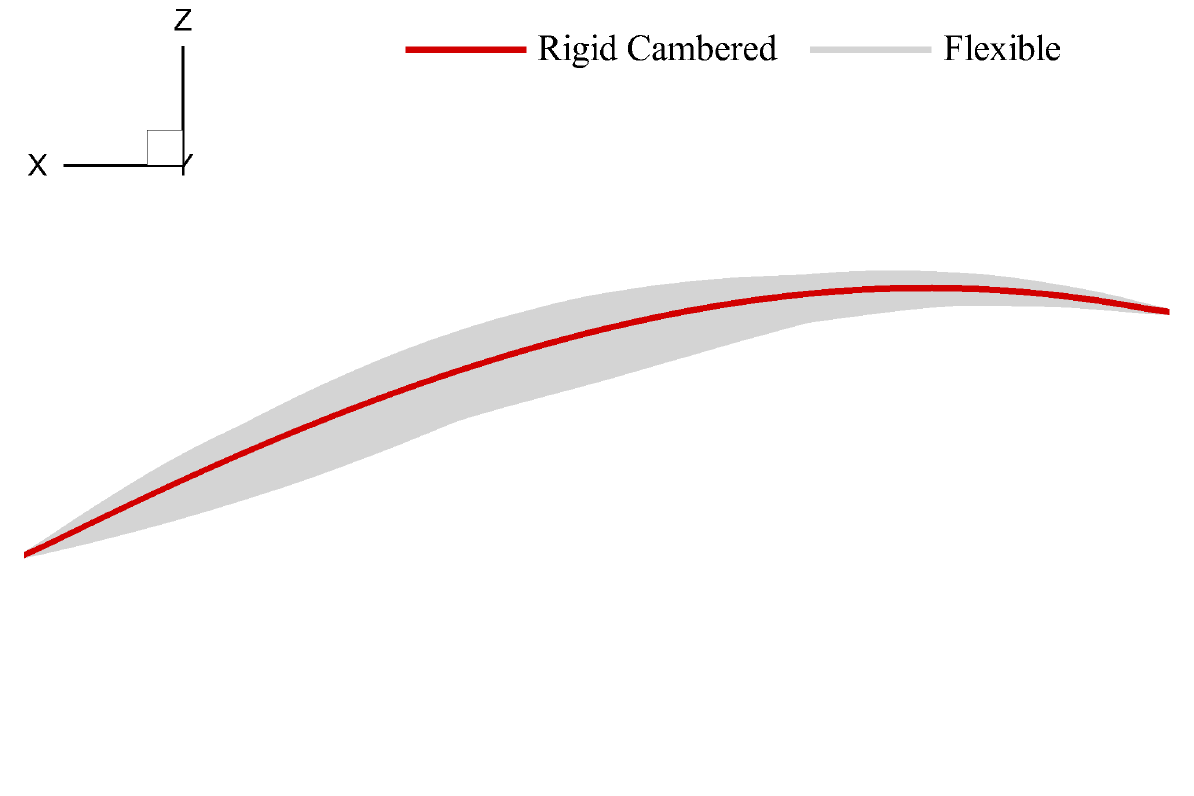}\label{fig:force_12d}} 
	\caption{\label{fig:force_12}  Comparison among a rigid flat wing, a rigid cambered wing and a flexible membrane at $\alpha=12^\circ$ for time-dependent aerodynamic forces and membrane deflection.}
\end{figure}

A comparison of the time-averaged pressure coefficient difference between the rigid flat wing and the rigid cambered wing is shown in \reffigs{fig:aoa_effect_dis_cp} \subref{fig:aoa_effect_dis_cpc} and \subref{fig:aoa_effect_dis_cpd}. It can be seen that the rigid flat wing produces the smallest suction force among the three wings. The rigid cambered wing generates larger suction forces on the wing surface within the transitional stall regime. Similar time-averaged pressure difference distributions between the membrane wing and its rigid cambered counterpart can be found in the pre-stall and deep stall regimes.

\begin{figure}[H]
	\centering
	\subfloat[][Rigid flat wing]{
		\includegraphics[width=0.25\textwidth]{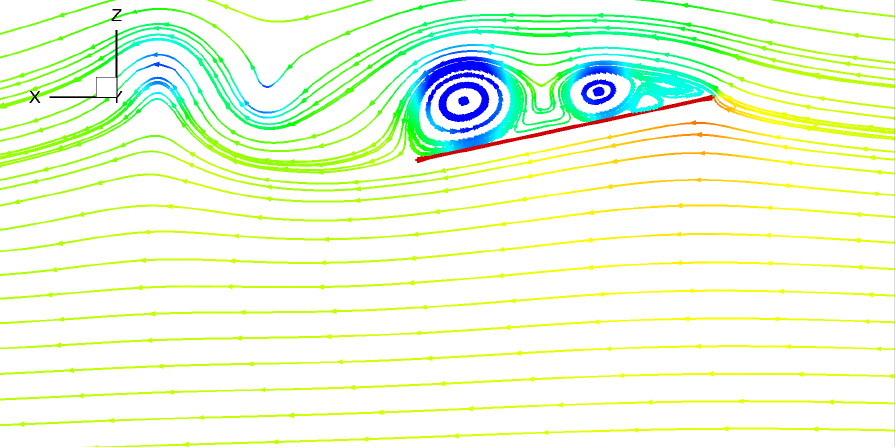}
		\includegraphics[width=0.25\textwidth]{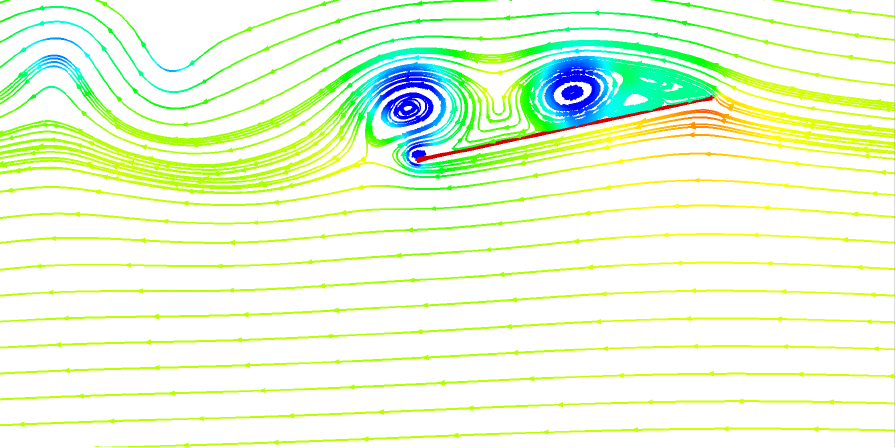}
		\includegraphics[width=0.25\textwidth]{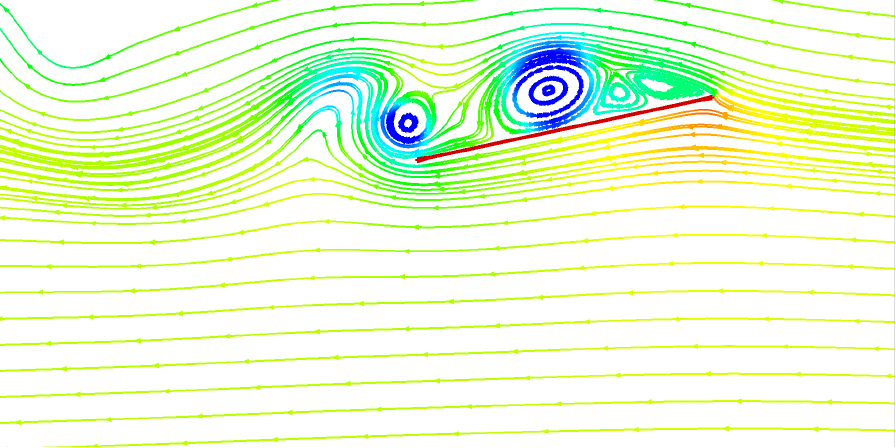}
		\includegraphics[width=0.25\textwidth]{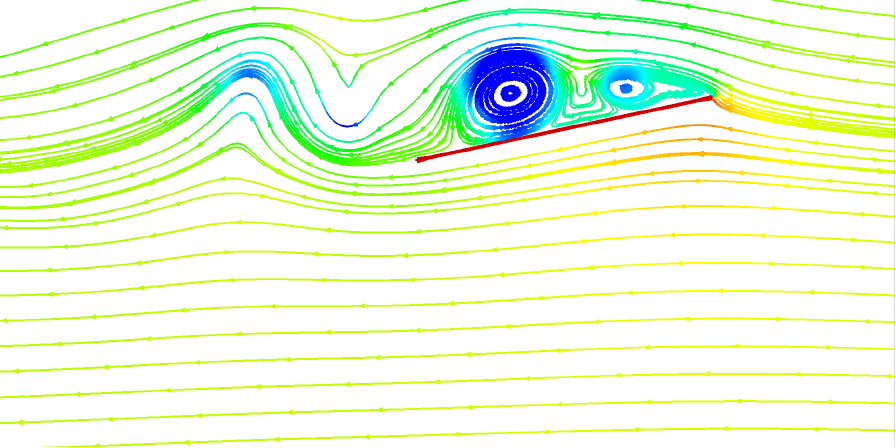}}
	\\
	\subfloat[][Rigid cambered wing]{
		\includegraphics[width=0.25\textwidth]{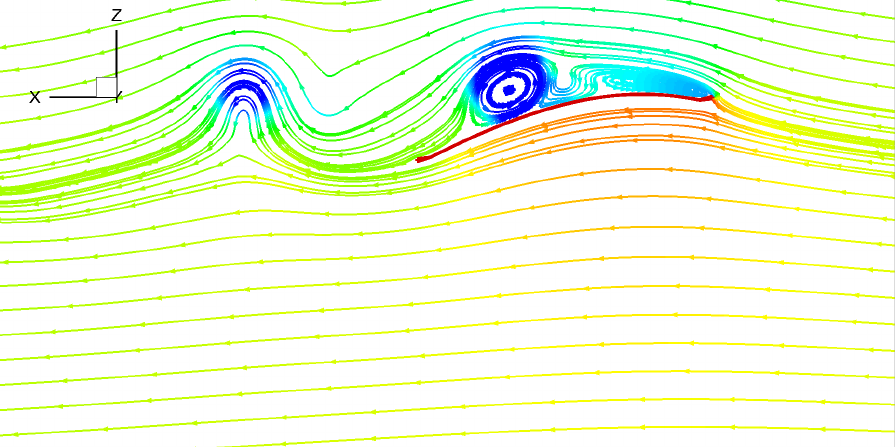}
		\includegraphics[width=0.25\textwidth]{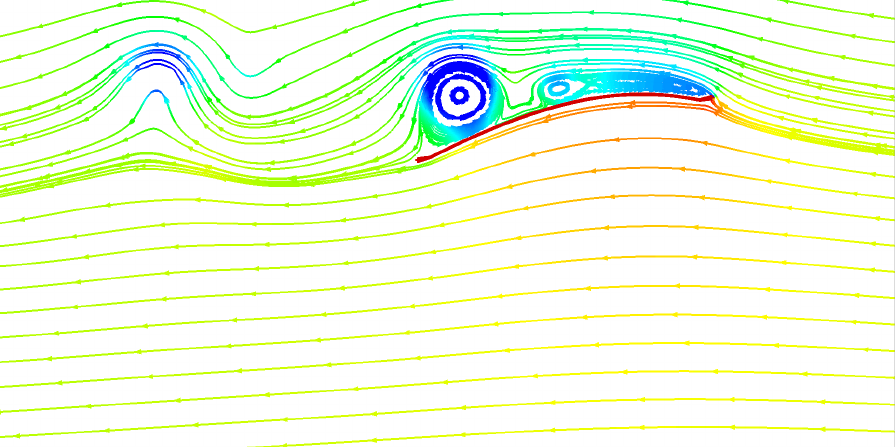}
		\includegraphics[width=0.25\textwidth]{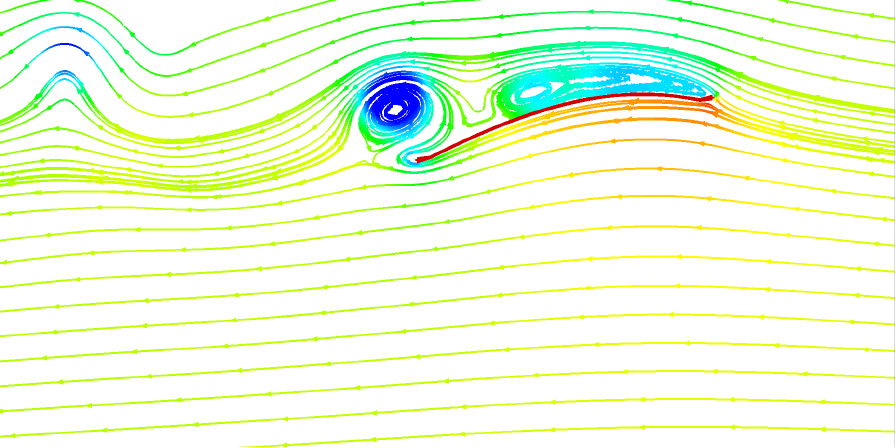}
		\includegraphics[width=0.25\textwidth]{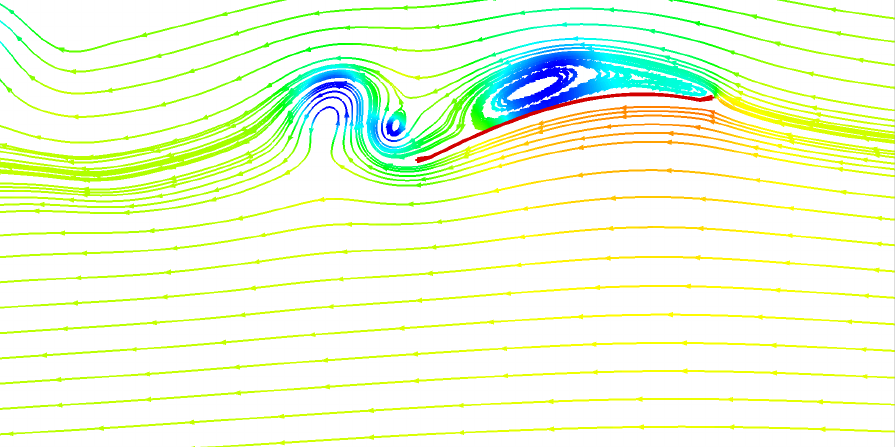}}
	\\
	\subfloat[][Flexible membrane wing]{
		\includegraphics[width=0.25\textwidth]{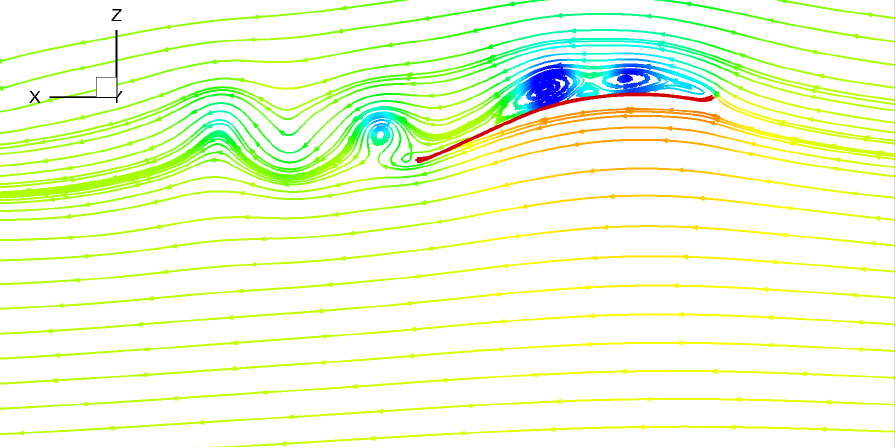}
		\includegraphics[width=0.25\textwidth]{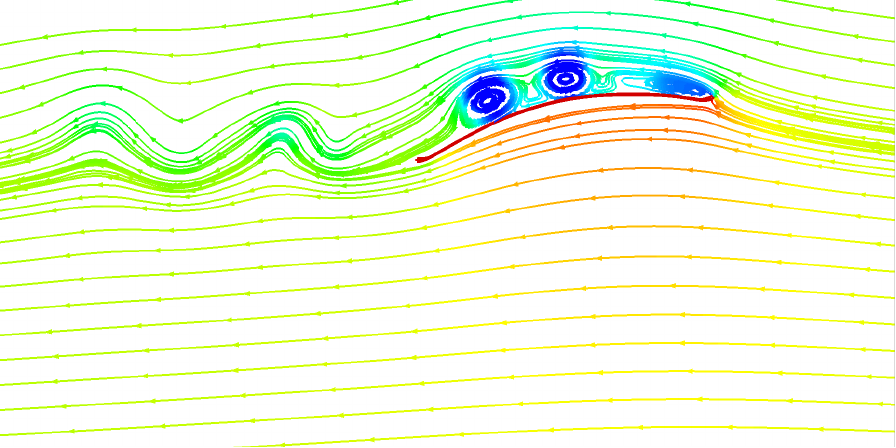}
		\includegraphics[width=0.25\textwidth]{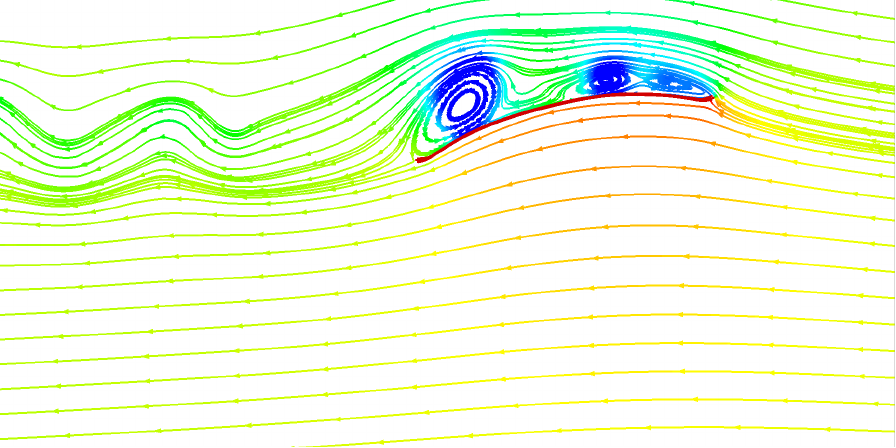}
		\includegraphics[width=0.25\textwidth]{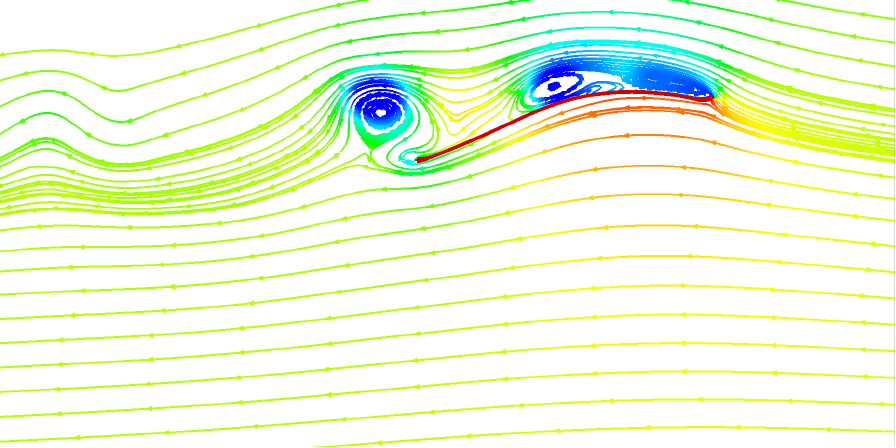}}
	\\
	\includegraphics[width=0.5\textwidth]{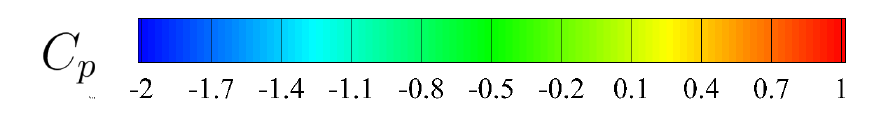}
	\caption{\label{fig:streamline_com_12}  Instantaneous streamlines colored by pressure coefficient at $\alpha$=$12^\circ$.}
\end{figure}

We select three typical angles of attack corresponding to the three regimes. The effect of the angle of attack and the role of flexibility is explored. Here, we consider $\alpha=12^\circ$, $30^\circ$ and $55^\circ$ for a comparison purpose. \refFigs{fig:force_12} \subref{fig:force_12a} and \subref{fig:force_12b} present a comparison of the time-dependent aerodynamic forces between the rigid flat wing, the rigid cambered wing and the flexible membrane at $\alpha=12^\circ$. The rigid cambered wing and the flexible membrane exhibit overall larger lift coefficients than the rigid flat wing. The drag force of the rigid flat wing shows the largest mean value but the smallest amplitude. When flexibility is introduced to the wing, the membrane produces much larger fluctuation amplitudes for both lift and drag forces. \refFigs{fig:force_12} \subref{fig:force_12c} and \subref{fig:force_12d} show the time histories of the maximum displacement and the deflection envelope of the flexible membrane. The camber and the shape of its rigid cambered counterpart are added to the plots as a baseline. The membrane vibration exhibits a synchronization with the variation of the lift force. The difference in the aerodynamic forces between the flexible membrane and its rigid cambered counterpart is caused by the flow-induced vibration around the mean camber. It can be seen from the instantaneous streamlines in \reffig{fig:streamline_com_12} that the separated flows are suppressed by the wing camber. The shedding vortices are coupled with the flexible membrane while yielding lift improvement, drag reduction and higher fluctuations.

\begin{figure}[H]
	\centering
	\subfloat[][Lift coefficient]{\includegraphics[width=0.5\textwidth]{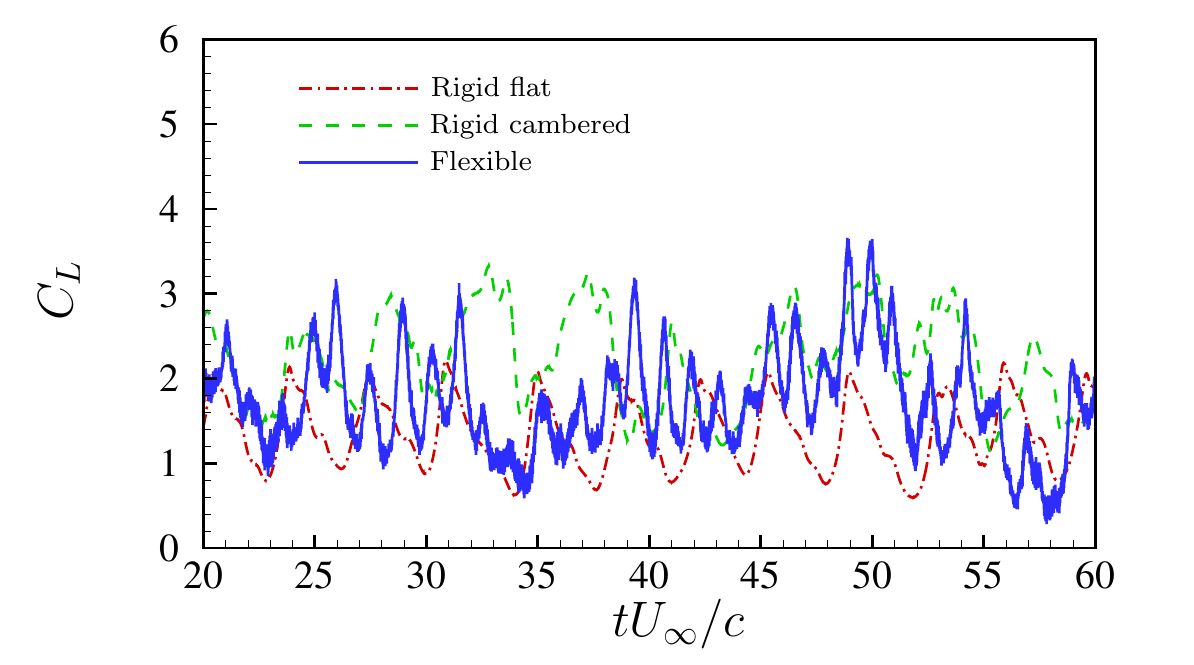}\label{fig:force_30a}}
	\subfloat[][Drag coefficient]{\includegraphics[width=0.5\textwidth]{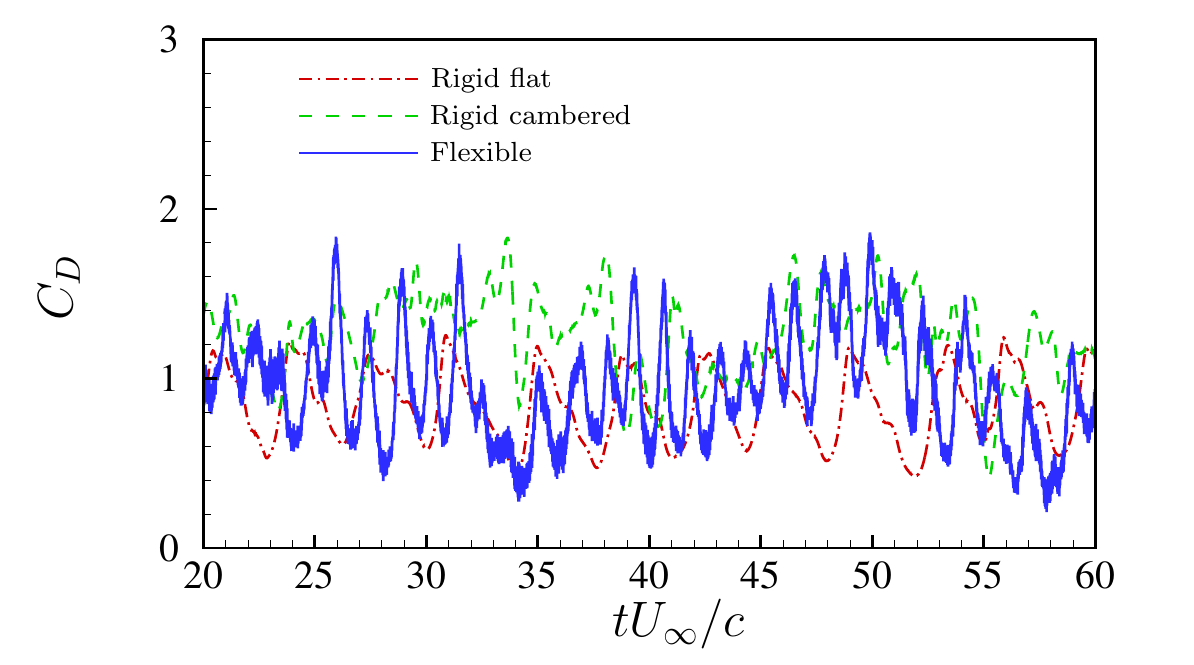}\label{fig:force_30b}}
	\\
    \subfloat[][Maximum displacement]{\includegraphics[width=0.5\textwidth]{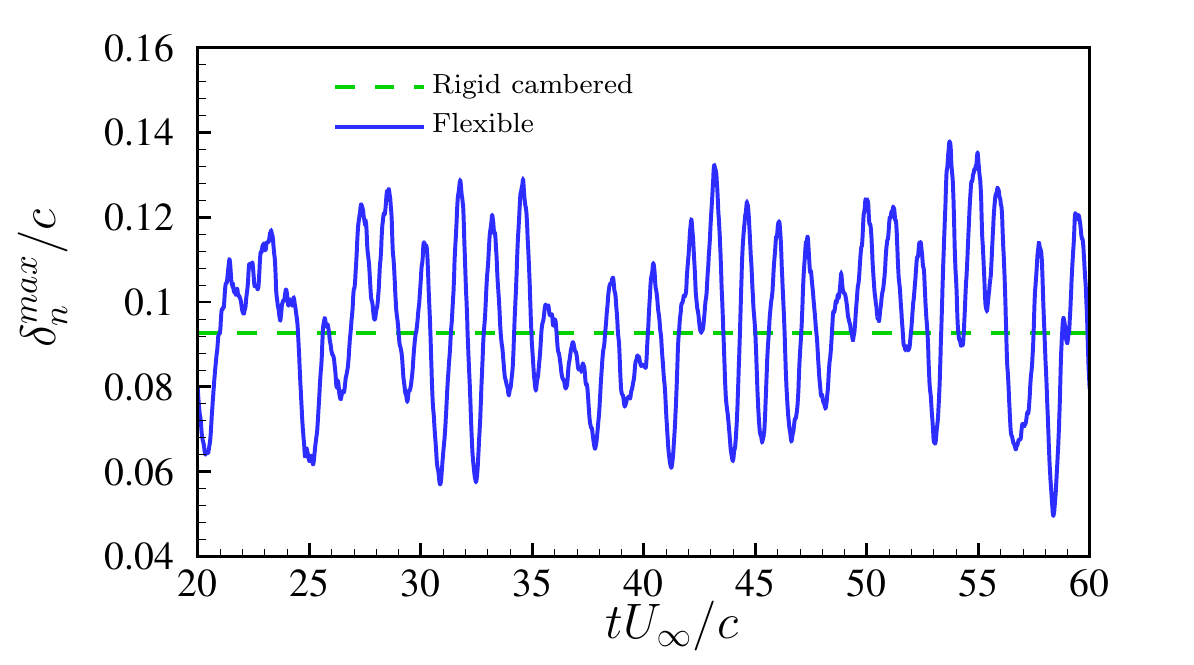}\label{fig:force_30c}}
    \quad \quad
	\subfloat[][Deflection envelope]{\includegraphics[width=0.44\textwidth]{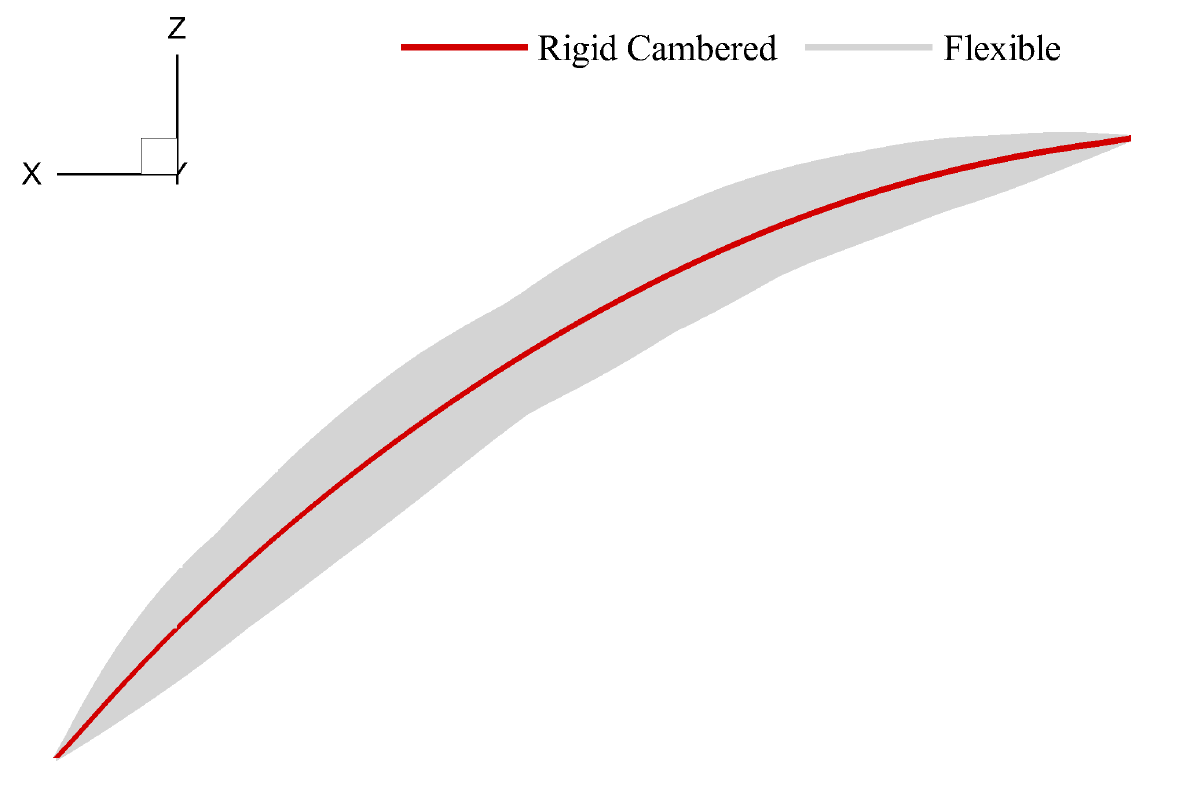}\label{fig:force_30d}} 
	\caption{\label{fig:force_30}  Comparison among a rigid flat wing, a rigid cambered wing and a flexible membrane at $\alpha=30^\circ$ for time-dependent aerodynamic forces and membrane deflection.}
\end{figure}

\begin{figure}[H]
	\centering
	\subfloat[][Rigid flat wing]{
		\includegraphics[width=0.25\textwidth]{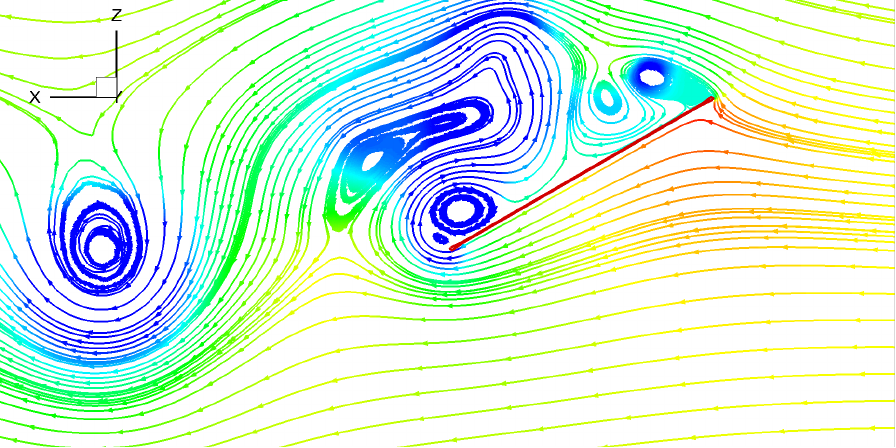}
		\includegraphics[width=0.25\textwidth]{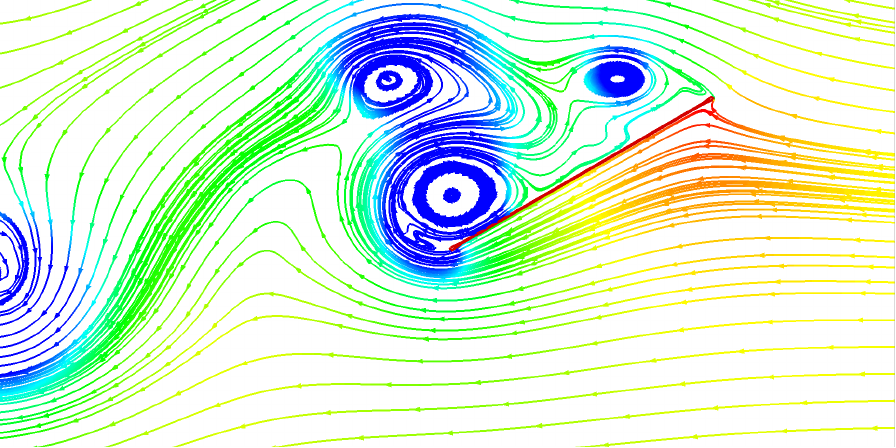}
		\includegraphics[width=0.25\textwidth]{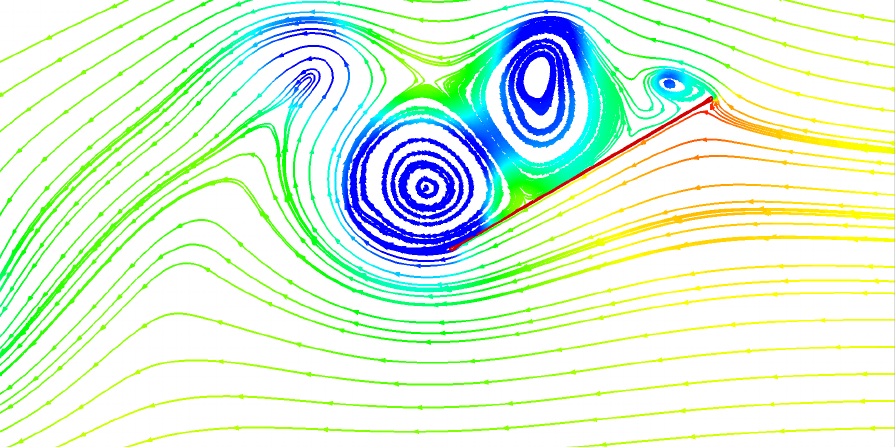}
		\includegraphics[width=0.25\textwidth]{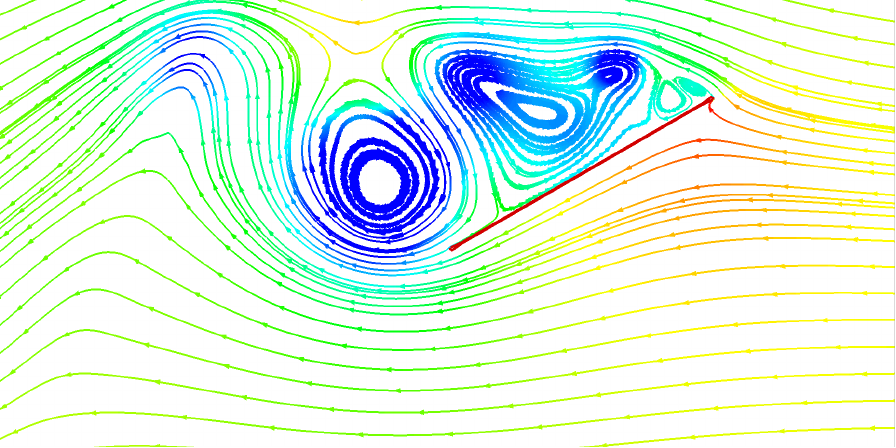}}
	\\
	\subfloat[][Rigid cambered wing]{
		\includegraphics[width=0.25\textwidth]{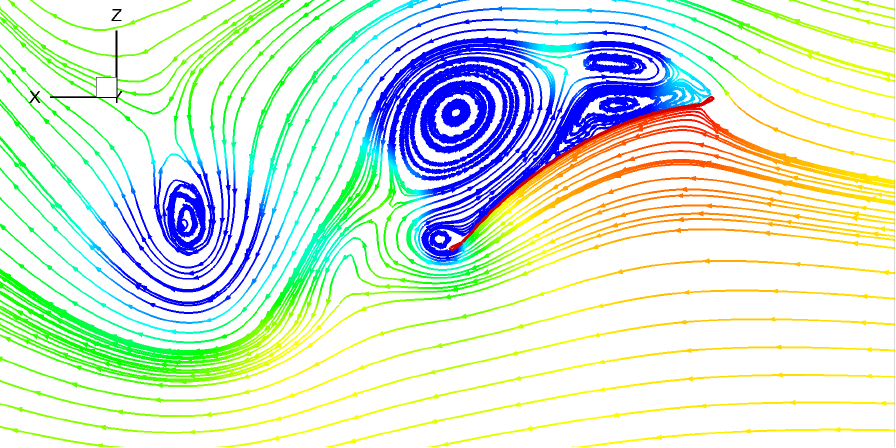}
		\includegraphics[width=0.25\textwidth]{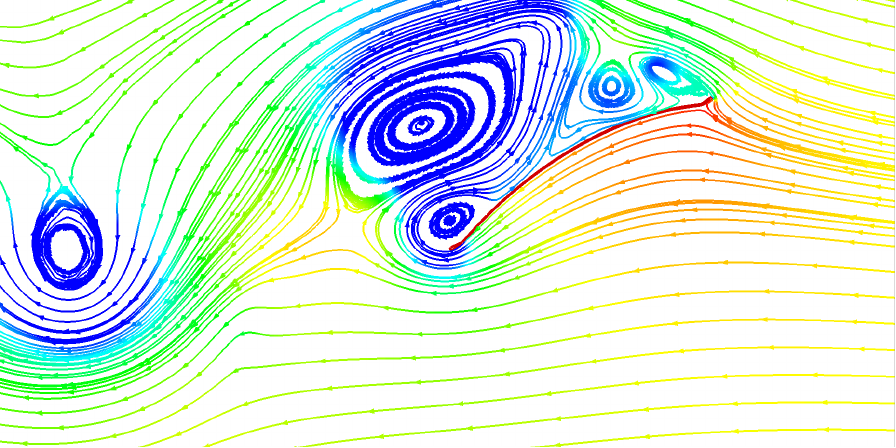}
		\includegraphics[width=0.25\textwidth]{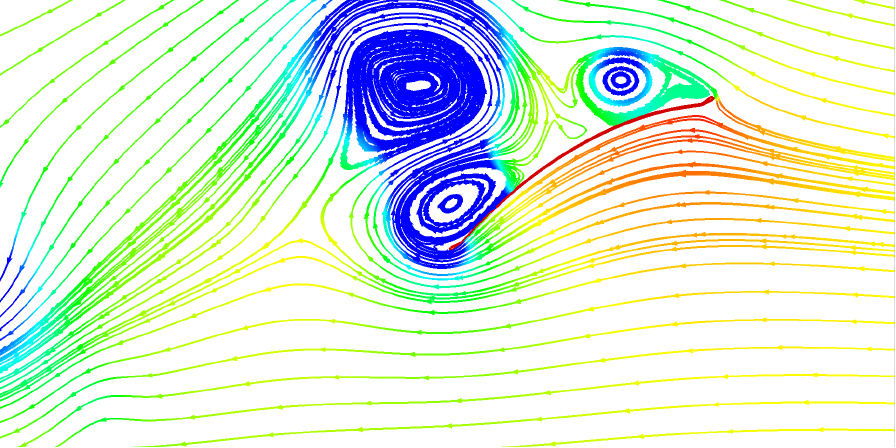}
		\includegraphics[width=0.25\textwidth]{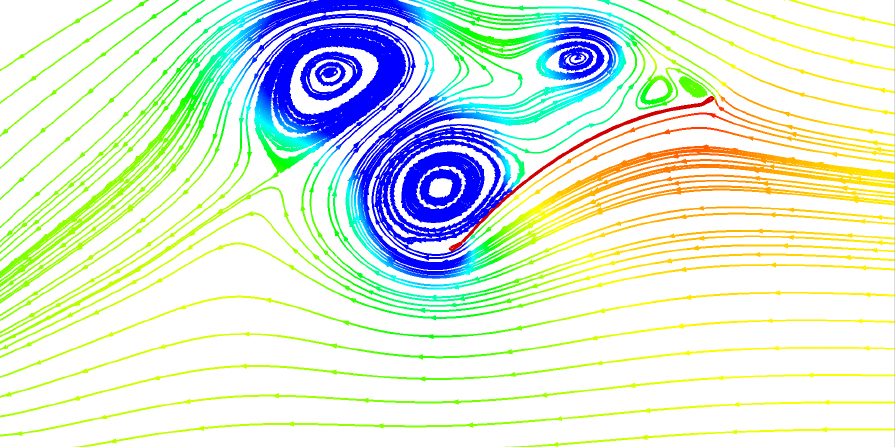}}
	\\
	\subfloat[][Flexible membrane wing]{
		\includegraphics[width=0.25\textwidth]{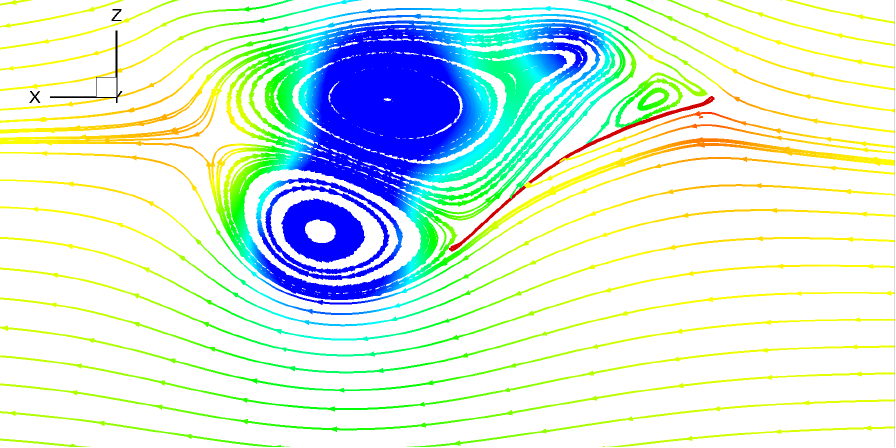}
		\includegraphics[width=0.25\textwidth]{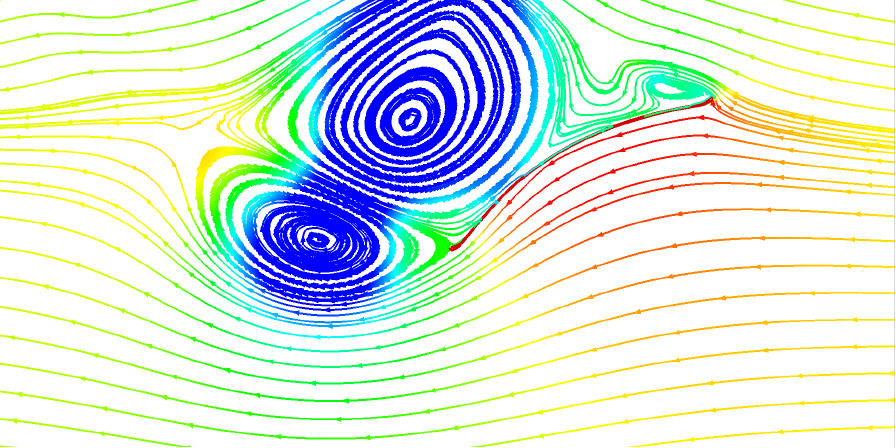}
		\includegraphics[width=0.25\textwidth]{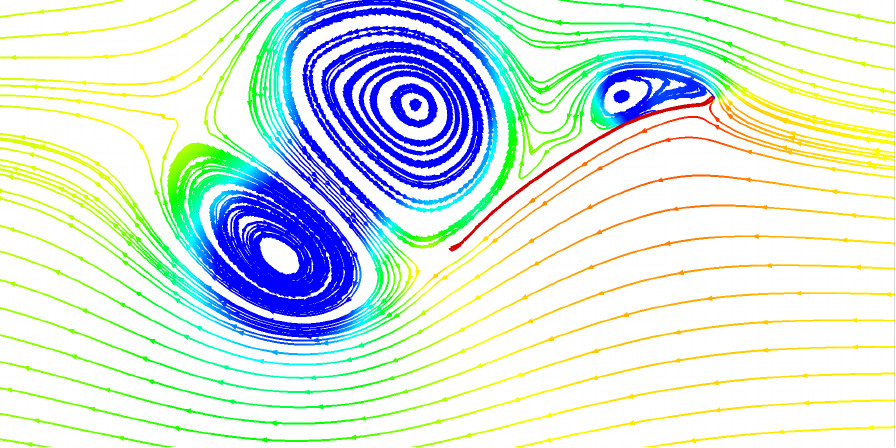}
		\includegraphics[width=0.25\textwidth]{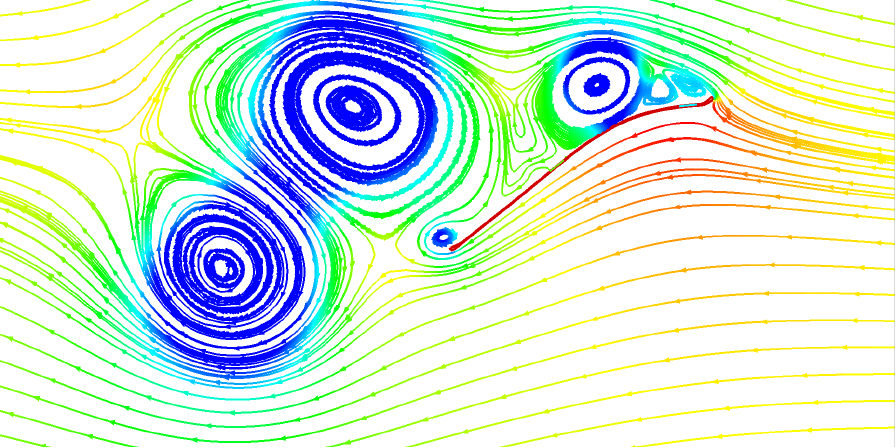}}
	\\
	\includegraphics[width=0.5\textwidth]{./streamline_leng.pdf}
	\caption{\label{fig:streamline_com_30}  Instantaneous streamlines colored by pressure coefficient at $\alpha$=$30^\circ$.}
\end{figure}

\refFigs{fig:force_30} \subref{fig:force_30a} and \subref{fig:force_30b} present the comparisons of the time-dependent aerodynamic forces for three wings. The rigid flat wing produces the overall lowest lift and drag forces. When the wing camber is considered, the lift and drag forces are greatly enhanced. This phenomenon is caused by the leading edge vortices with lower pressure as shown in \reffig{fig:streamline_com_30}. The membrane generates smaller lift and drag forces as well as comparable fluctuation amplitudes at this moderate angle of attack by comparing with its rigid cambered counterpart. \refFigs{fig:force_30} \subref{fig:force_30c} and \subref{fig:force_30d} show the time histories of the maximum displacement and the deflection envelope of the flexible membrane. The shape of the rigid cambered wing with a steady deflection in time is added for comparison purposes. The membrane vibration alleviates the strength of the leading edge vortex and reduces the suction forces on the leeward surface.

\begin{figure}[H]
	\centering
	\subfloat[][Lift coefficient]{\includegraphics[width=0.5\textwidth]{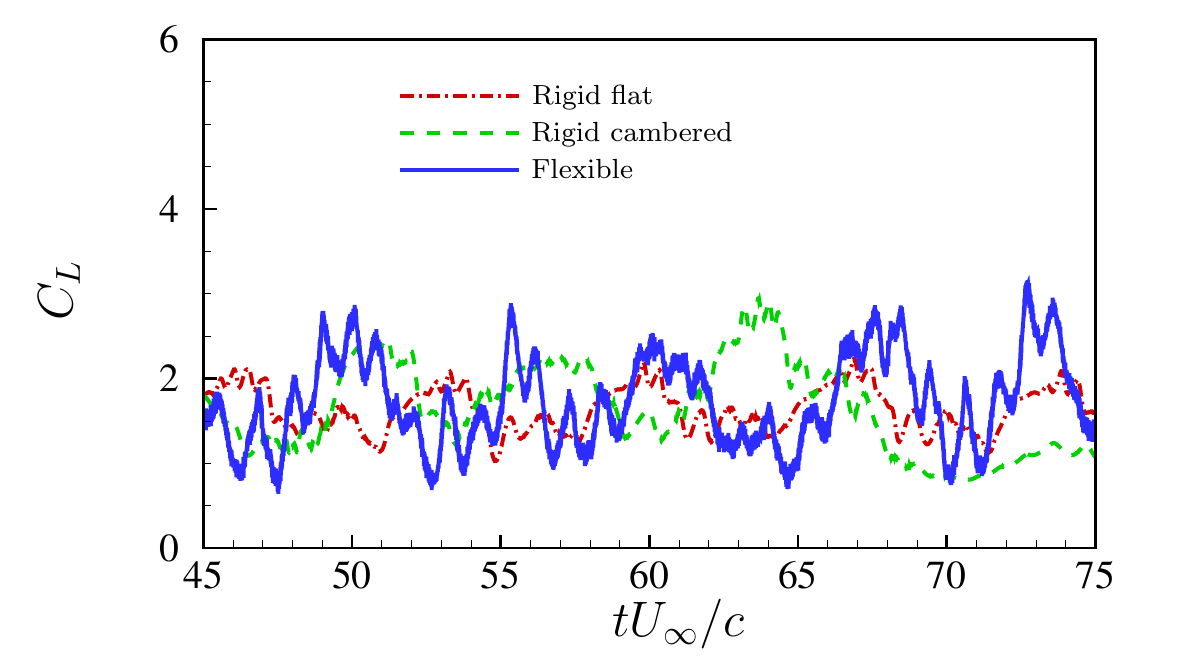}\label{fig:force_55a}}
	\subfloat[][Drag coefficient]{\includegraphics[width=0.5\textwidth]{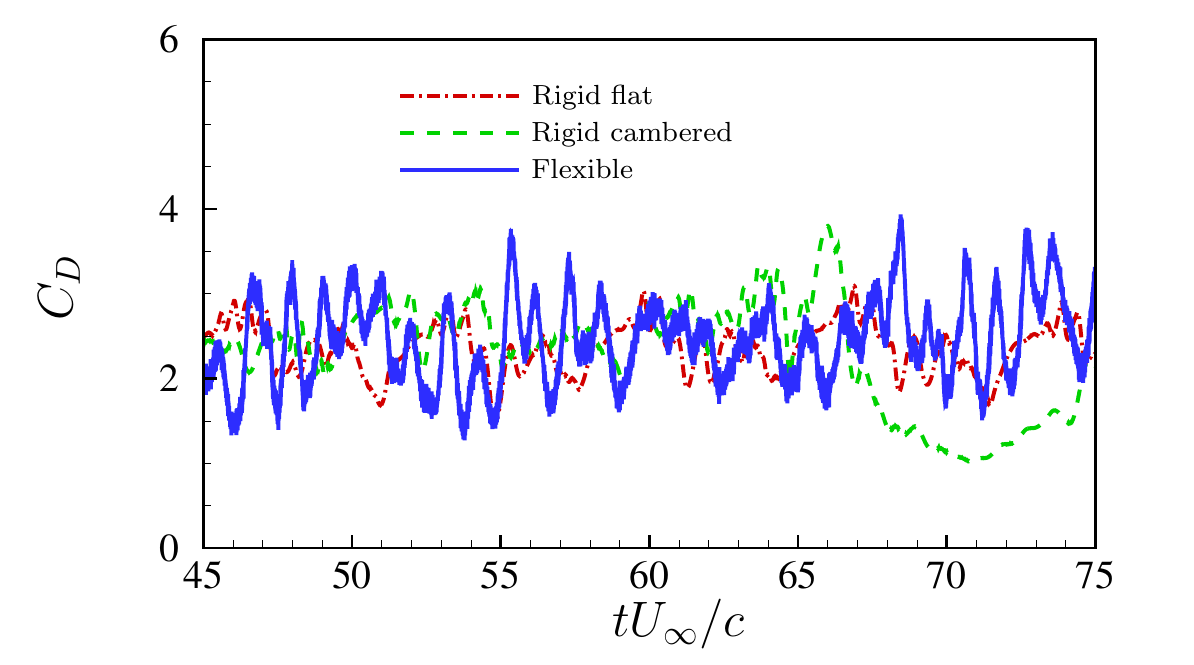}\label{fig:force_55b}}
	\\
    \subfloat[][Maximum displacement]{\includegraphics[width=0.5\textwidth]{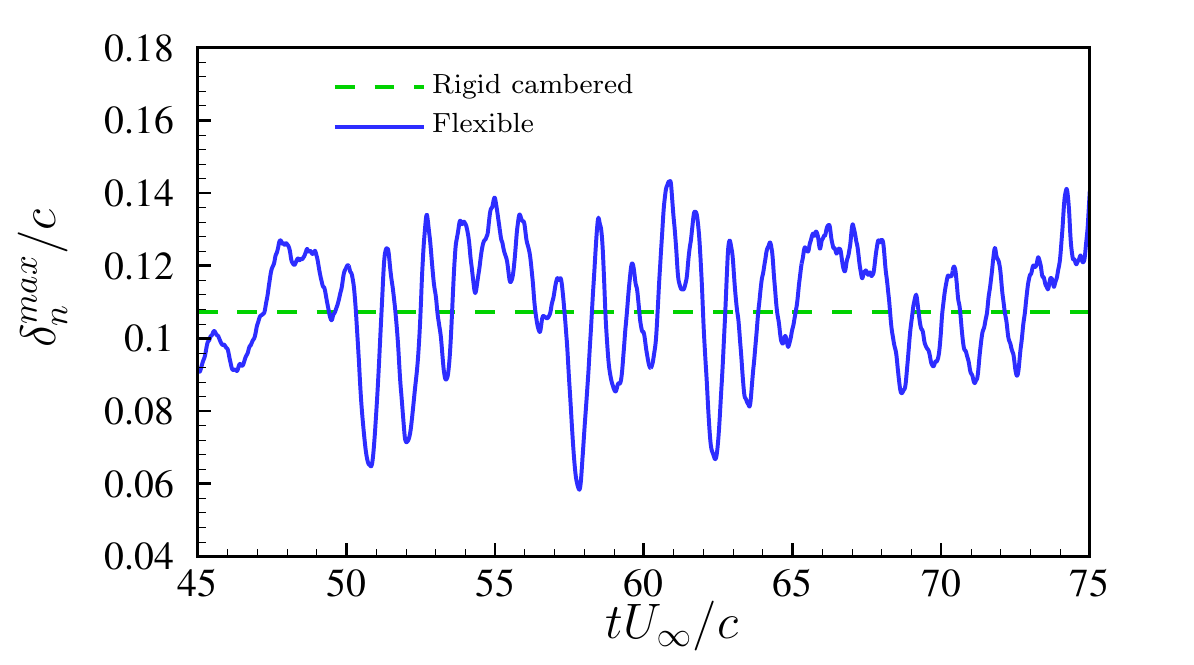}\label{fig:force_55c}}
    \quad \quad
	\subfloat[][Deflection envelope]{\includegraphics[width=0.44\textwidth]{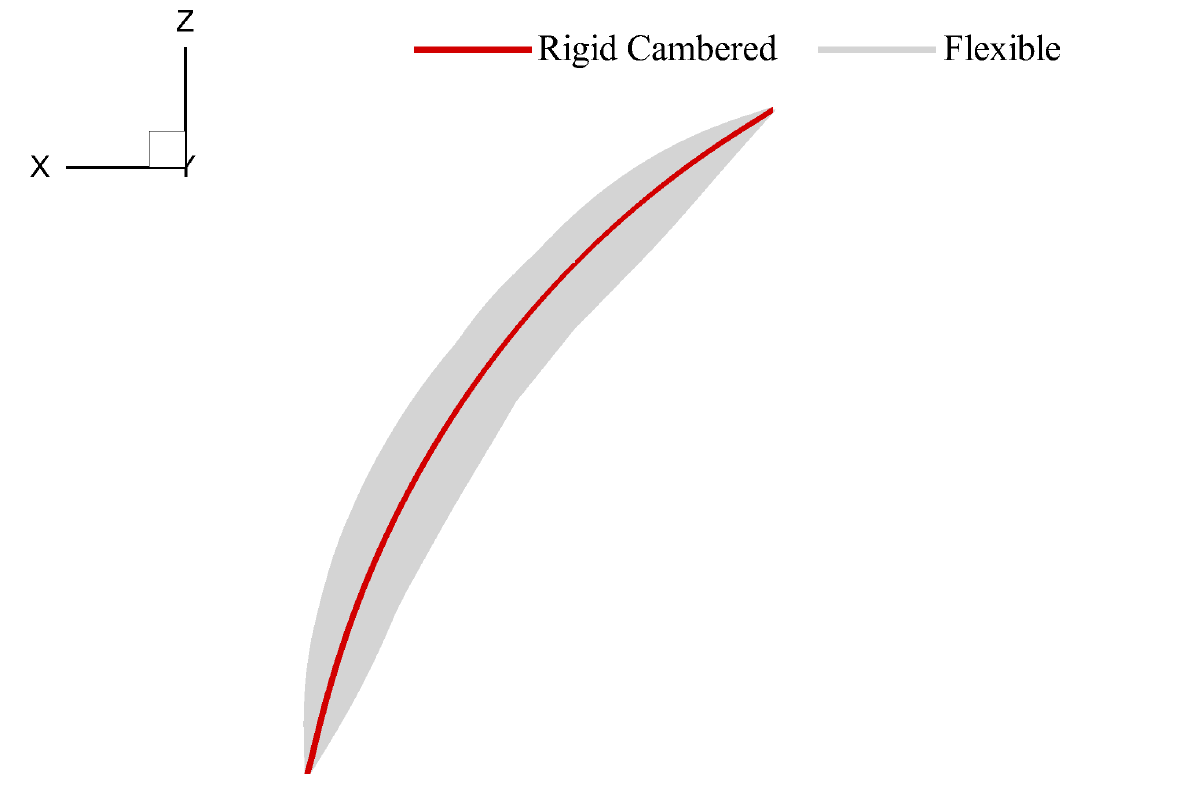}\label{fig:force_55d}} 
	\caption{\label{fig:force_55}  Comparison among a rigid flat wing, a rigid cambered wing and a flexible membrane at $\alpha=55^\circ$ for time-dependent aerodynamic forces and membrane deflection.}
\end{figure}

\begin{figure}[H]
	\centering
	\subfloat[][Rigid flat wing]{
		\includegraphics[width=0.25\textwidth]{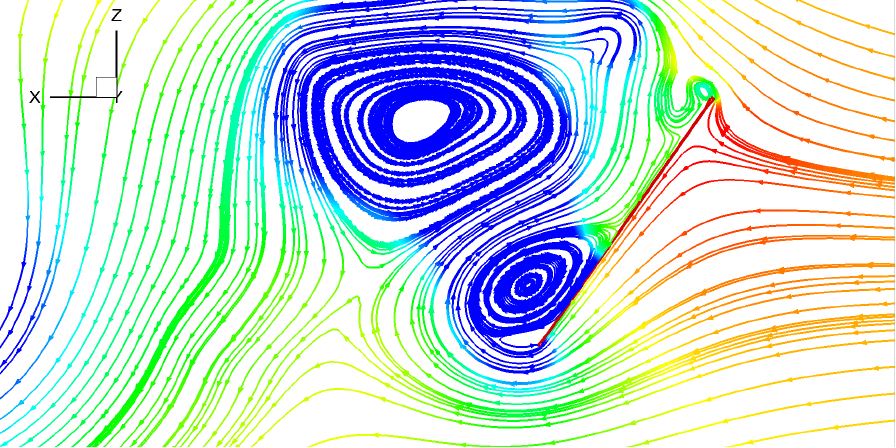}
		\includegraphics[width=0.25\textwidth]{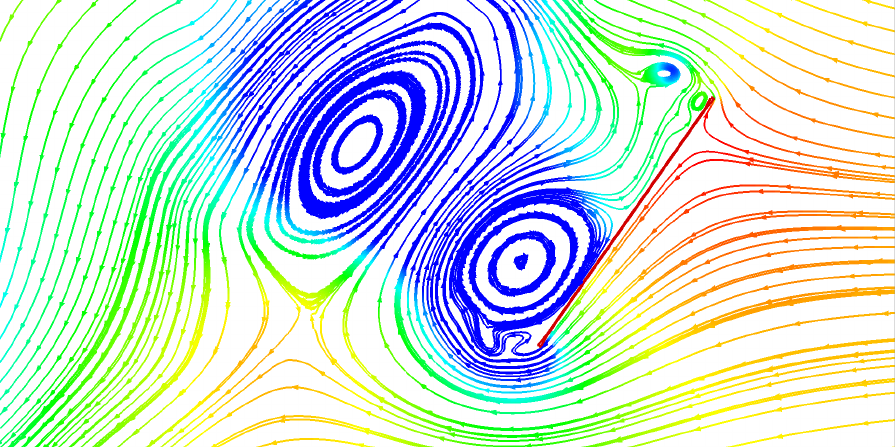}
		\includegraphics[width=0.25\textwidth]{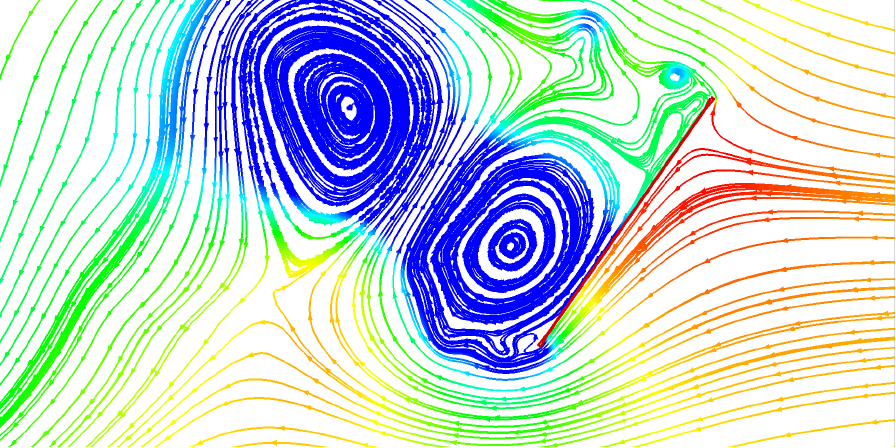}
		\includegraphics[width=0.25\textwidth]{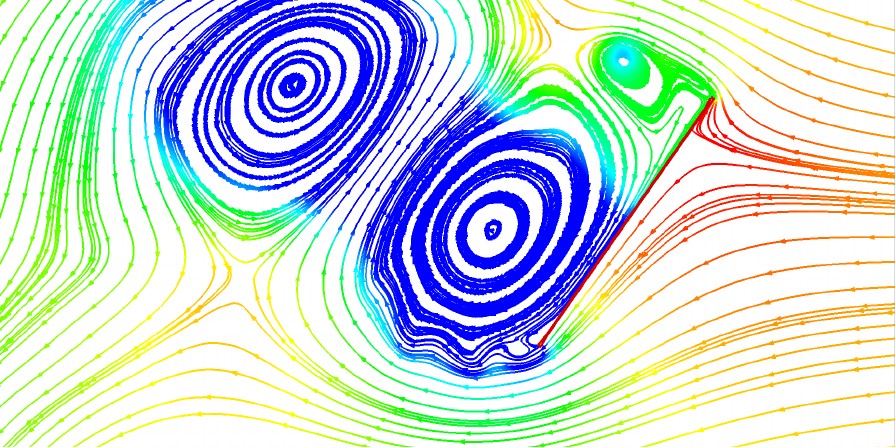}}
	\\
	\subfloat[][Rigid cambered wing]{
		\includegraphics[width=0.25\textwidth]{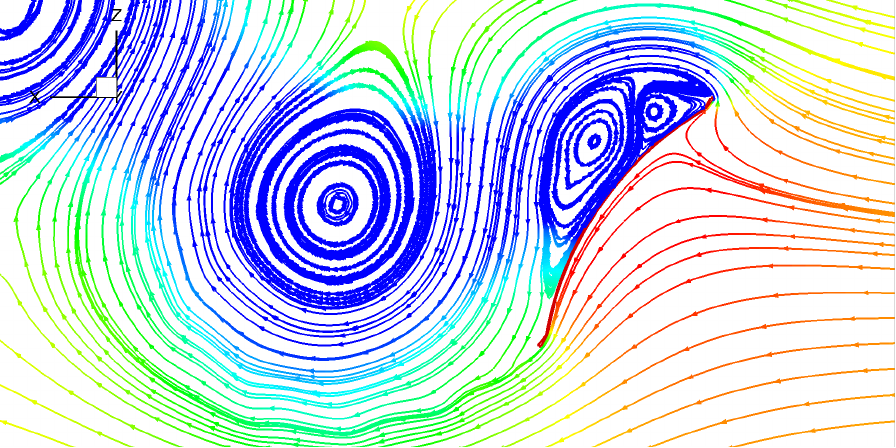}
		\includegraphics[width=0.25\textwidth]{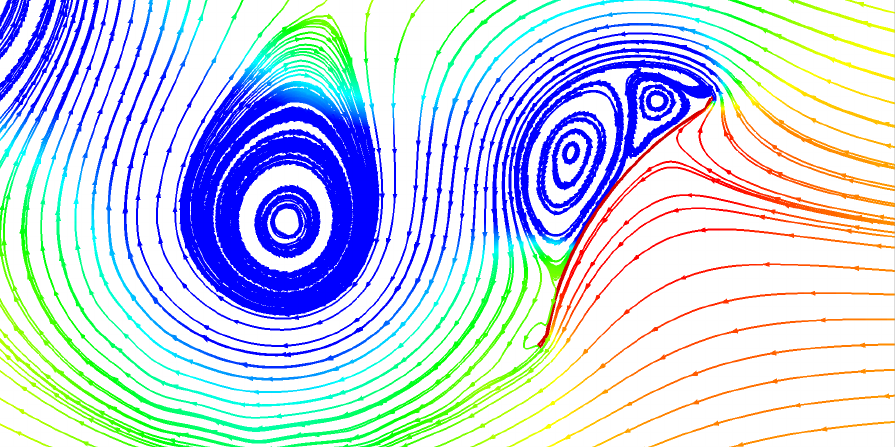}
		\includegraphics[width=0.25\textwidth]{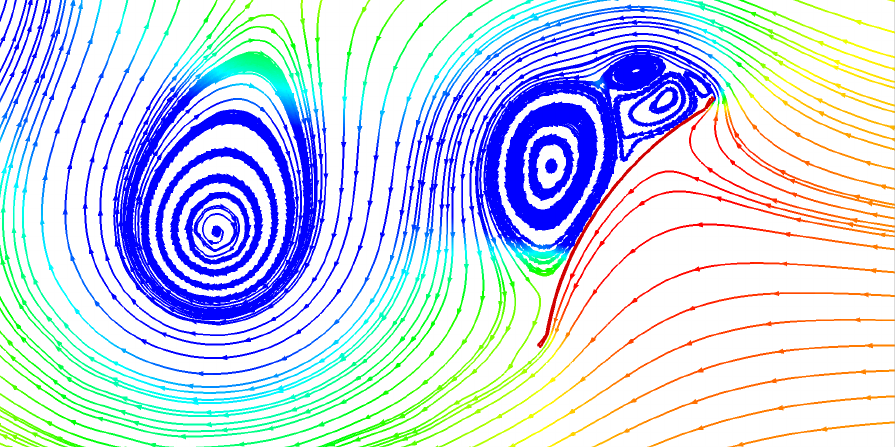}
		\includegraphics[width=0.25\textwidth]{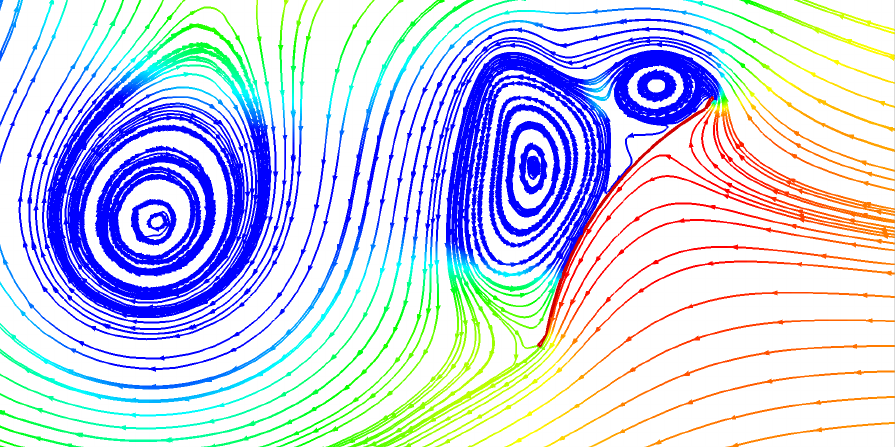}}
	\\
	\subfloat[][Flexible membrane wing]{
		\includegraphics[width=0.25\textwidth]{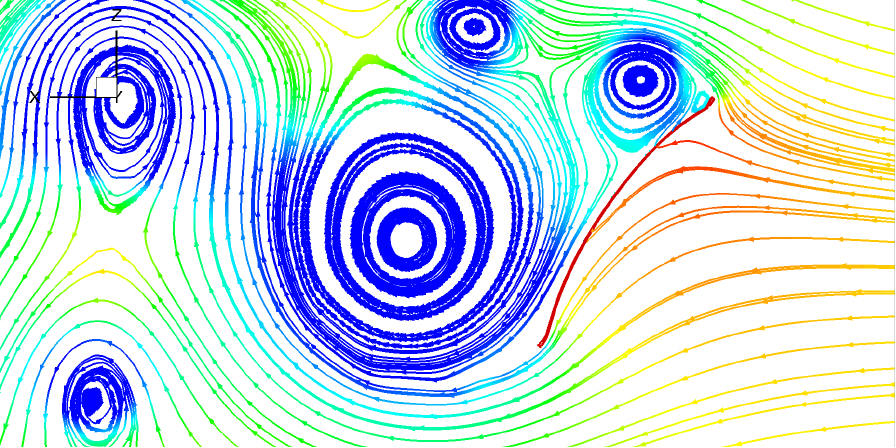}
		\includegraphics[width=0.25\textwidth]{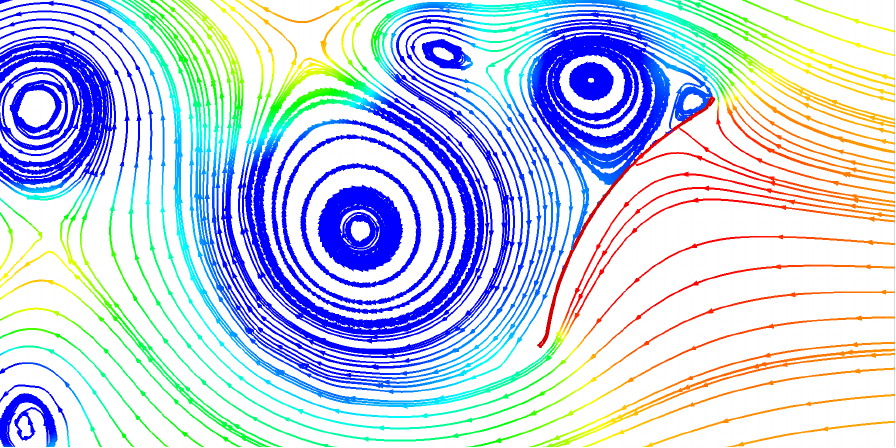}
		\includegraphics[width=0.25\textwidth]{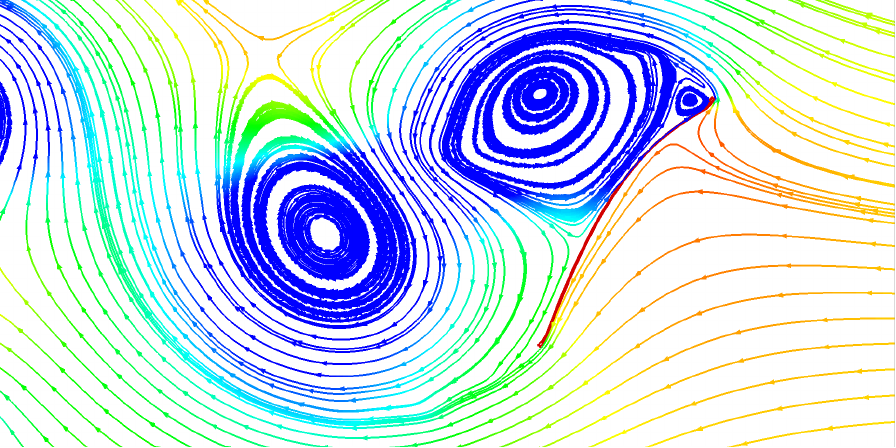}
		\includegraphics[width=0.25\textwidth]{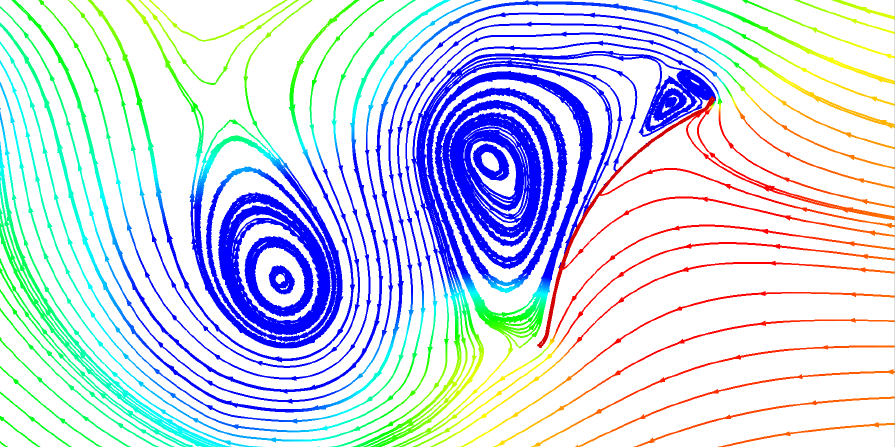}}
	\\
	\includegraphics[width=0.5\textwidth]{./streamline_leng.pdf}
	\caption{\label{fig:streamline_com_55}  Instantaneous streamlines colored by pressure coefficient at $\alpha$=$55^\circ$.}
\end{figure}

\refFigs{fig:force_55} \subref{fig:force_55a} and \subref{fig:force_55b} plot the aerodynamic forces at $\alpha=55^\circ$ in the deep stall regime. The rigid flat wing shows the poorest lift and drag forces. The rigid cambered wing and the flexible membrane produce aerodynamic forces with similar mean values and fluctuation amplitudes. The time histories of the maximum displacement and the deflection envelope of the flexible membrane are displayed in \reffigs{fig:force_55} \subref{fig:force_55c} and \subref{fig:force_55d}. We can observe from \reffig{fig:streamline_com_55} that the wing camber helps in generating stronger leading edge vortices by reducing the local angle of attack. The membrane vibrations caused by the fluid-membrane coupling effect show little influence on the mean aerodynamic forces and their amplitudes by comparing with the rigid cambered wing at this deep stall angle. The membrane flexibility effect mainly affects the frequency characteristics of the aerodynamic forces when the membrane is strongly coupled with unsteady flows. By comparing the aerodynamic forces and the flow features of the flexible membrane at smaller angles of attack, the vortex size becomes larger and the negative pressure region following the movement of the vortices gradually expands to the whole surface. As a result, the aerodynamic force normal to the membrane chord increases. However, the projected surface area along the $Z$-direction becomes smaller once the angle of attack exceeds $45^\circ$. Thus, the force component along the $Z$-direction, i.e., lift force, reduces at the deep stall conditions. The flow features at the deep stall conditions exhibit alternative shedding vortices from the leading and trailing edges, which share similarities with the vortical structures behind a bluff-body-like structure.

\begin{figure}[H]
	\centering
	\subfloat[][]{\includegraphics[width=1.0\textwidth]{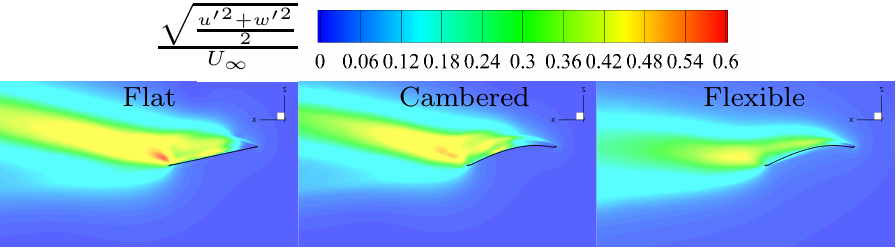}\label{fig:tkea}}
	\\
	\subfloat[][]{\includegraphics[width=1.0\textwidth]{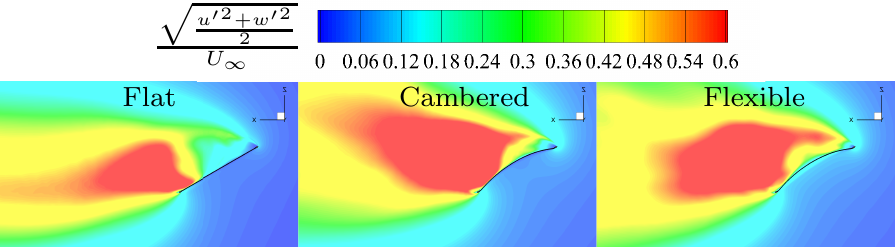}\label{fig:tkeb}}
	\\
	\subfloat[][]{\includegraphics[width=1.0\textwidth]{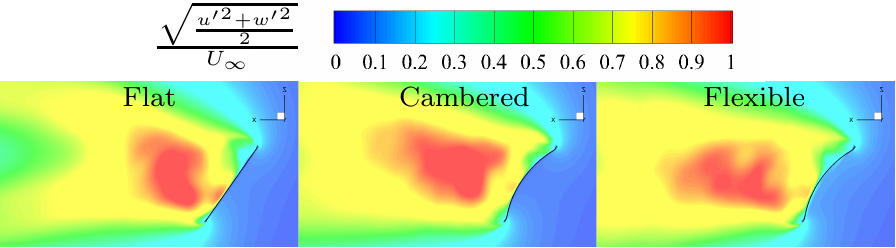}\label{fig:tkec}}
	\caption{\label{fig:tke}  Comparison of turbulent intensity between rigid flat wing, rigid cambered wing and flexible membrane at $\alpha$= (a) $12^\circ$, (b) $30^\circ$ and (c) $55^\circ$.}
\end{figure}

The turbulent intensities between the rigid flat wing, the rigid camber wing and the flexible membrane are compared in \reffig{fig:tke} at three typical angles of attack, respectively. The purpose is to gain further physical insight into the flow-induced vibration mechanism. In \reffig{fig:tke} \subref{fig:tkea}, the flexibility effect can suppress the flow fluctuations in the wake by comparing with its rigid counterpart in the pre-stall regime. \refFig{fig:tke} \subref{fig:tkeb} presents the turbulent intensity distributions for three wings at the transitional stall regime. The high turbulent intensity region gets closer to the wing surface when the camber effect and the flow-induced vibration are taken into account. The improvement of the aerodynamic forces is related to the energy transport in the boundary layer. Similar turbulent intensity distributions are observed from \reffig{fig:tke} \subref{fig:tkec} for three wings at the deep stall regime. As a result, the difference in the aerodynamic performance between these three wings becomes smaller when a deep stall occurs.

\subsection{Effect of Aeroelastic Number} \label{sec:aeroelastic}
\begin{figure}[H]
	\centering
	\subfloat[][]{\includegraphics[width=0.5\textwidth]{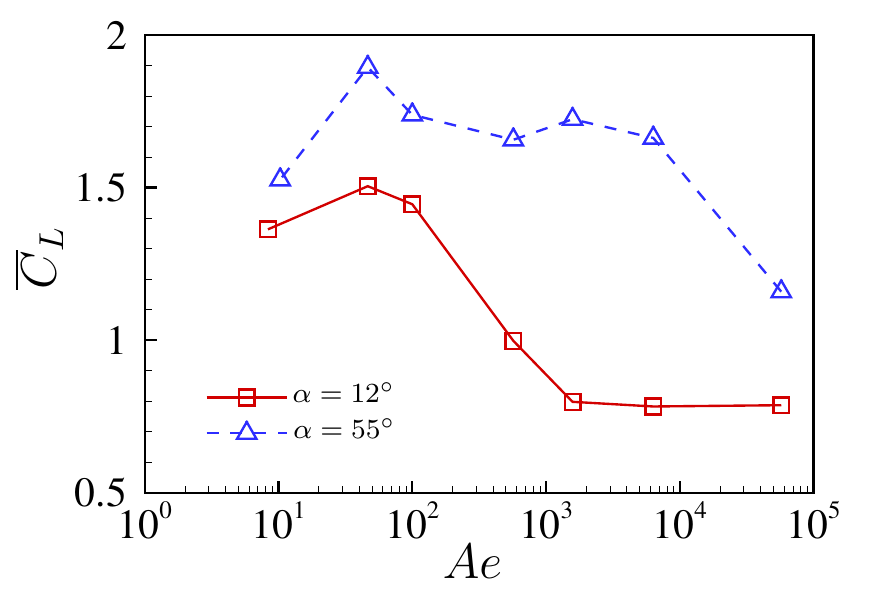}\label{fig:Ae_coma}}
	\subfloat[][]{\includegraphics[width=0.5\textwidth]{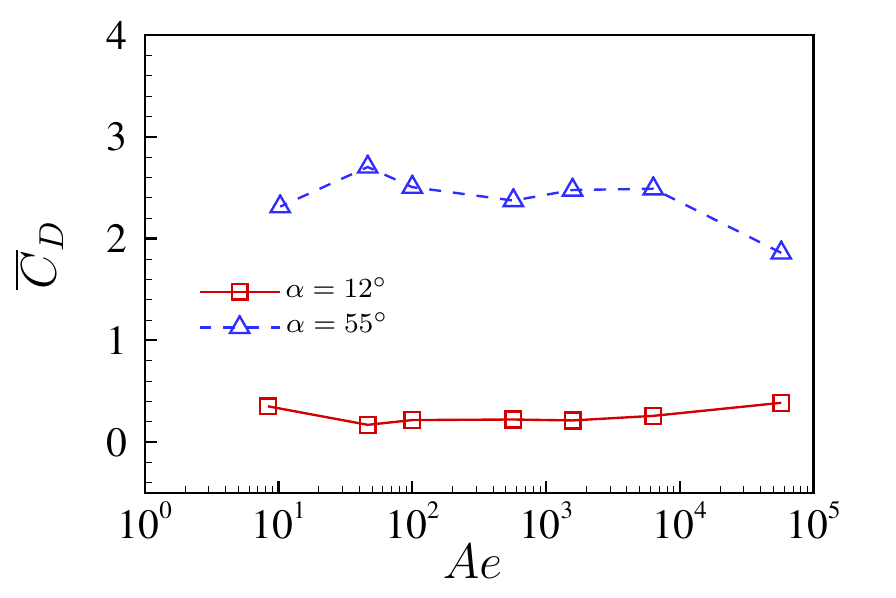}\label{fig:Ae_comb}}
	\\
	\subfloat[][]{\includegraphics[width=0.5\textwidth]{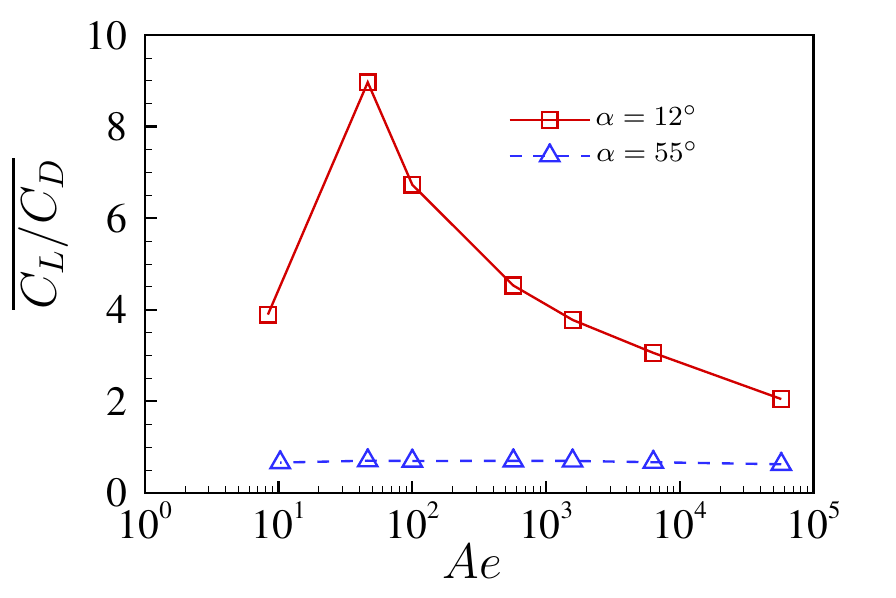}\label{fig:Ae_comc}}
     \subfloat[][]{\includegraphics[width=0.5\textwidth]{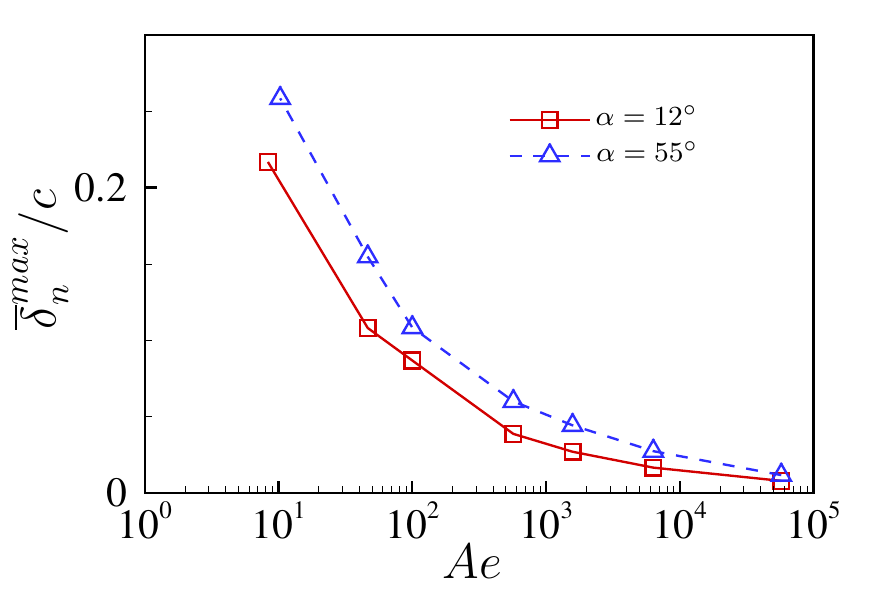}\label{fig:Ae_comd}}
     \\
     \subfloat[][]{\includegraphics[width=0.5\textwidth]{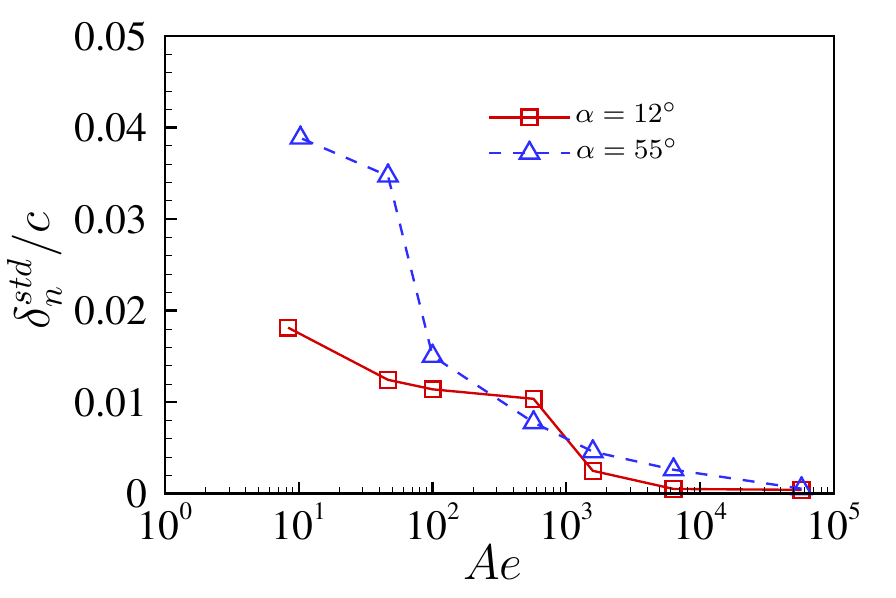}\label{fig:Ae_come}}
     \subfloat[][]{\includegraphics[width=0.5\textwidth]{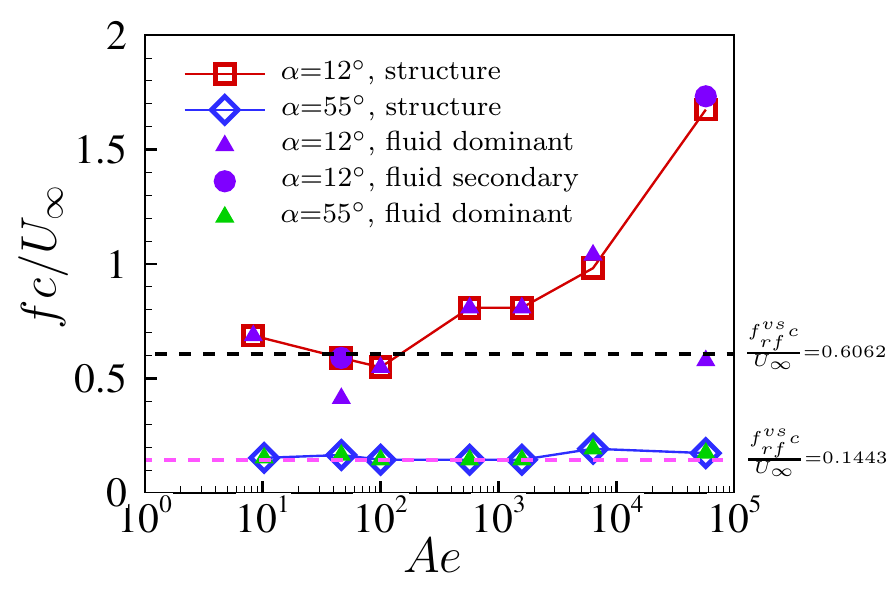}\label{fig:Ae_comf}}
	\caption{\label{fig:Ae_com}  Comparison of (a) time-averaged lift coefficient, (b) time-averaged drag coefficient, (c) time-averaged lift-to-drag ratio, (d) normalized maximum mean displacement, (e) standard deviation of normalized structural vibration and (f) characteristic frequency for flexible membrane wings at various aeroelastic numbers.}
\end{figure}

We further analyze the effect of aeroelastic numbers on membrane dynamics. Based on the discussion in \refse{sec:dynamics}, the flexible membrane exhibits an opposite drag variation as a function of the aeroelastic number below and above a critical angle of attack of $20^\circ$. Thus, we select an angle of attack at $12^\circ$ near the angle corresponding to the global maximum lift-to-drag. The angle of attack at $55^\circ$ above the critical value is chosen in the deep stall regime. \refFig{fig:Ae_com} presents a comparison of the time-averaged aerodynamic performance and membrane dynamic responses as a function of the aeroelastic number at two selected angles of attack. It can be seen from \reffig{fig:Ae_com} \subref{fig:Ae_coma} that the mean lift forces reach a peak value at $Ae$=46.46 and then show an overall downward trend as the aeroelastic number increases. The flexible membrane exhibits the drag enhancement at $\alpha=12^\circ$ and the drag reduction at $\alpha=55^\circ$ when the aeroelastic number is larger than 46.46 as shown in \reffig{fig:Ae_com} \subref{fig:Ae_comb}. The mean lift-to-drag ratio reaches a maximum value of 8.96 at $Ae$=46.46 and then decreases continuously to 2.05 at $\alpha=12^\circ$ in \reffig{fig:Ae_com} \subref{fig:Ae_comc}. The aeroelastic number has little effect on the mean lift-to-drag ratio in the deep stall condition.

\refFigs{fig:Ae_com} \subref{fig:Ae_comd} and \subref{fig:Ae_come} present the maximum mean displacement and the standard deviation of structural vibration as a function of aeroelastic number. Both mean displacement and its standard deviation show an overall downward trend as the aeroelastic number increases. The reduction of the membrane camber and vibration is correlated with the degradation of the lift performance. 

\refFig{fig:Ae_com} \subref{fig:Ae_comf} shows a comparison of the characteristic vibration frequency and vortex shedding frequency at two selected angles of attack. The frequency is normalized by the membrane chord and freestream velocity. In \reffig{fig:Ae_com} \subref{fig:Ae_comf}, the solid lines with labels denote the membrane vibration frequency. The solid labels in a triangle shape represent the dominant characteristic frequency of the fluid domain, while the solid label in a circle shape is the secondary characteristic frequency. These characteristic frequencies in the fluid domain are extracted from the whole coupled fluid-membrane system by using the Global Fourier mode decomposition (GFMD) method. More details about the GFMD method can be found in \cite{li2022aeroelastic}. The black dashed line is the normalized vortex shedding frequency of a rigid flat wing at $\alpha=12^\circ$, while the magenta dashed line denotes the characteristic frequency at $\alpha=55^\circ$. It can be seen from \reffig{fig:Ae_com} \subref{fig:Ae_comf} that the membrane vibration frequency is synchronized with the dominant or secondary frequencies of the fluid flows, resulting in the lock-in phenomenon \cite{li2020flow_accept}. The characteristic frequency of the coupled system exhibits similar values as the bluff-body vortex shedding frequency at $\alpha=12^\circ$ for the two smallest aeroelastic numbers. Similarly, the coupled dynamics of the flexible membrane are dominant by the bluff-body vortex shedding frequency at the post-stall regime of $\alpha=55^\circ$ in the examined parameter space. The coupled dynamics of the membrane at $\alpha=12^\circ$ are dominant by the structural natural frequency at higher aeroelastic numbers.

\begin{figure}[H]
	\centering
	\subfloat[][]{\includegraphics[width=0.5\textwidth]{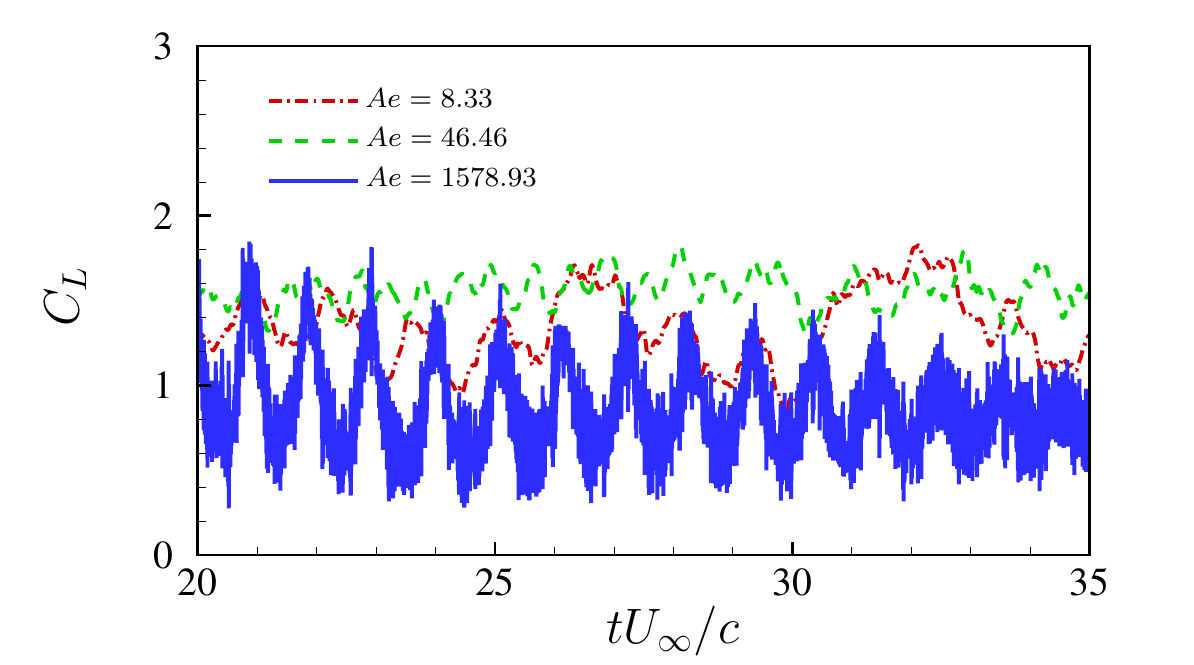}\label{fig:ae_force_12a}}
	\subfloat[][]{\includegraphics[width=0.5\textwidth]{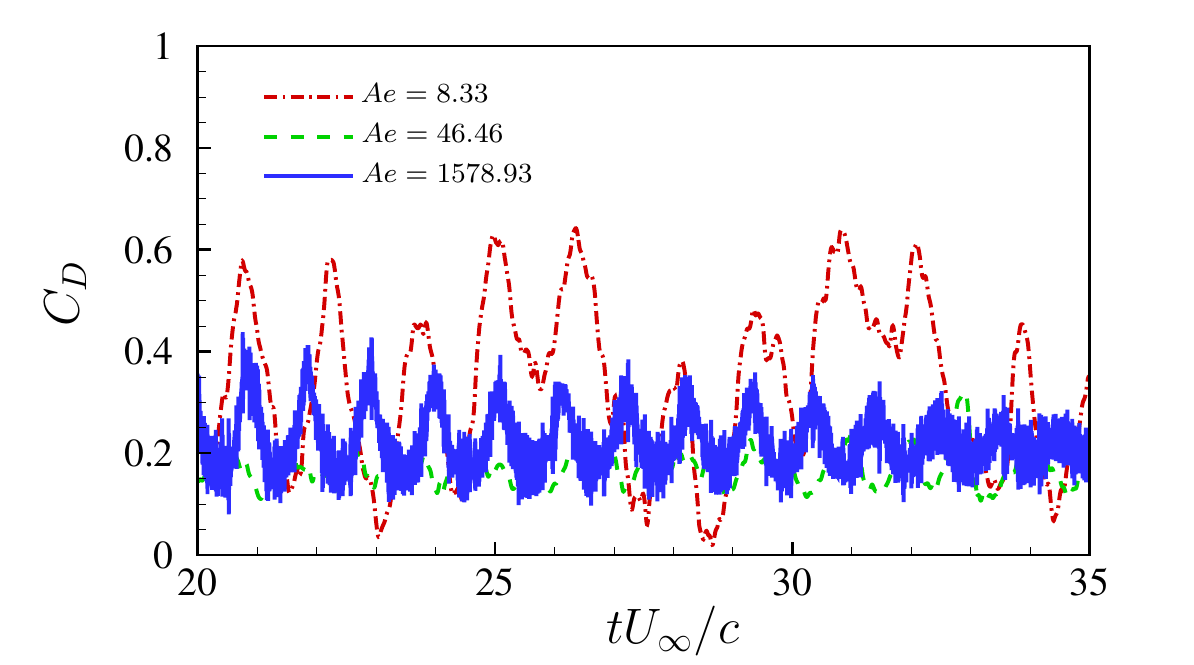}\label{fig:ae_force_12b}}
	\caption{\label{fig:ae_force_12}  Comparison of time-dependent (a) lift coefficient and (b) drag coefficient at $\alpha=12^\circ$.}
\end{figure}

\begin{figure}[H]
	\centering
	\subfloat[][]{
		\includegraphics[width=0.25\textwidth]{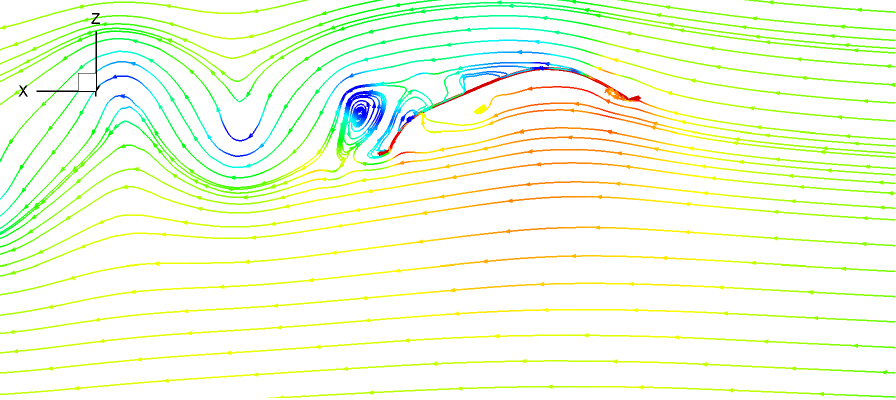}
		\includegraphics[width=0.25\textwidth]{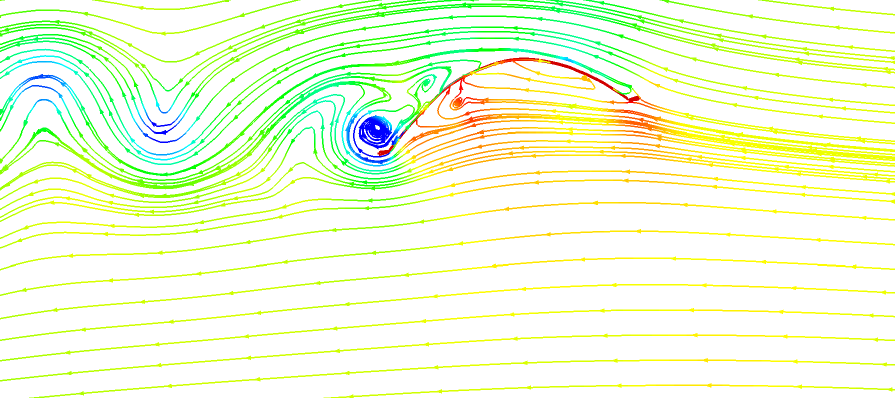}
		\includegraphics[width=0.25\textwidth]{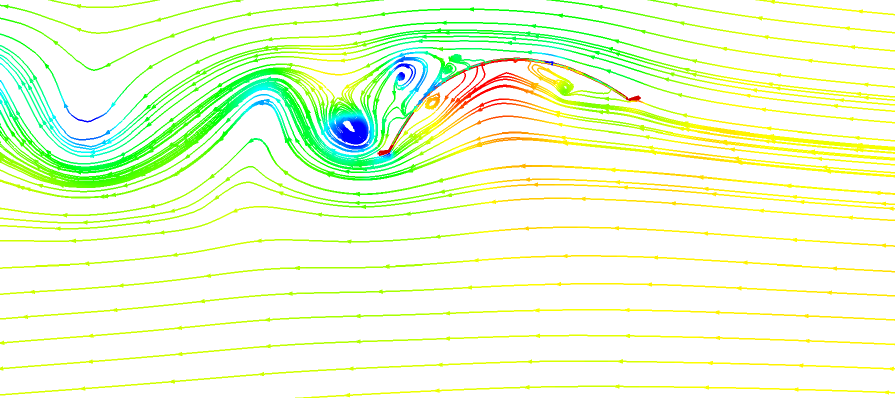}
		\includegraphics[width=0.25\textwidth]{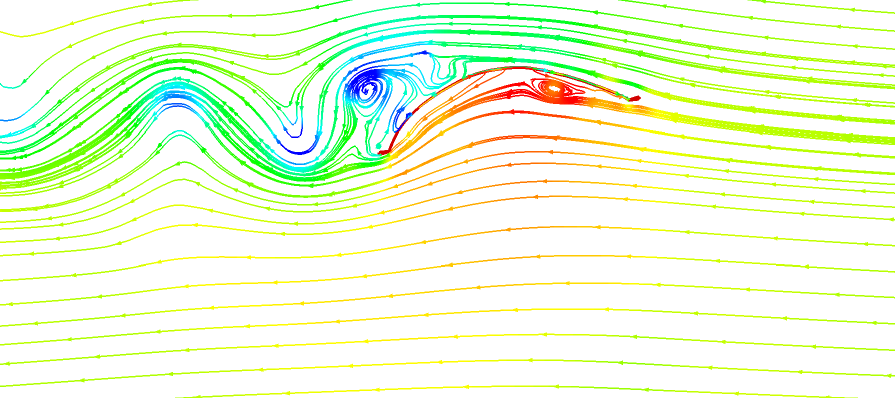}\label{fig:streamline_com_aoa12_aea}}
	\\
	\subfloat[][]{
		\includegraphics[width=0.25\textwidth]{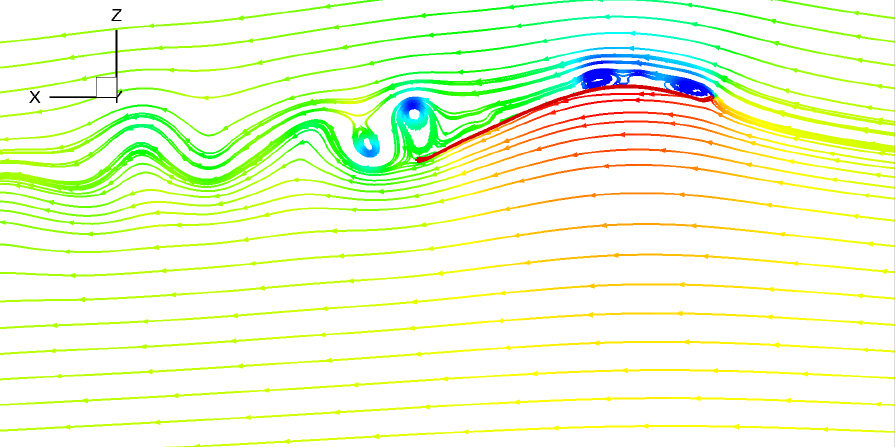}
		\includegraphics[width=0.25\textwidth]{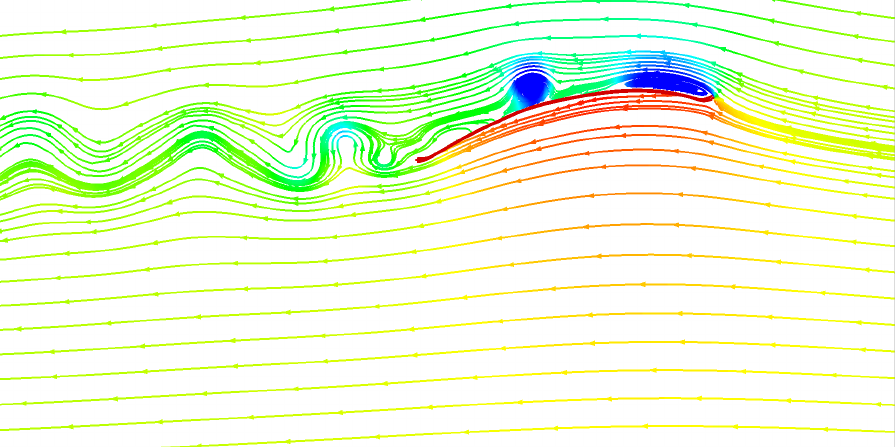}
		\includegraphics[width=0.25\textwidth]{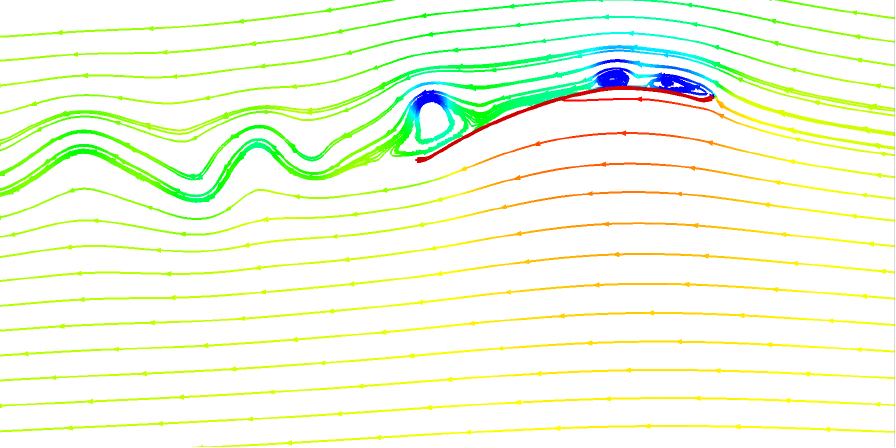}
		\includegraphics[width=0.25\textwidth]{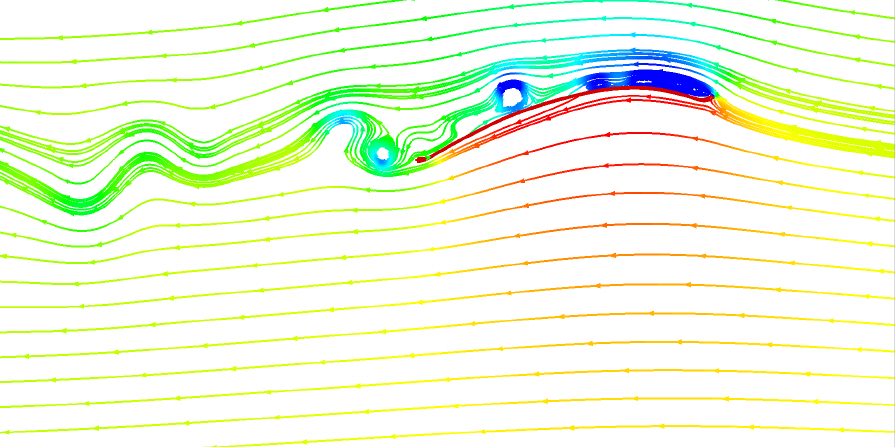}\label{fig:streamline_com_aoa12_aeb}}
	\\
        \subfloat[][]{
		\includegraphics[width=0.25\textwidth]{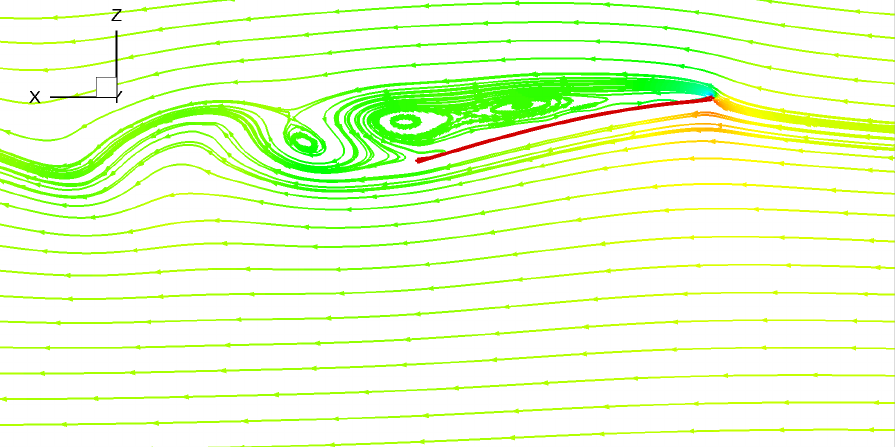}
		\includegraphics[width=0.25\textwidth]{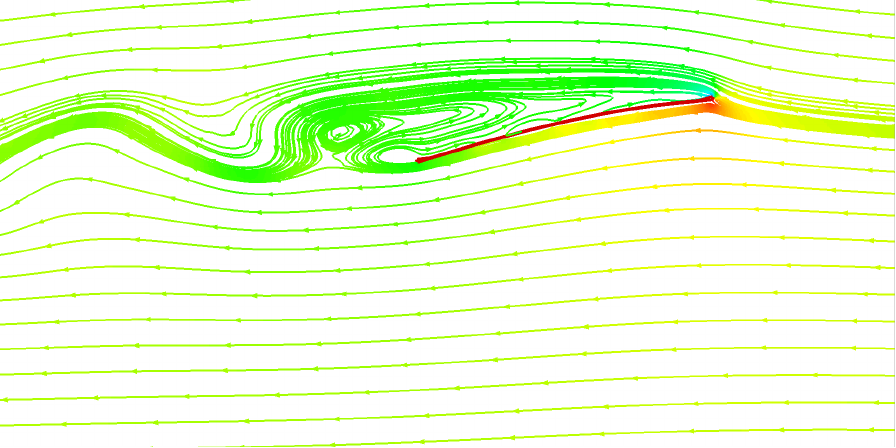}
		\includegraphics[width=0.25\textwidth]{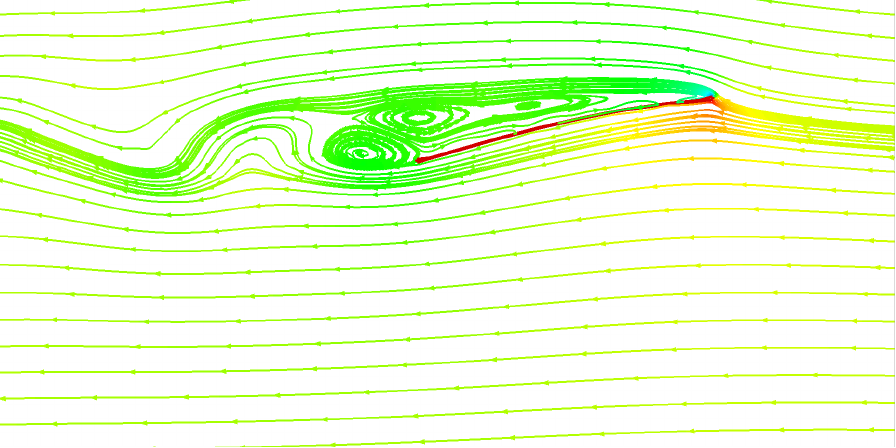}
		\includegraphics[width=0.25\textwidth]{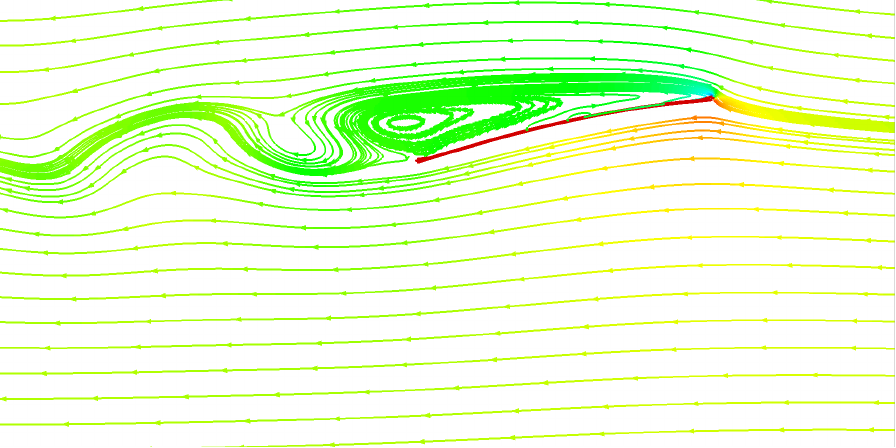}\label{fig:streamline_com_aoa12_aec}}
        \\
	\includegraphics[width=0.5\textwidth]{./streamline_leng.pdf}
	\caption{\label{fig:streamline_com_aoa12_ae}  Instantaneous streamlines colored by pressure coefficient of $Ae$= (a) 8.33, (b) 46.46 and (c) 1578.93 at $\alpha$=$12^\circ$.}
\end{figure}

\refFigs{fig:ae_force_12} and \ref{fig:streamline_com_aoa12_ae} present a comparison of the time-dependent aerodynamic forces and instantaneous streamlines at $\alpha=12^\circ$. The lift and drag coefficients exhibit large fluctuation amplitudes at a relatively small aeroelastic number of 8.33. In \reffig{fig:streamline_com_aoa12_ae} \subref{fig:streamline_com_aoa12_aea}, the membrane shows a large mean camber of nearly 20$\%$ of the chord. At this small angle of attack, the rear part of this cambered shape behaves like a bluff body and large-scale vortices are formed. As the aeroelastic number increases to 46.46, we observe an enhancement of the lift coefficient and a drag reduction from \reffigs{fig:ae_force_12}. The membrane becomes more rigid at a higher aeroelastic number. Thus, the membrane camber is reduced to form a streamlined shape in \reffig{fig:streamline_com_aoa12_ae} \subref{fig:streamline_com_aoa12_aeb}. The vortices are suppressed and better aerodynamic performance is achieved. The vortices get closer to the rear part of the membrane at $Ae$=1578.93, resulting in weaker suction force at the leading edge. Consequently, the aerodynamic performance is degraded when the membrane becomes more rigid.

\begin{figure}[H]
	\centering
	\subfloat[][]{\includegraphics[width=0.5\textwidth]{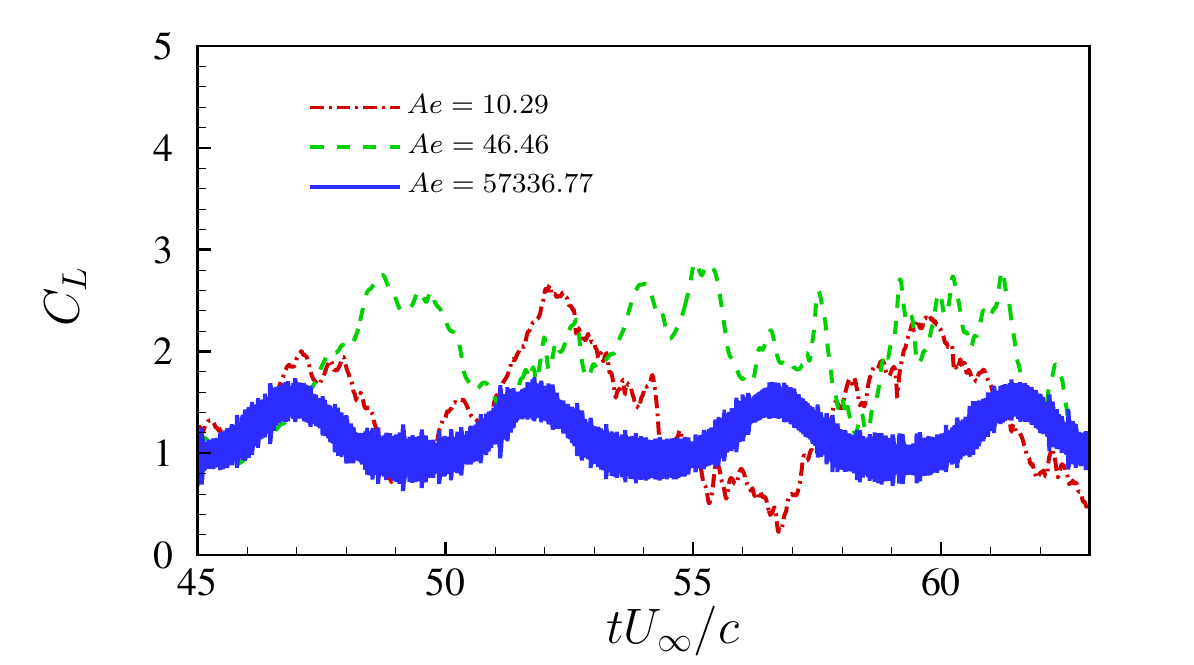}\label{fig:ae_force_55a}}
	\subfloat[][]{\includegraphics[width=0.5\textwidth]{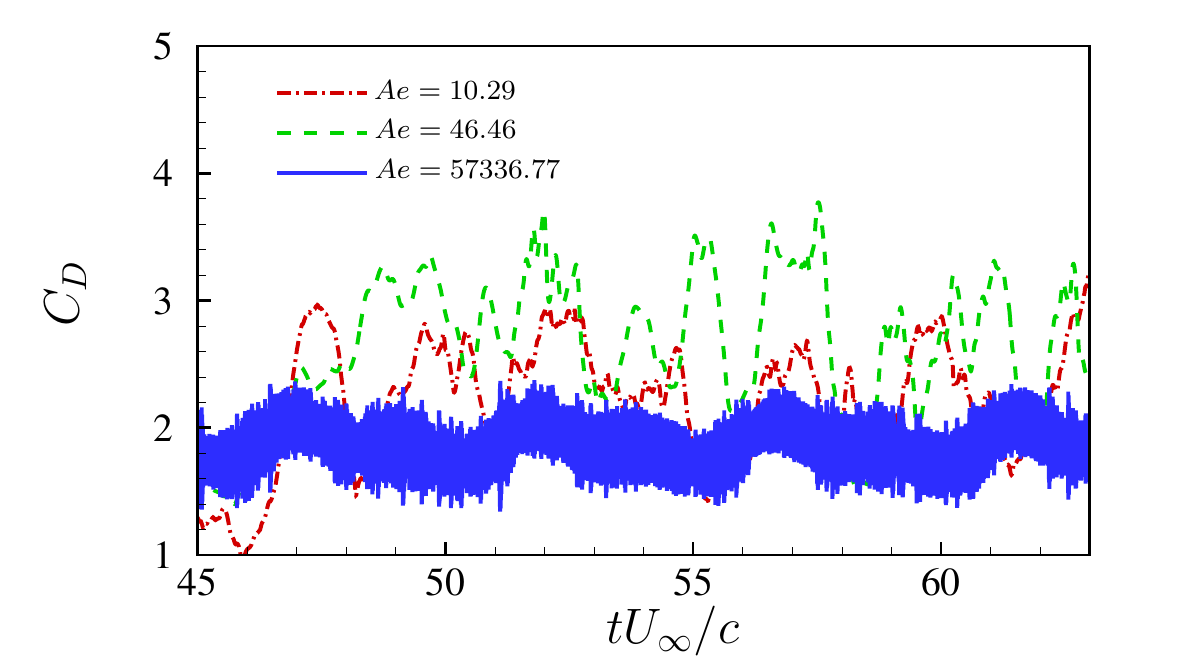}\label{fig:ae_force_55b}}
	\caption{\label{fig:ae_force_55}  Comparison of time-dependent (a) lift coefficient and (b) drag coefficient at $\alpha=55^\circ$.}
\end{figure}

\begin{figure}[H]
	\centering
        \subfloat[][]{
		\includegraphics[width=0.25\textwidth]{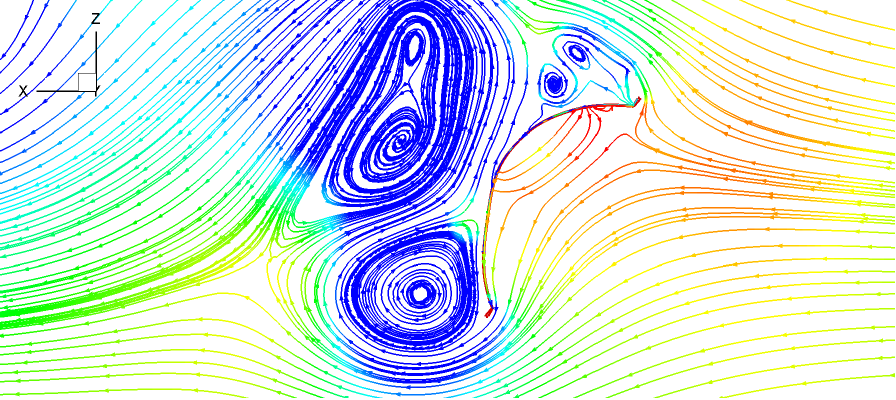}
		\includegraphics[width=0.25\textwidth]{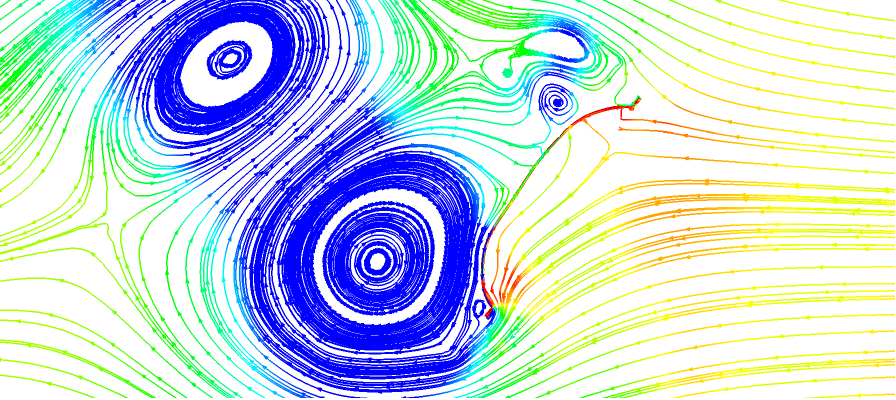}
		\includegraphics[width=0.25\textwidth]{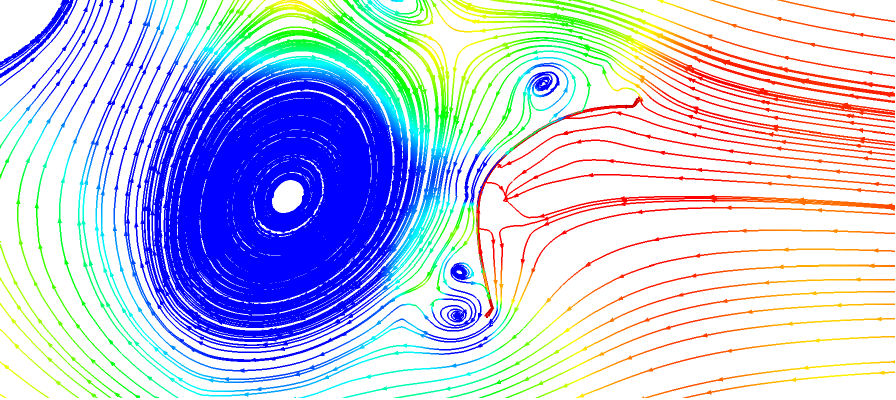}
		\includegraphics[width=0.25\textwidth]{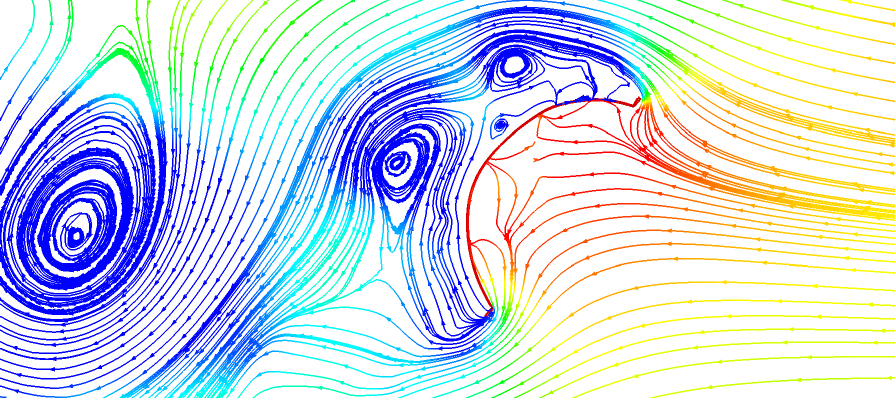}\label{fig:streamline_com_aoa55_aea}}
        \\
	\subfloat[][]{
		\includegraphics[width=0.25\textwidth]{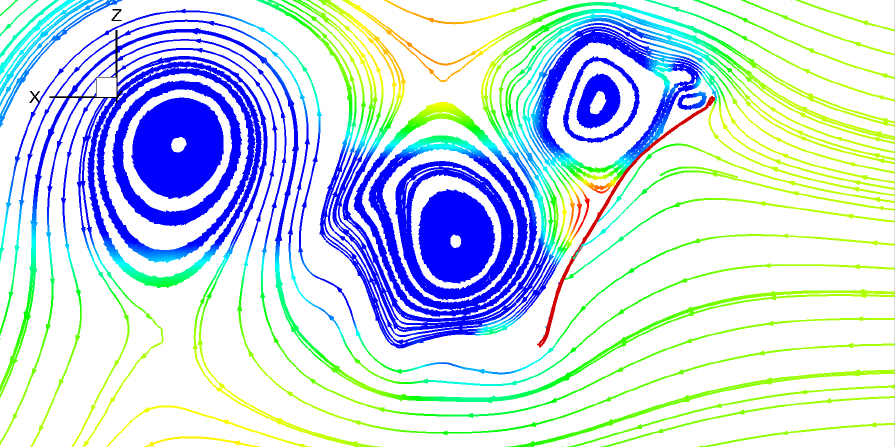}
		\includegraphics[width=0.25\textwidth]{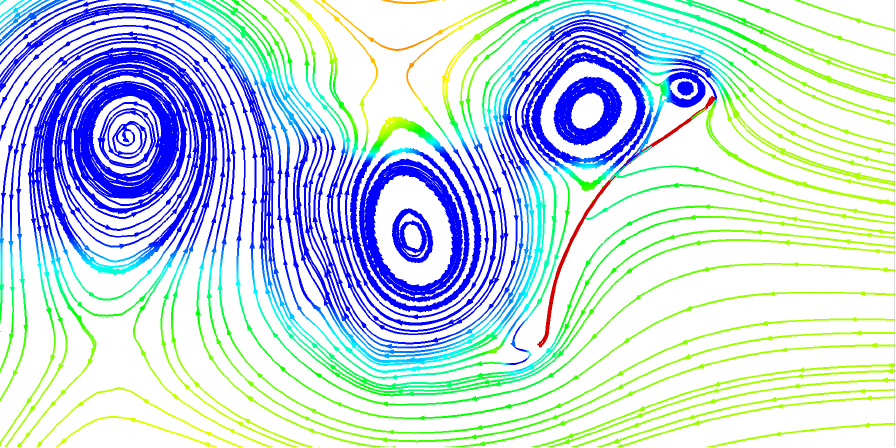}
		\includegraphics[width=0.25\textwidth]{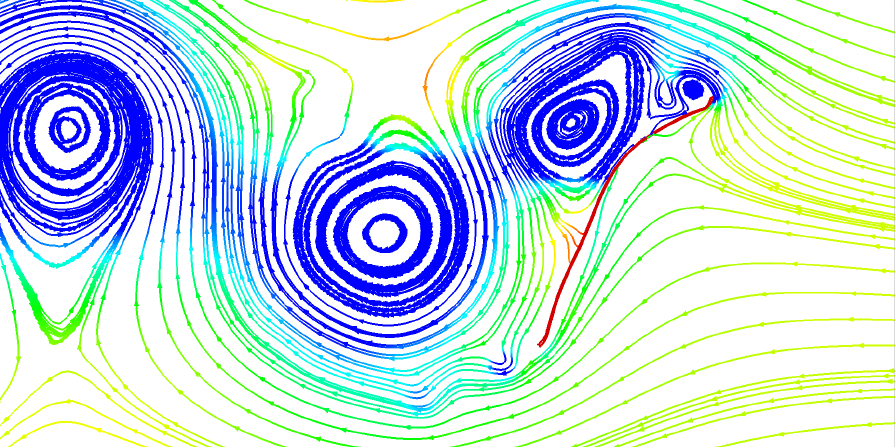}
		\includegraphics[width=0.25\textwidth]{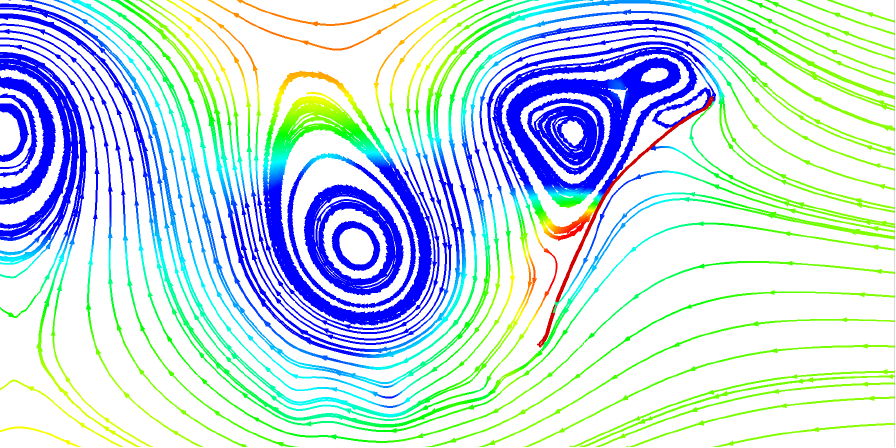}\label{fig:streamline_com_aoa55_aeb}}
	\\
	\subfloat[][]{
		\includegraphics[width=0.25\textwidth]{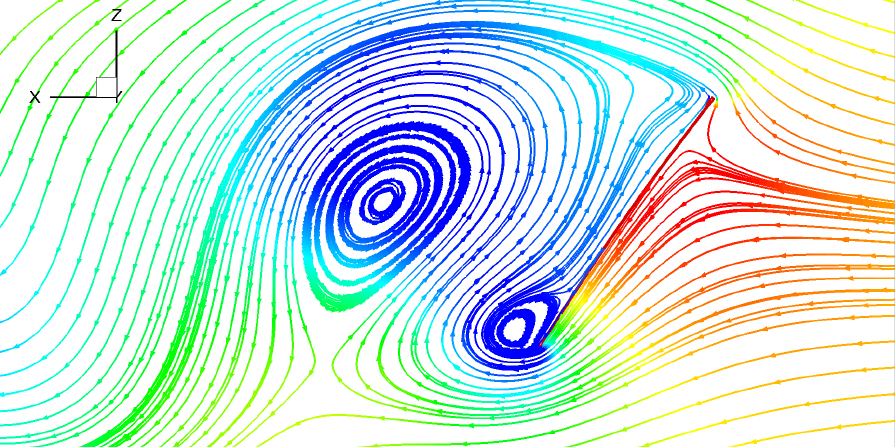}
		\includegraphics[width=0.25\textwidth]{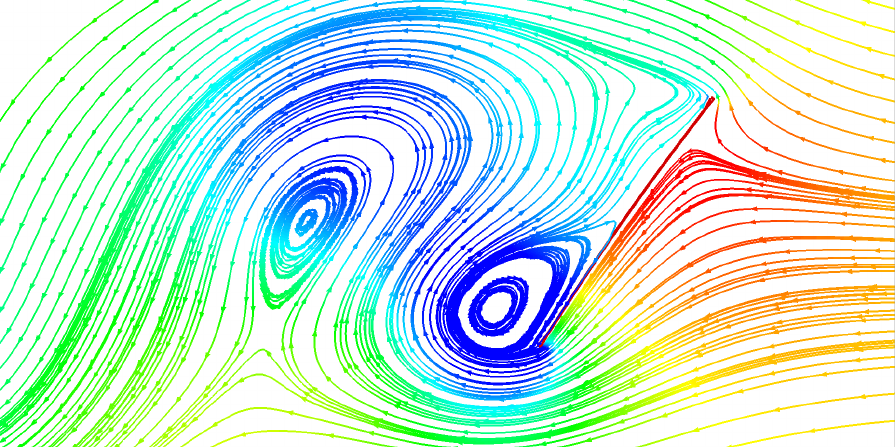}
		\includegraphics[width=0.25\textwidth]{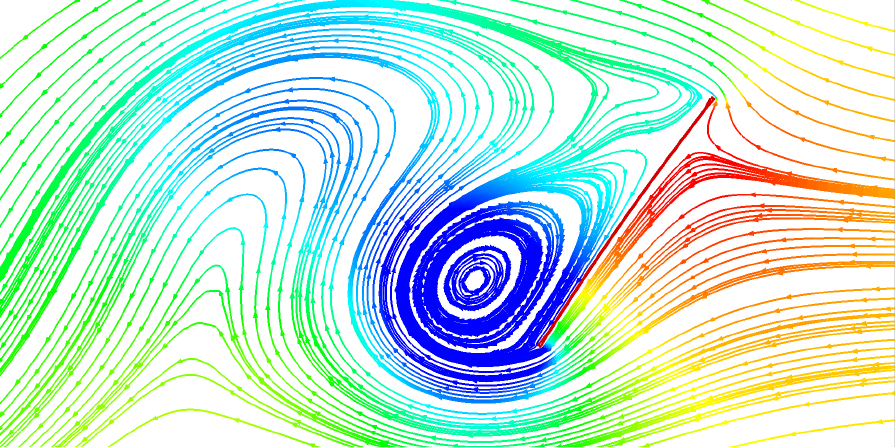}
		\includegraphics[width=0.25\textwidth]{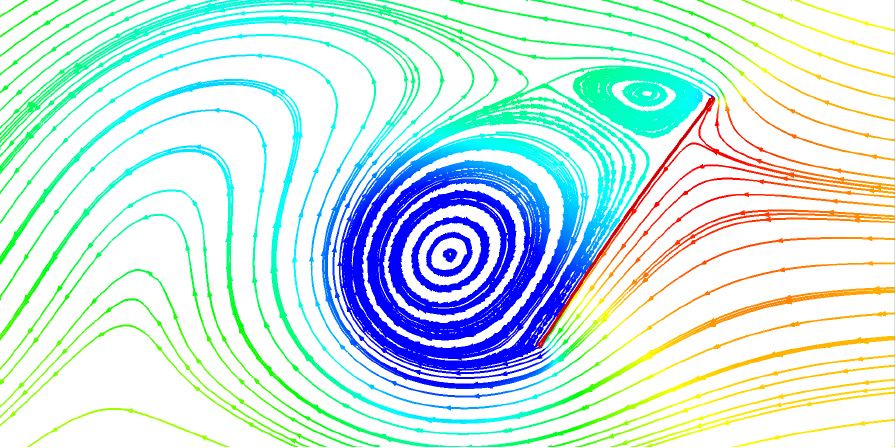}\label{fig:streamline_com_aoa55_aec}}
	\\
	\includegraphics[width=0.5\textwidth]{./streamline_leng.pdf}
	\caption{\label{fig:streamline_com_aoa55_ae}  Instantaneous streamlines colored by pressure coefficient of $Ae$= (a) 10.29, (b) 46.46 and (c) 57336.77 at $\alpha$=$55^\circ$.}
\end{figure}

\refFig{fig:ae_force_55} plots the time-dependent lift and drag forces at three selected aeroelastic numbers in the deep stall conditions. The optimal lift coefficient is achieved at $Ae$=46.46. The mean values and amplitudes of the lift and drag coefficients are greatly reduced at a larger aeroelastic number. As shown in \reffig{fig:streamline_com_aoa55_ae} \subref{fig:streamline_com_aoa55_aea}, the membrane exhibits a large camber of 25$\%$ of the chord at $Ae$=10.29. The flow features behind the cambered membrane are similar to those of the rear part of a cylinder. As $Ae$ increases to 46.46, vortices stay attached to the leading edge longer, resulting in better lift performance and larger drag. The coupling between the unsteady flow and the flexible membrane becomes weaker at $Ae$=57336.77.

\subsection{Scaling Relations for Mean Membrane Response} \label{sec:scaling1}
As discussed above, the mean aerodynamic performance is strongly related to the membrane deformation at various angles of attack and aeroelastic numbers. It indicates that some underlying mechanisms between the membrane deformation and the aerodynamic forces governed by the aeroelastic parameters must exist. The membrane deformation is governed by the balance between the aerodynamic loads on the membrane surface and the elastic tension force. Thus, a non-dimensional parameter,  the so-called Weber number \cite{song2008aeromechanics}, is defined as the ratio between the aerodynamic force and the membrane tension to characterize the membrane deformation, which can be given as
\begin{equation}
	We = \frac{F_n}{E^s h} = C_n \frac{\frac{1}{2} \rho^f U_{\infty}^2 c}{E^s h} = \frac{C_n}{Ae}
	\label{We} 
\end{equation}
where $F_n$ denotes the normal aerodynamic forces on the membrane surface. Different from the definition of Weber number in \cite{waldman2017camber} for low angles of attack, we employ the normal force instead of the lift force considered in \cite{waldman2017camber} to characterize the membrane dynamics at high angles of attack.

\begin{figure}[H]
	\centering
	\subfloat[][Cartesian coordinates]{\includegraphics[width=1.0\textwidth]{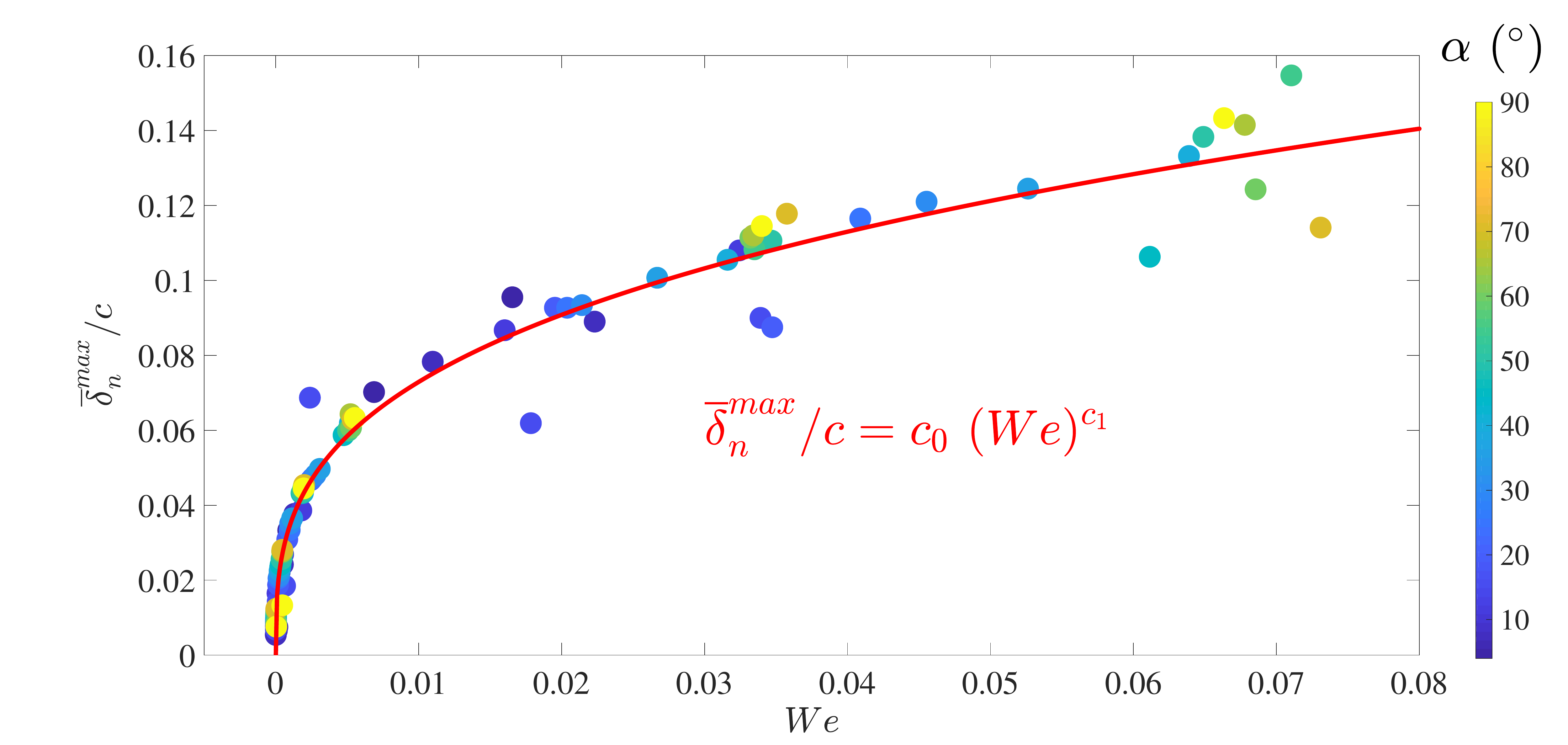}\label{fig:scalinga}}
	\\
	\subfloat[][Logarithmic coordinates]{\includegraphics[width=1.0\textwidth]{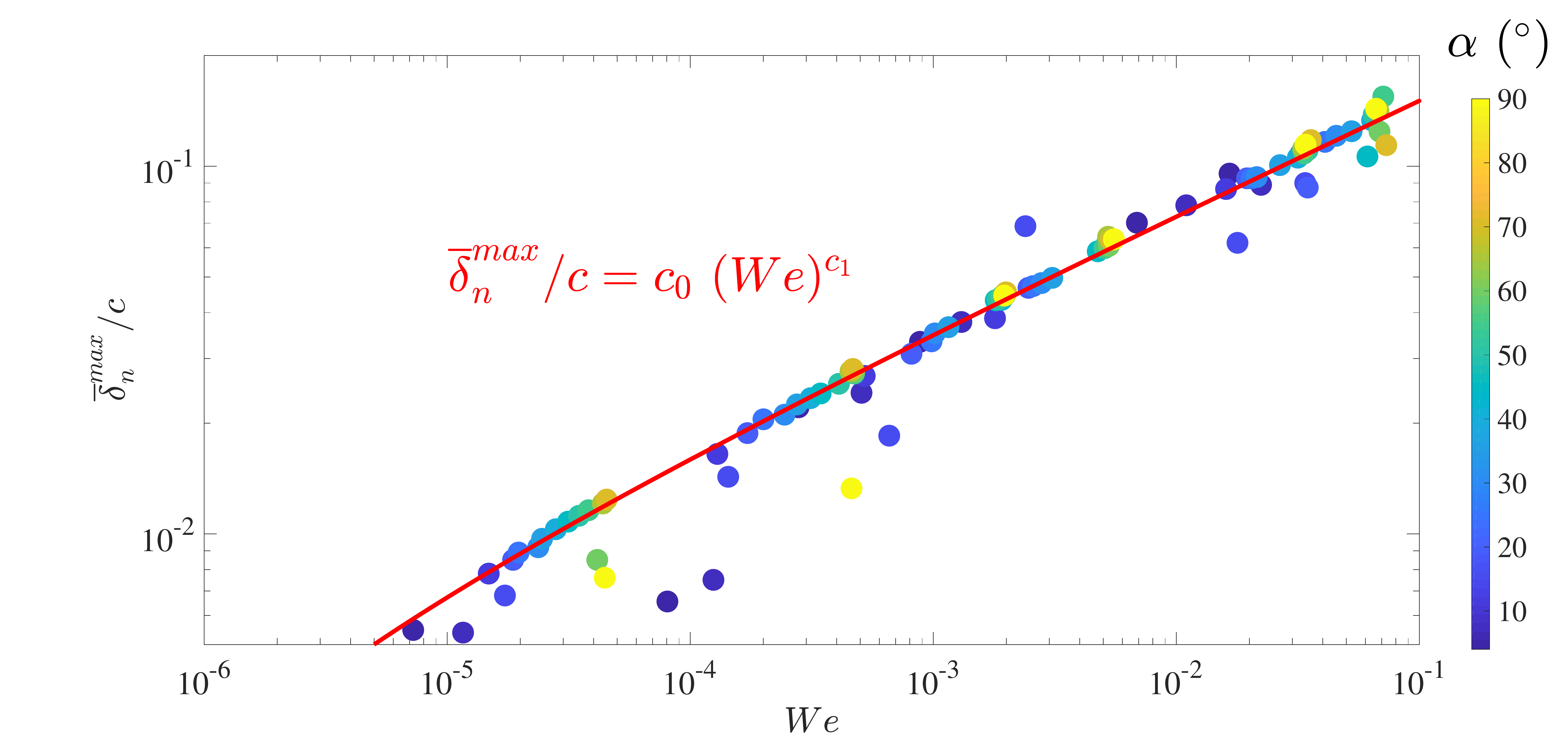}\label{fig:scalingb}}
	\caption{\label{fig:scaling} Time-averaged non-dimensional maximum normal membrane displacement $\overline{\delta}_n^{max}/c$ as a function of Weber number at different angles of attack. The coefficients of the fitting line in red color is $c_0=0.3178$ and $c_1=0.3212$.}
\end{figure}

We quantify the relationship between the maximum normal membrane displacement $\overline{\delta}_n^{max}/c$ and the Weber number $We$ in \reffig{fig:scaling}. All the numerical simulation results are summarized and plotted herein. The data label is colored by the angle of attack for each case. \refFig{fig:scaling} \subref{fig:scalingb} is plotted in the logarithmic scale. We can observe from \reffig{fig:scaling} that the data is reasonably collapsed together onto a specific power function curve. Similar scaling relations have been presented in \cite{song2008aeromechanics,waldman2017camber}. The empirical solution of the scaling relation can be expressed as
\begin{equation}
	\frac{\overline{\delta}_n^{max}}{c}=c_0 \ (We)^{c_1} = c_0 \ \left(\frac{C_n}{Ae}\right)^{c_1},
	\label{We2} 
\end{equation}
where $c_0$ and $c_1$ are the coefficients determined by a fitting power function. Based on our numerical simulation results, the coefficients can be determined as $c_0=0.3178$ and $c_1=0.3212$, respectively. For untensioned membranes, Waldman et al. \cite{waldman2017camber} suggested a relationship between the Weber number and the membrane displacement as
\begin{equation}
	We \sim \frac{64}{3} z^3 + \mathcal{O}(z^5), 
	\label{We3} 
\end{equation}
where $z$ is the membrane displacement defined in \cite{waldman2017camber}. By neglecting the higher order term $\mathcal{O}(z^5)$, \refeq{We3} can be rewritten as
\begin{equation}
	z \sim 0.3606 (We)^{\frac{1}{3}}.
	\label{We4} 
\end{equation}

It is worth mentioning that the coefficients in \refeq{We4} are similar to the coefficients determined from our numerical simulations in \refeq{We2}.

It can be seen from \reffig{fig:scaling} that the data show a good match with the empirical solution at lower Weber numbers. The data at higher Weber numbers with moderate to high angles of attack has a poorer agreement. This is attributed to the stronger nonlinearity effect at a higher Weber number (a relatively lower aeroelastic number). The proposed scaling relation provides a good way to explain how the aerodynamic performance behaves by correlating with the membrane deformation at different aeroelastic numbers. For flexible membrane wings with a fixed aeroelastic number, the increase of the membrane displacement from 0 to 0.06 results in an improvement of the Weber number of 0.0056. When the membrane displacement changes from 0.08 to 0.14, the Weber number has an increase of 0.0643. It means that the membrane wing with displacements larger than a critical value can produce more aerodynamic loads on the membrane surface rather than relatively small displacements. 

\subsection{Scaling Relations for Unsteady Membrane Dynamics} \label{sec:scaling2}
\begin{figure}[H]
	\centering
	\includegraphics[width=0.8\textwidth]{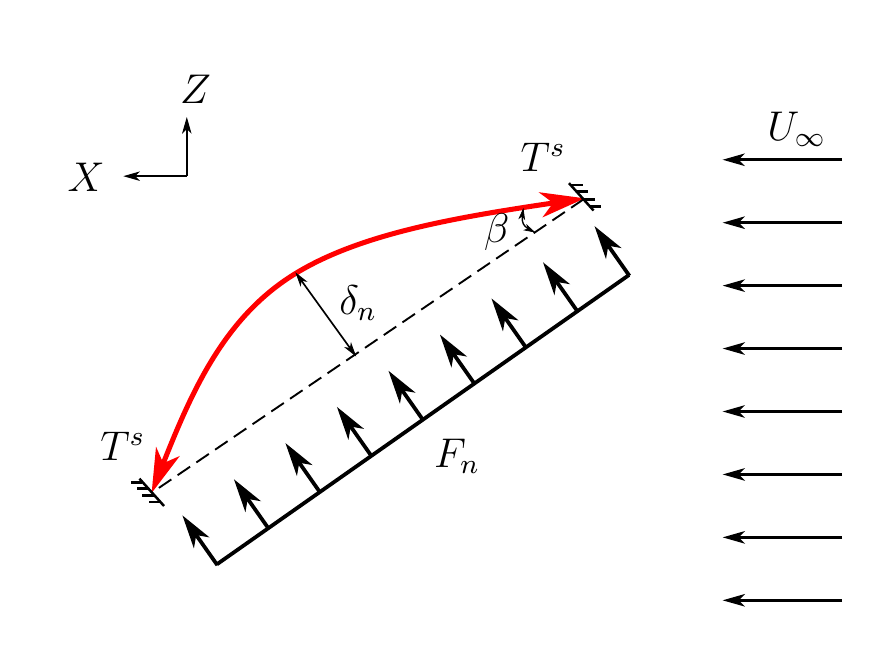}
	\caption{\label{fig:schematic}  Schematic of a two-dimensional membrane model with uniform aerodynamic load $F_n$ along the chord.}
\end{figure}

In \refse{sec:scaling1}, we studied the relationship between the mean membrane displacement and the mean aerodynamic loads. Except for the underlying mechanisms of the mean membrane responses, we also find that the membrane vibrations are correlated with the unsteady aerodynamic forces on the membrane surface in some underlying ways. We next explore the scaling relation for the unsteady membrane dynamics in this section.
It is known that vibrating flexible structures carry the oscillating fluid mass which makes the structure vibrate at a relatively lower frequency than in a vacuum. Effectively this process increases the apparent mass of the flexible structure which is generally referred to as the added mass effect. Hence it is of prime importance to estimate the added mass effects for the coupled fluid-structure dynamics. Herein, we consider a two-dimensional flexible membrane model with fixed leading and trailing edges immersed in a uniform flow as shown in \reffig{fig:schematic}. The aerodynamic load $p(\xi,t)$ is assumed to be uniformly distributed along the membrane chord $\xi$. The governing equation of a two-dimensional membrane wing is given as \cite{waldman2017camber,das2020deformation}
\begin{equation}
	\rho^s h \frac{\partial^2 \delta_n}{\partial t^2} - T^s \kappa =p(\xi,t),
	\label{unsteady_membrane0} 
\end{equation}
where $T^s=E^s h (\lambda-1)$ denotes the tension force along the membrane and $\lambda$ is the stretch ratio of the total length of the deformed membrane to the chord length. $\kappa$ is the curvature of the deformed wing, which can be written as $\kappa=\sin(\beta)/ \frac{c}{2}$. Here, $\beta$ is the contact angle between the deformed membrane and the chord at the leading and trailing edges. $\sin(\beta)$ can be further represented by $\delta_n / \frac{c}{2}$ based on a linearization assumption \cite{song2008aeromechanics,das2020compliant}. The second term on the left-hand side of \refeq{unsteady_membrane0} can be further written as
\begin{equation}
	{T^s} \kappa = 2 {T^s} \frac{\sin(\beta)}{c} =     
        \frac{ 4 E^s h (\lambda -1) \delta_n}{c^2},
	\label{unsteady_membrane1_1} 
\end{equation}

Substituting \refeq{unsteady_membrane1_1} into \refeq{unsteady_membrane0}, we have
\begin{equation}
	\rho^s h \frac{\partial^2 \delta_n}{\partial t^2} - \frac{ 4 E^s h (\lambda -1) \delta_n}{c^2} =p(\xi,t).
	\label{unsteady_membrane} 
\end{equation}

 Based on a simplified incompressible aerodynamic model \cite{chen2015research}, the aerodynamic load $p(\xi,t)$ can be decomposed into the aerodynamic acoustic pressure component $p_a(\xi,t)$ and the quasi-static wind pressure component $p_w(\xi,t)$ as
\begin{equation}
    p(\xi,t) = p_a(\xi,t) + p_w(\xi,t),
    \label{pressure_model} 
\end{equation}
where the aerodynamic acoustic pressure component represents the compressive action of the vibrating membrane on the airflow, and the quasi-static load contribution considers the mean aerodynamic pressure which changes with the shape change of the membrane. The expression of these two pressure components are given as
\begin{equation}
    p_a(\xi,t) = -\frac{2 \rho^f c}{i \pi} \frac{\partial^2 \delta_n}{\partial t^2}
    \label{pressure_model2} 
\end{equation}
\begin{equation}
    p_w(\xi,t) = - \frac{1}{2} \rho^f U_{\infty}^2 \left[ \frac{a^i}{A_{\delta_n^{\prime}} (\omega_s^i)^2} \frac{\partial^2 \delta_n}{\partial t^2} + \frac{b^i}{A_{\delta_n^{\prime}} \omega_s^i} \frac{\partial \delta_n}{\partial t} - a_0 \sin{\left( \frac{i \pi x}{c} \right)} \right]
    \label{pressure_model3} 
\end{equation}
where $\omega_s^i$ represents the vibration circular frequency of the $i$ mode. Here, $a^i$, $b^i$ and $a_0$ denote the shape-change coefficients of the amplitude of the pressure difference. By substituting \refeq{pressure_model}-(\ref{pressure_model3}) into \refeq{unsteady_membrane} and neglecting the damping term, \refeq{unsteady_membrane} can be written as
\begin{equation}
	(\rho^s h + m^{am}) \frac{\partial^2 \delta_n}{\partial t^2} - \frac{ 4 E^s h (\lambda -1) \delta_n}{c^2} = p_0(\xi,t),
	\label{pressure_model4} 
\end{equation}
where $ m^{am} $ is the added mass induced by the membrane vibration
\begin{equation}
	m^{am} = \frac{2 \rho^f c}{i \pi} + \frac{\rho^f U_{\infty}^2 a^i}{2 A_{\delta_n^{\prime}} (\omega_s^i)^2} 
	\label{pressure_model5} 
\end{equation}

The time-varying membrane displacement, the pressure difference and the tension force can be decomposed into the time-averaged and fluctuating components
\begin{equation}
	\delta_n = \overline{\delta_n} + \delta_n^{\prime}, \quad \quad p_0(\xi,t) = \overline{p_0(\xi,t)} + p_0^{\prime}(\xi,t)
    \label{unsteady_membrane2}
\end{equation}
By substituting \refeq{unsteady_membrane2}  into \refeq{pressure_model4}, we have
\begin{equation}
	(\rho^s h + m^{am}) \frac{\partial^2 (\overline{\delta_n} + \delta_n^{\prime})}{\partial t^2} - \frac{ 4 E^s h (\lambda -1) (\overline{\delta_n} + \delta_n^{\prime})}{c^2} = \overline{p_0(\xi,t)} + p_0^{\prime}(\xi,t)
	\label{unsteady_membrane3} 
\end{equation}
For the time-averaged membrane shape, the static force balance law satisfies the following  relation
\begin{equation}
    \frac{ 4 E^s h (\lambda -1) \overline{\delta_n} }{c^2} + \overline{p_0(\xi,t)}= 0.
	\label{unsteady_membrane4} 
\end{equation}
The second-order time partial derivative term of the mean displacement $\frac{\partial^2 \overline{\delta_n} }{\partial t^2}$ is zero. Thus, \refeq{unsteady_membrane3} can be rewritten as
\begin{equation}
	(\rho^s h + m^{am}) \frac{\partial^2 \delta_n^{\prime}}{\partial t^2} - \frac{ 4 E^s h (\lambda -1)  \delta_n^{\prime}}{c^2} 
    =  p_0^{\prime}(\xi,t)
	\label{unsteady_membrane5} 
\end{equation}

The time-dependent aerodynamic load fluctuation can be written into a sum of harmonic functions with an amplitude of $A^i_{p_0^{\prime}}$ and an oscillating circular frequency of $\omega^i_f$ at $i$-th order by the Fourier series, which is given as
\begin{equation}
	p_0^{\prime}(\xi,t) = \sum_{i=1}^{N} A^i_{p_0^{\prime}} \sin{(\omega^i_f t)}
    \label{unsteady_membrane6} 
\end{equation}
Similarly, the unsteady membrane vibration fluctuation can be written as
\begin{equation}
	\delta_n^{\prime} = \sum_{i=1}^{N} A^i_{\delta_n^{\prime}} \sin{(\omega^i_f t)}
    \label{unsteady_membrane7} 
\end{equation}
where $A^i_{\delta_n^{\prime}}$ is the membrane oscillating amplitude. Using \refeq{unsteady_membrane6}-(\ref{unsteady_membrane7}) into \refeq{unsteady_membrane5}, the force balance equation for each oscillating circular frequency $\omega^i_f$ can be written as
\begin{equation}
	(\rho^s h + m^{am}) ({\omega^{i}_f})^2 A^i_{\delta_n^{\prime}} \sin{(\omega^i_f t)} + \frac{4 E^s h (\lambda -1)}{c^2} A^{i}_{\delta_n^{\prime}}  \sin{(\omega^i_f t)} =  A^i_{p_0^{\prime}} \sin{(\omega^i_f t)},
	\label{unsteady_membrane9} 
\end{equation}

Next we can consider the force balance at a dominant frequency $\omega^{d}_f$ as follows
\begin{equation}
	(\rho^s h + m^{am}) ({\omega^{d}_f})^2 A^{d}_{\delta_n^{\prime}} + A^{d}_{\delta_n^{\prime}} \frac{4 E^s h (\lambda -1)}{c^2}  =  A^{d}_{p_0^{\prime}}
	\label{unsteady_membrane10} 
\end{equation}
By integrating along the whole membrane surface $S$, \refeq{unsteady_membrane10} can give the following force balance:
\begin{equation}
    \int_S \left[(\rho^s h + m^{am}) ({\omega^{d}_f})^2 A^{d}_{\delta_n^{\prime}} + A^{d}_{\delta_n^{\prime}} \frac{4 E^s h (\lambda -1)}{c^2} \right] \text{d}S = \int_S \left( A^{d}_{p_0^{\prime}} \right) \text{d}S
	\label{unsteady_membrane11} 
\end{equation}
Subsequently, \refeq{unsteady_membrane11} can be further written into a general form by removing the superscript $d$
\begin{equation}
	M_{eff} ({\omega_f})^2 A_{\delta_n^{\prime}} + A_{\delta_n^{\prime}} \frac{4 E^s h b (\lambda -1)}{c} =  A_{F_n^{\prime}}
	\label{unsteady_membrane12} 
\end{equation}
where $A_{F_n^{\prime}}$ represents the normal force amplitude and $M_{eff}=(\rho^s h + m^{am})bc$ is the effective mass of the oscillating membrane. The left-hand side of \refeq{unsteady_membrane12} represents the oscillating amplitude of the inertial force and the tension force driven by the membrane vibrations. We define this force amplitude of the combined inertia-elastic fluctuation as $A_{F_{(I+E)}^{\prime}}=A_{F_I^{\prime}}+A_{F_E^{\prime}}=M_{eff} ({\omega_f})^2 A_{\delta_n^{\prime}} + A_{\delta_n^{\prime}} \frac{4 E^s h b (\lambda -1)}{c}$. The right-hand side of the equation is the oscillating amplitude of the aerodynamic loads on the membrane surface. The scaling relation shown in \refeq{unsteady_membrane12} reveals the relationship between the unsteady aerodynamic loads and the membrane vibrations. \refeq{unsteady_membrane12} can be further written into a non-dimensionalized form as
\begin{equation}
	\left( 8\pi^2 m^* + \frac{16 \pi}{i}  \right) {St}^2 A^*_{\delta_n^{\prime}} + a^i + 4 Ae (\lambda -1) A^*_{\delta_n^{\prime}} =  A_{C_n^{\prime}}
	\label{unsteady_membrane13} 
\end{equation}
where $m^*$ is the mass ratio. Herein, $St$ is the Strouhal number defined as
\begin{equation}
    St = \frac{fc}{U_{\infty}},
    \label{unsteady_membrane14}
\end{equation}
and the chord-normalized membrane displacement amplitude $A^*_{\delta_n^{\prime}}$ is given as
\begin{equation}
    A^*_{\delta_n^{\prime}} = \frac{A_{\delta_n^{\prime}}}{c}.
    \label{unsteady_membrane15}
\end{equation}
The left-hand side of \refeq{unsteady_membrane13} is the non-dimensionalized form of $A_{F_{(I+E)}^{\prime}}$, which can be represented by
\begin{equation}
    A_{C_{(I+E)}^{\prime}} = \frac{A_{F_{(I+E)}^{\prime}}}{\frac{1}{2} \rho^f U_{\infty}^2 S} = \left( 8\pi^2 m^* + \frac{16 \pi}{i}  \right) {St}^2 A^*_{\delta_n^{\prime}} + a^i + 4 Ae (\lambda -1) A^*_{\delta_n^{\prime}} 
    \label{unsteady_membrane16}
\end{equation}
The amplitude of the normal force coefficient $A_{C_n^{\prime}}$ is defined as
\begin{equation}
     A_{C_n^{\prime}} = \frac{A_{F_n^{\prime}}}{\frac{1}{2} \rho^f U_{\infty}^2 S}
     \label{unsteady_membrane17}
\end{equation}

\begin{figure}[H]
	\centering
	\includegraphics[width=1.0\textwidth]{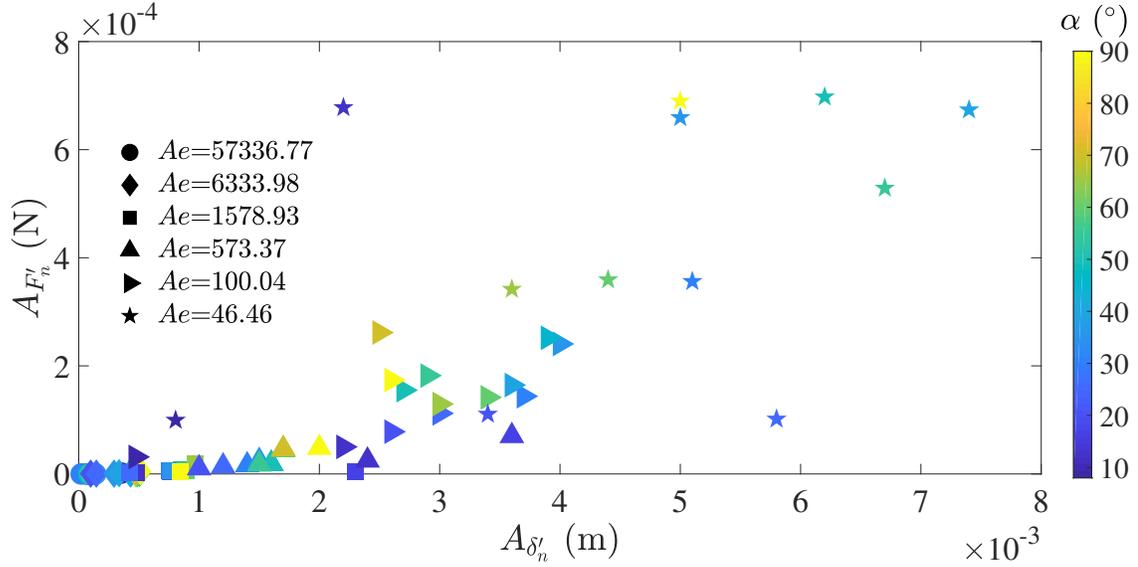}
	\caption{\label{fig:force_scaling}  Amplitudes of the normal force fluctuation as a function of amplitudes of the membrane displacement fluctuation at different angles of attack.}
\end{figure}

\refFig{fig:force_scaling} presents all the simulated data in the $A_{\delta_n^{\prime}} - A_{F_n^{\prime}}$ phase plane. The data label is colored by the angle of attack and the shape is indicated by the aeroelastic number. It can be seen from the plot that the membrane displacement fluctuation shows an overall positive correlation with the normal force fluctuation. The flexible membrane can produce significantly more normal force fluctuation when the membrane displacement fluctuation amplitude exceeds 1.3$\%$ chord length.
 
\begin{figure}[H]
	\centering
	\subfloat[][]{\includegraphics[width=1.0\textwidth]{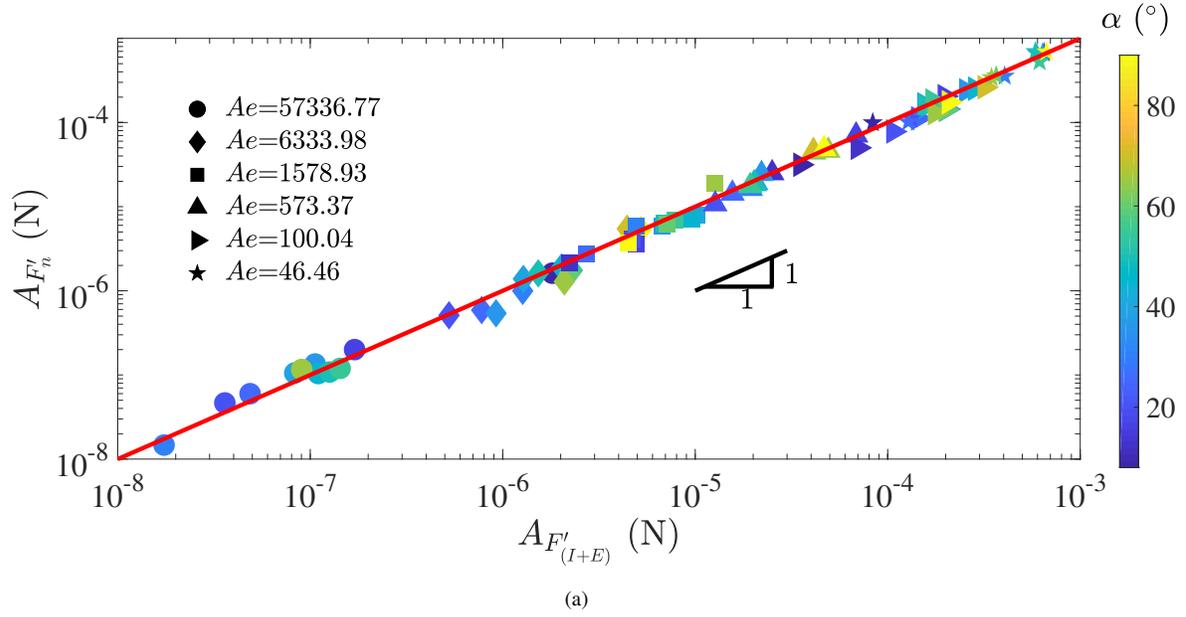}\label{fig:force_scaling2a}}
	\\
	\subfloat[][]{\includegraphics[width=1.0\textwidth]{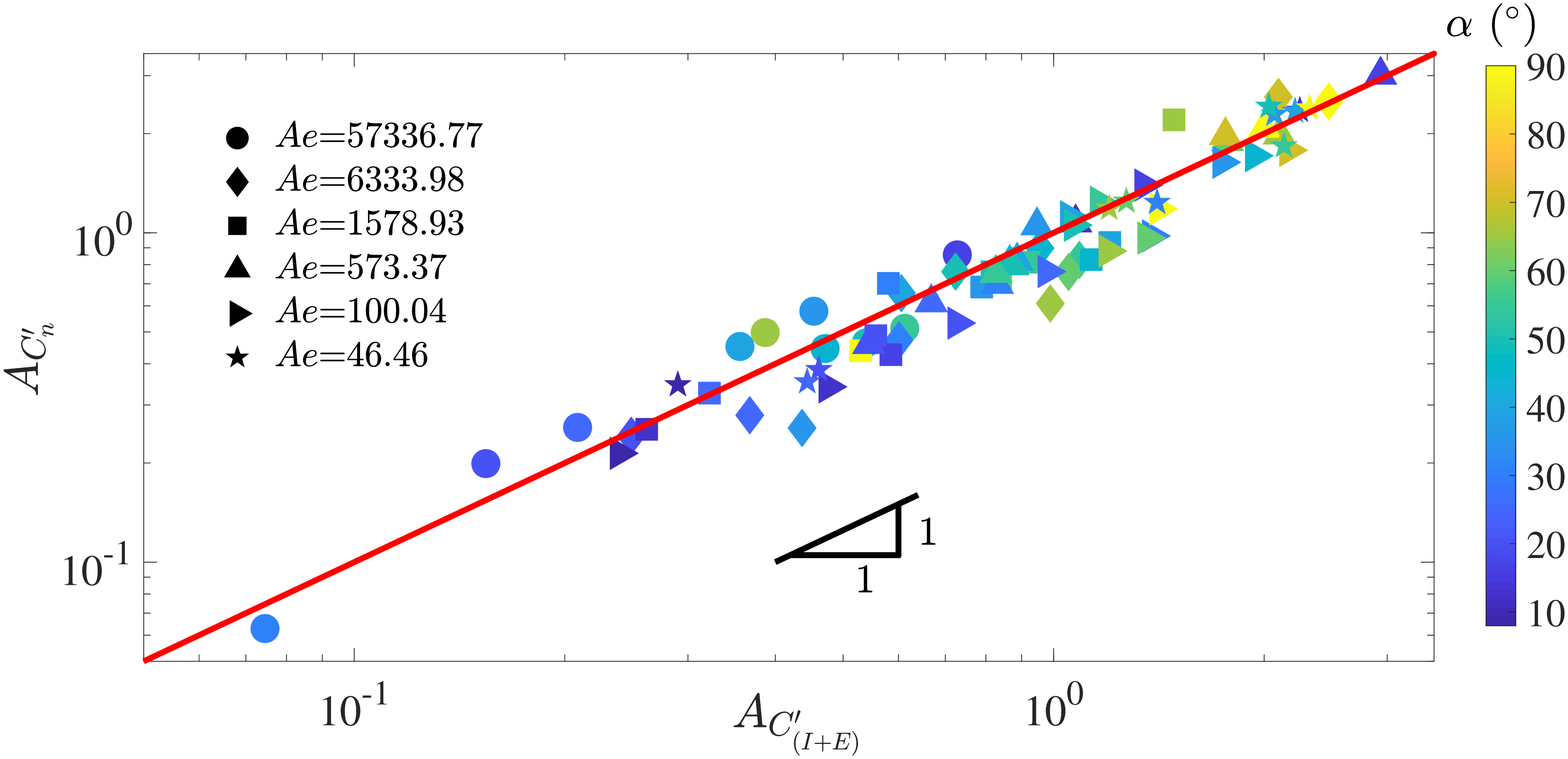}\label{fig:force_scaling2b}}
	\caption{\label{fig:force_scaling2}  Scaling relations for (a) amplitudes of unsteady force and kinematics fluctuations and (b) non-dimensional amplitudes of unsteady force and kinematics fluctuations.}
\end{figure}

\refFig{fig:force_scaling2} \subref{fig:force_scaling2a} shows the relationship between the normal force fluctuation and the inertial force fluctuation for all simulation data. It can be seen that all the data collapse together onto a straight line with slope 1. The finding observed in \reffig{fig:force_scaling2} is consistent with the force-kinematics scaling relation derived by the linearization assumption in \refeq{unsteady_membrane7}. We can observe from \reffig{fig:force_scaling2} that the forces produced by the membrane wing show a growing trend as the aeroelastic number decreases. The dot labels with $Ae=57336.77$ are located on the left bottom corner and the star labels with $Ae=46.46$ occupy the right top corner. The results indicate that a flexible wing can produce larger force fluctuation than a rigid wing. \refFig{fig:force_scaling2} \subref{fig:force_scaling2b} presents the non-dimensional force-kinematics scaling relations evaluated from the numerical simulation data based on \refeq{unsteady_membrane13}. Because all the simulated cases in this paper have the same membrane structural mass, the generated forces with larger amplitudes are mainly contributed by the inertial effect related to the fluid-added mass  and the membrane tension.

Except for the displacement fluctuation amplitude, we can see from \refeq{unsteady_membrane12} that the aerodynamic force fluctuation amplitude is also governed by the dominant oscillating frequency of the surrounding fluid flows. This dominant frequency is influenced by the coupling between the vortex shedding frequency and the natural frequency of the deformed membrane \cite{li2020flow_accept}. This finding provides another way to design active control methods for changing the unsteady aerodynamic force characteristics by manipulating the frequency of the coupled fluid-membrane system. For example, one can stretch or release the membrane like a bat to change the membrane natural frequency. Active flow control methods, like jet flow into the wake, can be applied to alter the surrounding flow features \cite{yao2017feedback}. The flexible membrane can also be forced to oscillate in a specific frequency by applying external periodic loads \cite{sun2016nonlinear} or smart materials \cite{curet2014aerodynamic,bohnker2019control,huang2021fluid}.

The findings given by the two scaling relations can offer effective guidelines to design active control methods for producing optimal aerodynamic performance. Similar to the investigations in our previous studies \cite{li2020flow_accept,li2022aeroelastic} on a three-dimensional membrane wing, the aerodynamic performance is governed by two effects, namely (i) camber effect and (ii) flow-induced vibration. Some active control manners can be designed and applied to the flexible membrane and produce desirable aerodynamic forces by governing the wing camber and the flow-induced vibration. Smart sensors can be installed in the membrane wing to monitor and measure the surrounding flow features and the membrane deformation characteristics. The collected data can be used as the input of the scaling relations shown in \refeq{We2} and \refeq{unsteady_membrane13} to predict the aerodynamic forces for active control purposes.

\section{Conclusion} \label{sec:section5}
In this study, we presented a systematic numerical study on the aeroelastic characteristics of two-dimensional flexible wings. The membrane aeroelasticity was simulated in a wide parameter space to characterize the effects of the angle of attack and the aeroelastic number. Three distinct regimes can be classified from the variation of the aerodynamic performance as a function of the angle of attack, namely the pre-stall, the transitional stall and the deep stall. The flexible membrane reached its maximum time-averaged lift coefficient at the stall angle near $\alpha=50^\circ$ and then quickly reduced to zero at $90^\circ$. The largest time-averaged lift-to-drag ratio was observed near $\alpha=8^\circ$. A smaller aeroelastic number, which means a relatively softer membrane, was beneficial to improve the time-averaged lift coefficient and the lift-to-drag ratio as well as producing larger mean membrane deformation. A transition line at $20^\circ$ was determined from the phase diagram of the mean drag coefficient. The mean drag can be enhanced below the line while it is suppressed above the line when the aeroelastic number became larger. The optimal time-averaged lift and lift-to-drag ratio regions were determined by contour lines in the phase diagram. Empirical solutions of the contour line envelop were constructed based on exponent curve fitting functions. These empirical solutions provided design guidelines for selecting the optimal combination of the angle of attack and aeroelastic number when designing highly efficient morphing air vehicles. The role of flexibility was assessed by comparing the aerodynamic forces between a rigid flat wing, a rigid cambered wing and a flexible membrane over a wide range of angles of attack. By comparing the turbulent intensities, we found that the high energy transport in the boundary layer caused by the flow-induced vibration was beneficial to improve the aerodynamic performance. 

The non-dimensional parameter (Weber number) was considered to describe the connection between the aerodynamic loads on the membrane surface and the aeroelastic number. Based on our high-fidelity numerical simulations, we suggested a scaling relation between the Weber number and the mean membrane displacement in the studied parameter space. The membrane displacement was positively correlated with the rough one-third power law of the Weber number over a wide range of angles of attack. Based on the proposed scaling relation, increasing the wing camber or releasing the membrane rigidity was the key to improving the aerodynamic force. Additionally, the aerodynamic force can be significantly enhanced when the membrane displacement exceeded a critical value, 6$\%$ of the membrane chord length. By deriving from the governing equations of the two-dimensional membrane and a simplified incompressible aerodynamic model, we further proposed a scaling relation between the aerodynamic force fluctuation and the membrane vibration fluctuation. This new scaling relation revealed that the aerodynamic force fluctuation showed a positive correlation with the mass ratio, the Strouhal number, the aeroelastic number and the membrane vibration fluctuation, which formed the combined inertia-elastic force. These two scaling relations offered mathematical models to adjust the aerodynamic forces according to the flight environment through the adjustment of the membrane deformation, the structural properties and the frequency spectra. These findings can facilitate the development of active flow control strategies toward optimal aerodynamic performance.

\section*{Acknowledgements}
The first author would like to acknowledge the support from the National Natural Science Foundation of China (NSFC) (Grant Number 12202362). The second author would like to acknowledge the support from the University of British Columbia and the Natural Sciences and Engineering Research Council of Canada (NSERC). We greatly acknowledge Professor Kenny Breuer for providing valuable feedback about the non-dimensional scaling relations.

\section*{Declaration of interests}
The authors report no conflict of interest.

\section*{Appendix. Comparative Study and Validation} \label{sec:appendixa}
\begin{figure}[H]
	\centering
	\subfloat[][]{\includegraphics[width=0.5\textwidth]{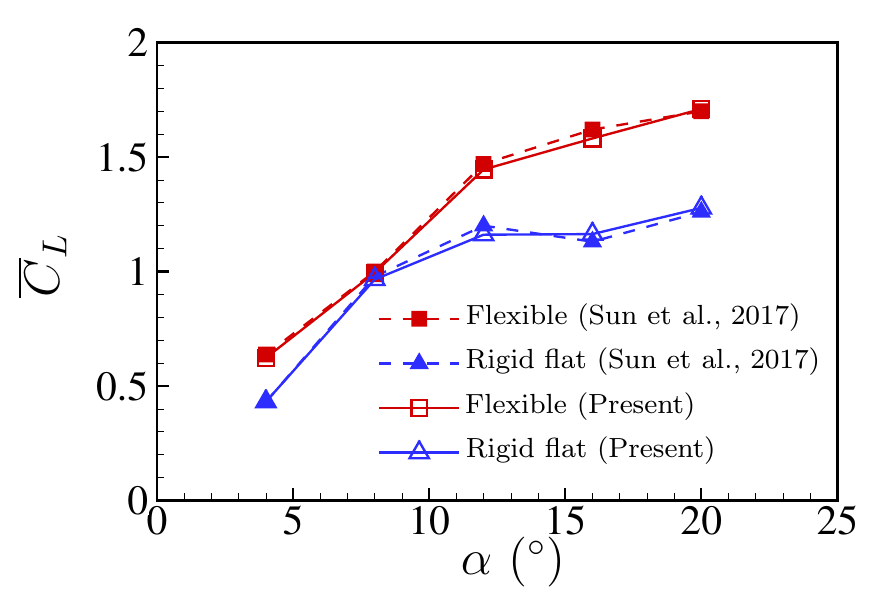}\label{fig:verficationa}}
	\subfloat[][]{\includegraphics[width=0.5\textwidth]{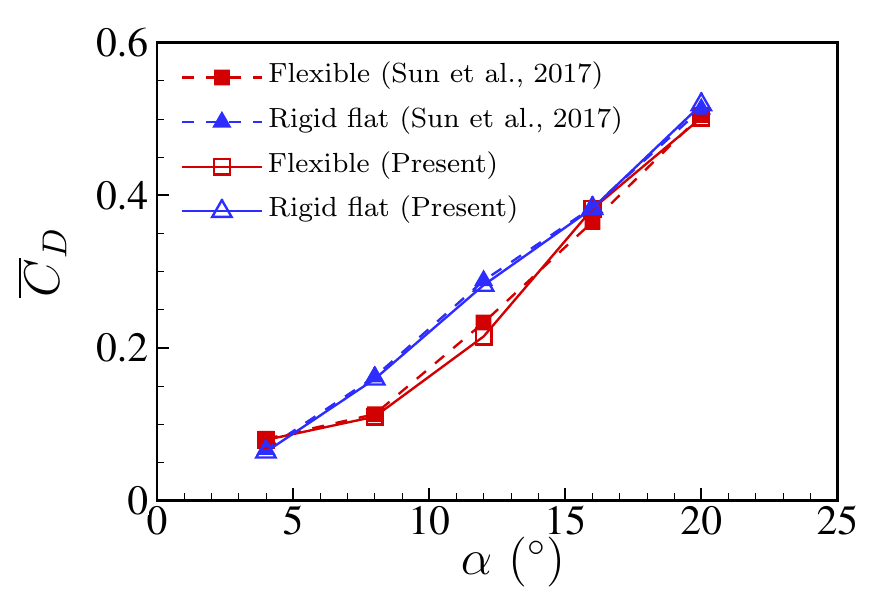}\label{fig:verficationb}}
	\caption{\label{fig:comparative}  Comparison of time-averaged (a) lift coefficient and (b) drag coefficient between the present data and the numerical simulation results in Sun et al. \cite{sun2017effect}.}
\end{figure}

The two-dimensional flexible membrane model established in \refse{sec:section3} is considered for a comparative study with other numerical simulation results. The membrane dynamics is simulated in the range of $\alpha \in [4^\circ, 20^\circ]$ with an interval of $4^\circ$ at $Re$=2500, $m^*$=0.589 and $Ae$=100.04. The time-averaged lift and drag coefficients of the flexible membrane and its rigid flat counterpart are compared with the numerical simulation results from Sun et al. \cite{sun2017effect} in \reffig{fig:comparative}. It can be seen that the aerodynamic characteristics of both wings obtained from our high-fidelity solver show a good agreement with the aeroelastic solver presented in Sun et al. \cite{sun2017effect}.

\begin{figure}[H]
	\centering
	\includegraphics[width=0.8\textwidth]{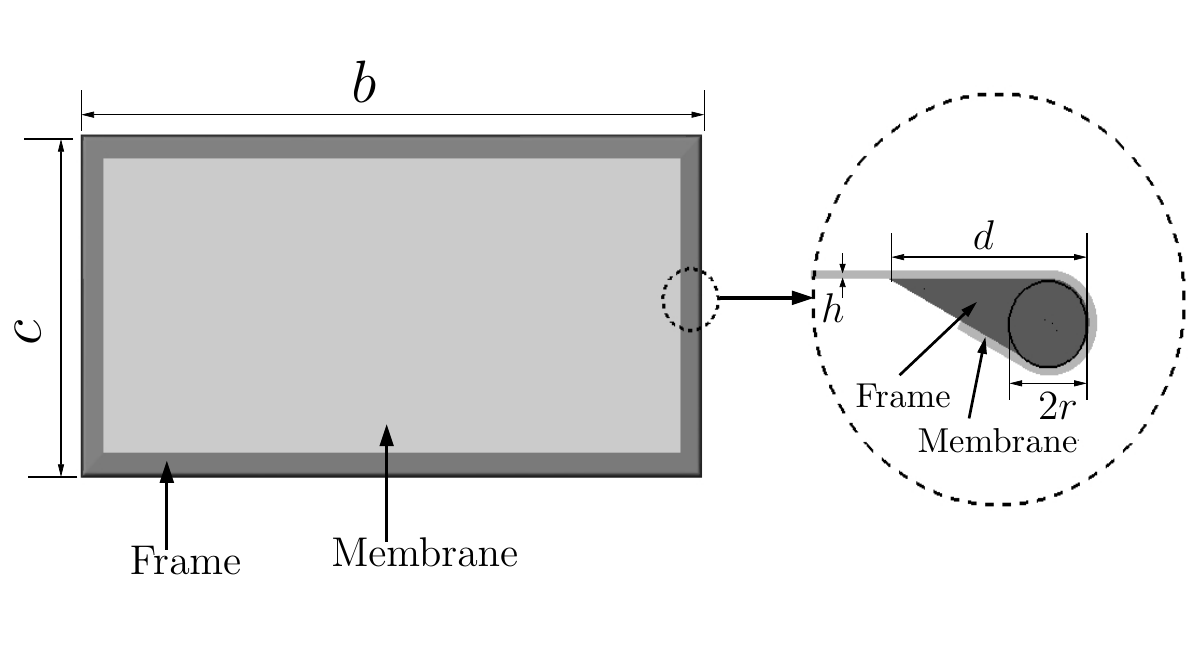}
	\caption{Three-dimensional rectangular membrane wing geometry.}
	\label{membranegeo}
\end{figure}

A three-dimensional flexible membrane model conducted in experimental studies \cite{rojratsirikul2010unsteady} is considered for comparative study and validation purposes. This rectangular membrane has a chord length of $c$=68.75mm and a span length of $b$=137.5mm, resulting in an aspect ratio of two. \refFig{membranegeo} presents a geometry schematic of the flexible wing. The membrane component is made of latex rubber with a thickness of $h$=0.2mm, Young's modulus of $E^s$=2.2MPa and material density of $\rho^s$=1000${\rm{kg \cdot m^{-3}}}$. Three key non-dimensional physical parameters that govern the membrane dynamic responses are given as $Re$=24300, $Ae$=423.14 and $m^*$=2.4. 

\begin{figure}[H]
	\centering
	\subfloat[][]{\includegraphics[width=0.5\textwidth]{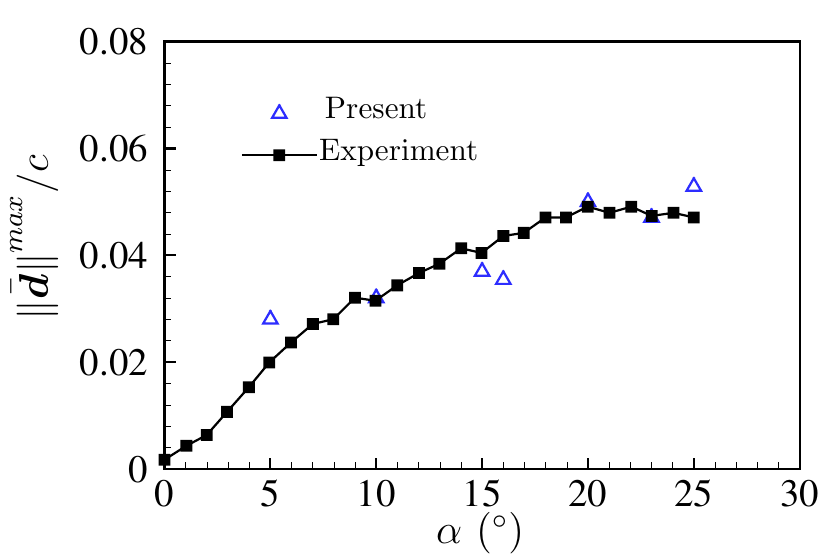}\label{averagediscna}}
	\subfloat[][]{\includegraphics[width=0.5\textwidth]{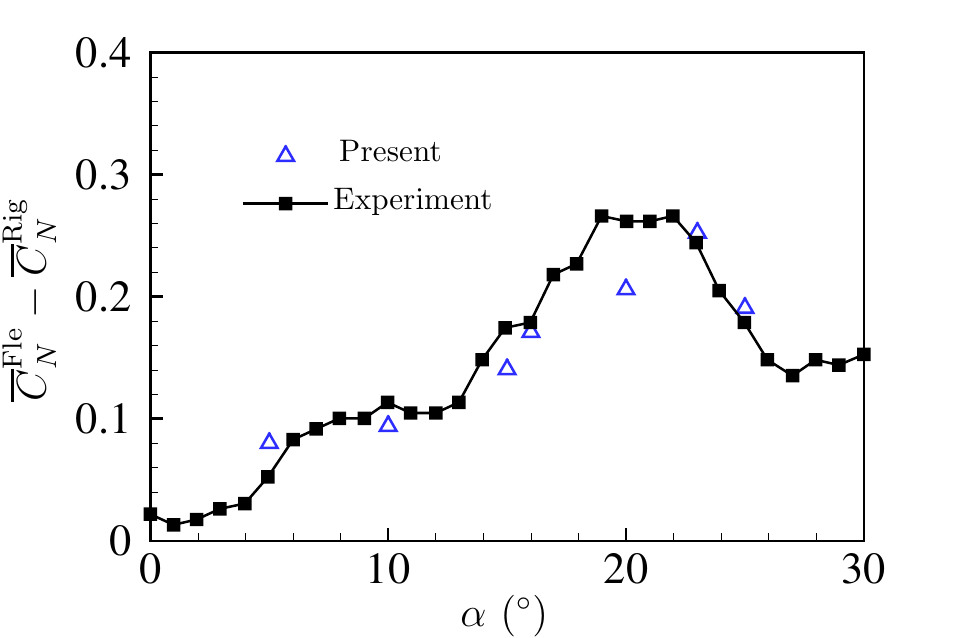}\label{averagediscnb}}
	\\
       \subfloat[][]{\includegraphics[width=0.5\textwidth]{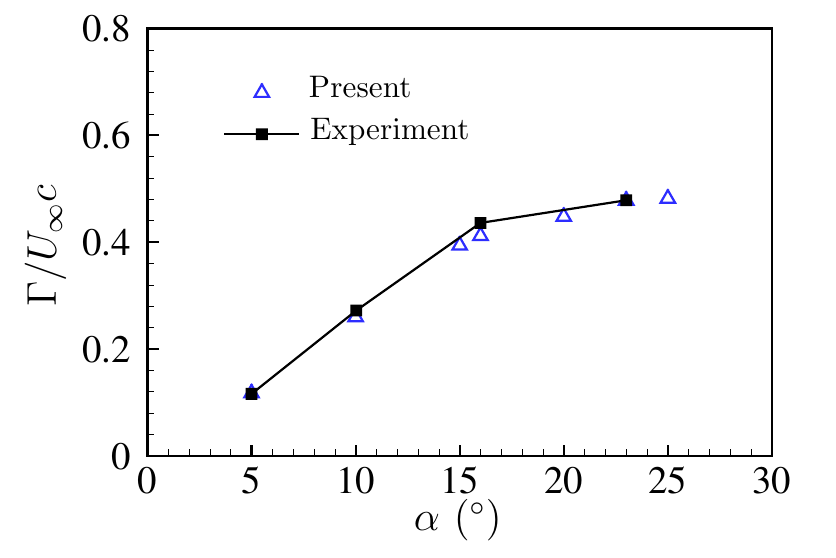}\label{averagediscnc}}
	\subfloat[][]{\includegraphics[width=0.5\textwidth]{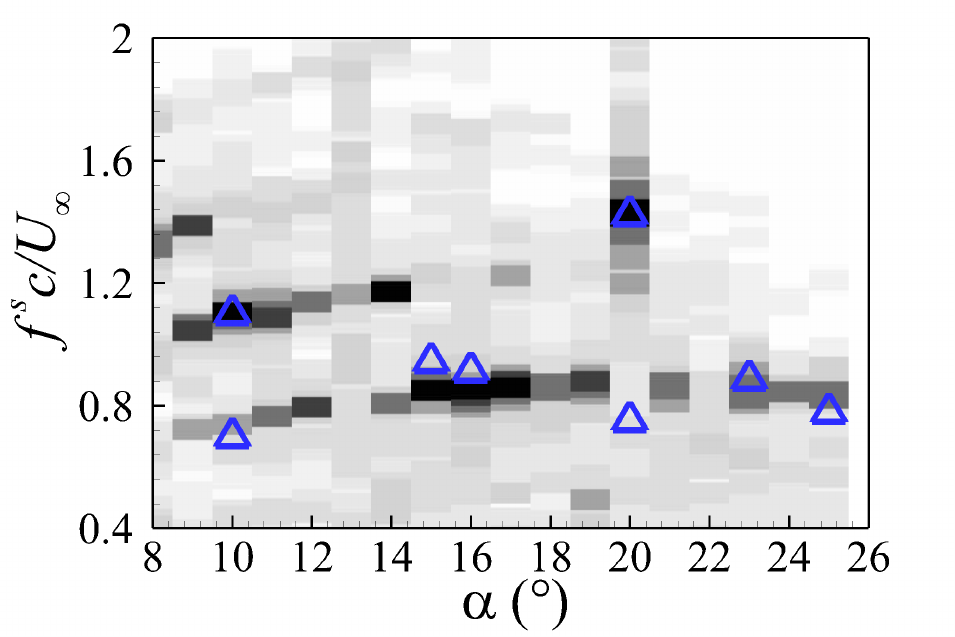}\label{averagediscnd}}
	\caption{Comparison between the present data and the experimental results \cite{rojratsirikul2010unsteady,rojratsirikul2011flow} for: (a) magnitude of time-averaged normalized maximum membrane deformation, (b)  time-averaged normal force coefficient  difference between the flexible membrane wing and rigid wing, (c) normalized circulation of the vortices at the wingtip on a plane normal to the freestream direction and (d) membrane vibration frequency spectra at the location with maximum standard deviation of the membrane deflection.}
	\label{averagediscn}
\end{figure}

Our high-fidelity aeroelastic solver is employed to simulate the membrane dynamic responses for validation purposes. \refFig{averagediscn} presents a comparison of the mean membrane displacement, the time-averaged aerodynamic forces, the circulation of tip vortices and the vibration frequency spectra with the experimental results \cite{rojratsirikul2010unsteady,rojratsirikul2011flow}. The overall trends of the membrane dynamics are reasonably captured by our simulation results. More details of the investigations of the three-dimensional rectangular membrane can be found in previous studies \cite{li2022aeroelastic,li2020flow_accept}.

\bibliography{sample}

\begin{thebibliography}{43}
\newcommand{\enquote}[1]{``#1''}
\providecommand{\natexlab}[1]{#1}
\providecommand{\url}[1]{\texttt{#1}}
\providecommand{\urlprefix}{URL }
\expandafter\ifx\csname urlstyle\endcsname\relax
  \providecommand{\doi}[1]{\discretionary{}{}{}https://doi.org/#1}\else
  \providecommand{\doi}[1]{\discretionary{}{}{}\urlstyle{rm}\url{https://doi.org/#1}}\fi

\bibitem[{Shyy et~al.(2010)Shyy, Aono, Chimakurthi, Trizila, Kang, Cesnik, and
  Liu}]{shyy2010recent}
Shyy, W., Aono, H., Chimakurthi, S.~K., Trizila, P., Kang, C.-K., Cesnik,
  C.~E., and Liu, H., \enquote{Recent progress in flapping wing aerodynamics
  and aeroelasticity,} \emph{Progress in Aerospace Sciences}, Vol.~46, No.~7,
  2010, pp. 284--327.

\bibitem[{Jiakun et~al.(2021)Jiakun, Zhe, Fangbao, and Gang}]{jiakun2021review}
Jiakun, H., Zhe, H., Fangbao, T., and Gang, C., \enquote{Review on bio-inspired
  flight systems and bionic aerodynamics,} \emph{Chinese Journal of
  Aeronautics}, Vol.~34, No.~7, 2021, pp. 170--186.

\bibitem[{Hefler et~al.(2021)Hefler, Kang, Qiu, and Shyy}]{hefler2021distinct}
Hefler, C., Kang, C.-k., Qiu, H., and Shyy, W., \emph{Distinct Aerodynamics of
  Insect-Scale Flight}, Cambridge University Press, 2021.

\bibitem[{Tiomkin and Raveh(2021)}]{tiomkin2021review}
Tiomkin, S., and Raveh, D.~E., \enquote{A review of membrane-wing
  aeroelasticity,} \emph{Progress in Aerospace Sciences}, Vol. 126, 2021, p.
  100738.

\bibitem[{Hubel et~al.(2010)Hubel, Hristov, Swartz, and Breuer}]{hubel2010time}
Hubel, T.~Y., Hristov, N.~I., Swartz, S.~M., and Breuer, K.~S.,
  \enquote{Time-resolved wake structure and kinematics of bat flight,}
  \emph{Animal Locomotion}, Springer, 2010, pp. 371--381.

\bibitem[{Wang et~al.(2015)Wang, Zhang, He, and Liu}]{wang2015lift}
Wang, S., Zhang, X., He, G., and Liu, T., \enquote{Lift enhancement by bats'
  dynamically changing wingspan,} \emph{Journal of the Royal Society
  Interface}, Vol.~12, No. 113, 2015, p. 20150821.

\bibitem[{Wu(2011)}]{wu2011fish}
Wu, T.~Y., \enquote{Fish swimming and bird/insect flight,} \emph{Annual Review
  of Fluid Mechanics}, Vol.~43, 2011, pp. 25--58.

\bibitem[{Alexander(2015)}]{alexander2015wing}
Alexander, D.~E., \emph{On the wing: insects, pterosaurs, birds, bats and the
  evolution of animal flight}, Oxford University Press, USA, 2015.

\bibitem[{Hedenstr{\"o}m and Johansson(2015)}]{hedenstrom2015bat}
Hedenstr{\"o}m, A., and Johansson, L.~C., \enquote{Bat flight: aerodynamics,
  kinematics and flight morphology,} \emph{The Journal of experimental
  biology}, Vol. 218, No.~5, 2015, pp. 653--663.

\bibitem[{Rojratsirikul et~al.(2009)Rojratsirikul, Wang, and
  Gursul}]{rojratsirikul2009unsteady}
Rojratsirikul, P., Wang, Z., and Gursul, I., \enquote{Unsteady fluid--structure
  interactions of membrane airfoils at low Reynolds numbers,} \emph{Experiments
  in fluids}, Vol.~46, No.~5, 2009, p. 859.

\bibitem[{Rojratsirikul et~al.(2010{\natexlab{a}})Rojratsirikul, Wang, and
  Gursul}]{rojratsirikul2010effect}
Rojratsirikul, P., Wang, Z., and Gursul, I., \enquote{Effect of pre-strain and
  excess length on unsteady fluid--structure interactions of membrane
  airfoils,} \emph{Journal of Fluids and Structures}, Vol.~26, No.~3,
  2010{\natexlab{a}}, pp. 359--376.

\bibitem[{Tregidgo et~al.(2013)Tregidgo, Wang, and
  Gursul}]{tregidgo2013unsteady}
Tregidgo, L., Wang, Z., and Gursul, I., \enquote{Unsteady fluid--structure
  interactions of a pitching membrane wing,} \emph{Aerospace Science and
  Technology}, Vol.~28, No.~1, 2013, pp. 79--90.

\bibitem[{Tzezana and Breuer(2019)}]{tzezana2019thrust}
Tzezana, G.~A., and Breuer, K.~S., \enquote{Thrust, drag and wake structure in
  flapping compliant membrane wings,} \emph{Journal of Fluid Mechanics}, Vol.
  862, 2019, pp. 871--888.

\bibitem[{Bleischwitz et~al.(2018)Bleischwitz, De~Kat, and
  Ganapathisubramani}]{bleischwitz2018near}
Bleischwitz, R., De~Kat, R., and Ganapathisubramani, B., \enquote{Near-wake
  characteristics of rigid and membrane wings in ground effect,} \emph{Journal
  of Fluids and Structures}, Vol.~80, 2018, pp. 199--216.

\bibitem[{Serrano-Galiano et~al.(2018)Serrano-Galiano, Sandham, and
  Sandberg}]{serrano2018fluid}
Serrano-Galiano, S., Sandham, N.~D., and Sandberg, R.~D.,
  \enquote{Fluid--structure coupling mechanism and its aerodynamic effect on
  membrane aerofoils,} \emph{Journal of Fluid Mechanics}, Vol. 848, 2018, pp.
  1127--1156.

\bibitem[{Mavroyiakoumou and Alben(2021)}]{mavroyiakoumou2021dynamics}
Mavroyiakoumou, C., and Alben, S., \enquote{Dynamics of tethered membranes in
  inviscid flow,} \emph{arXiv preprint arXiv:2106.08219}, 2021.

\bibitem[{Song et~al.(2008)Song, Tian, Israeli, Galvao, Bishop, Swartz, and
  Breuer}]{song2008aeromechanics}
Song, A., Tian, X., Israeli, E., Galvao, R., Bishop, K., Swartz, S., and
  Breuer, K., \enquote{Aeromechanics of membrane wings with implications for
  animal flight,} \emph{AIAA journal}, Vol.~46, No.~8, 2008, pp. 2096--2106.

\bibitem[{Gordnier and Attar(2014)}]{gordnier2014impact}
Gordnier, R.~E., and Attar, P.~J., \enquote{Impact of flexibility on the
  aerodynamics of an aspect ratio two membrane wing,} \emph{Journal of Fluids
  and Structures}, Vol.~45, 2014, pp. 138--152.

\bibitem[{Mavroyiakoumou and Alben(2020)}]{mavroyiakoumou2020large}
Mavroyiakoumou, C., and Alben, S., \enquote{Large-amplitude membrane flutter in
  inviscid flow,} \emph{Journal of Fluid Mechanics}, Vol. 891, 2020.

\bibitem[{Li et~al.(2021{\natexlab{a}})Li, Jaiman, and
  Khoo}]{li2020flow_accept}
Li, G., Jaiman, R.~K., and Khoo, B.~C., \enquote{Flow-excited membrane
  instability at moderate Reynolds numbers,} \emph{Journal of Fluid Mechanics},
  Vol. 929, 2021{\natexlab{a}}.

\bibitem[{Li et~al.(2022)Li, Jaiman, and Khoo}]{li2022aeroelastic}
Li, G., Jaiman, R.~K., and Khoo, B.~C., \enquote{Aeroelastic mode decomposition
  framework and mode selection mechanism in fluid--membrane interaction,}
  \emph{Journal of Fluids and Structures}, Vol. 108, 2022, p. 103428.

\bibitem[{Gordnier(2009)}]{gordnier2009high}
Gordnier, R.~E., \enquote{High fidelity computational simulation of a membrane
  wing airfoil,} \emph{Journal of Fluids and Structures}, Vol.~25, No.~5, 2009,
  pp. 897--917.

\bibitem[{Rojratsirikul et~al.(2011)Rojratsirikul, Genc, Wang, and
  Gursul}]{rojratsirikul2011flow}
Rojratsirikul, P., Genc, M., Wang, Z., and Gursul, I., \enquote{Flow-induced
  vibrations of low aspect ratio rectangular membrane wings,} \emph{Journal of
  Fluids and Structures}, Vol.~27, No.~8, 2011, pp. 1296--1309.

\bibitem[{Swartz and Konow(2015)}]{swartz2015advances}
Swartz, S., and Konow, N., \enquote{Advances in the study of bat flight: the
  wing and the wind,} \emph{Canadian Journal of Zoology}, Vol.~93, No.~12,
  2015, pp. 977--990.

\bibitem[{Bleischwitz et~al.(2015)Bleischwitz, de~Kat, and
  Ganapathisubramani}]{bleischwitz2015aspect}
Bleischwitz, R., de~Kat, R., and Ganapathisubramani, B., \enquote{Aspect-ratio
  effects on aeromechanics of membrane wings at moderate Reynolds numbers,}
  \emph{AIAA Journal}, Vol.~53, No.~3, 2015, pp. 780--788.

\bibitem[{Newman(1987)}]{newman1987aerodynamic}
Newman, B.~G., \enquote{Aerodynamic theory for membranes and sails,}
  \emph{Progress in aerospace sciences}, Vol.~24, No.~1, 1987, pp. 1--27.

\bibitem[{Fan and Xia(2014)}]{fan2014simulation}
Fan, Y., and Xia, J., \enquote{Simulation of 3D parachute fluid--structure
  interaction based on nonlinear finite element method and preconditioning
  finite volume method,} \emph{chinese Journal of Aeronautics}, Vol.~27, No.~6,
  2014, pp. 1373--1383.

\bibitem[{Das et~al.(2020{\natexlab{a}})Das, Mathai, and
  Breuer}]{das2020compliant}
Das, A., Mathai, V., and Breuer, K., \enquote{Compliant membranes exhibit
  enhanced drag due to membrane fluctuations,} \emph{APS Division of Fluid
  Dynamics Meeting Abstracts}, 2020{\natexlab{a}}, pp. Q02--016.

\bibitem[{Sun et~al.(2017)Sun, Ren, and Zhang}]{sun2017nonlinear}
Sun, X., Ren, X., and Zhang, J., \enquote{Nonlinear dynamic responses of a
  perimeter-reinforced membrane wing in laminar flows,} \emph{Nonlinear
  Dynamics}, Vol.~88, No.~1, 2017, pp. 749--776.

\bibitem[{Tiomkin and Raveh(2019)}]{tiomkin2019membrane}
Tiomkin, S., and Raveh, D., \enquote{On membrane-wing stability in laminar
  flow,} \emph{Journal of Fluids and Structures}, Vol.~91, 2019, p. 102694.

\bibitem[{Waldman and Breuer(2017)}]{waldman2017camber}
Waldman, R.~M., and Breuer, K.~S., \enquote{Camber and aerodynamic performance
  of compliant membrane wings,} \emph{Journal of Fluids and Structures},
  Vol.~68, 2017, pp. 390--402.

\bibitem[{Jaiman et~al.(2016)Jaiman, Pillalamarri, and Guan}]{jaiman2016stable}
Jaiman, R., Pillalamarri, N., and Guan, M., \enquote{A stable second-order
  partitioned iterative scheme for freely vibrating low-mass bluff bodies in a
  uniform flow,} \emph{Computer Methods in Applied Mechanics and Engineering},
  Vol. 301, 2016, pp. 187--215.

\bibitem[{Li et~al.(2019)Li, Law, and Jaiman}]{li2018novel}
Li, G., Law, Y.~Z., and Jaiman, R.~K., \enquote{A novel 3D variational
  aeroelastic framework for flexible multibody dynamics: Application to
  bat-like flapping dynamics,} \emph{Computers \& Fluids}, Vol. 180, 2019, pp.
  96--116.

\bibitem[{Li et~al.(2021{\natexlab{b}})Li, Kemp, Jaiman, and Khoo}]{li2021high}
Li, G., Kemp, G., Jaiman, R.~K., and Khoo, B.~C., \enquote{A high-fidelity
  numerical study on the propulsive performance of pitching flexible plates,}
  \emph{Physics of Fluids}, Vol.~33, No.~5, 2021{\natexlab{b}}, p. 051901.

\bibitem[{Sun and Zhang(2017)}]{sun2017effect}
Sun, X., and Zhang, J., \enquote{Effect of the reinforced leading or trailing
  edge on the aerodynamic performance of a perimeter-reinforced membrane wing,}
  \emph{Journal of Fluids and Structures}, Vol.~68, 2017, pp. 90--112.

\bibitem[{Das et~al.(2020{\natexlab{b}})Das, Mathai, and
  Breuer}]{das2020deformation}
Das, A., Mathai, V., and Breuer, K., \enquote{Deformation, forces, and flows
  associated with extremely compliant membrane disks,} \emph{AIAA Scitech 2020
  Forum}, 2020{\natexlab{b}}, p. 1049.

\bibitem[{Chen et~al.(2015)Chen, Wu, and Sun}]{chen2015research}
Chen, Z., Wu, Y., and Sun, X., \enquote{Research on the added mass of open-type
  one-way tensioned membrane structure in uniform flow,} \emph{Journal of Wind
  Engineering and Industrial Aerodynamics}, Vol. 137, 2015, pp. 69--77.

\bibitem[{Yao and Jaiman(2017)}]{yao2017feedback}
Yao, W., and Jaiman, R., \enquote{Feedback control of unstable flow and
  vortex-induced vibration using the eigensystem realization algorithm,}
  \emph{Journal of Fluid Mechanics}, Vol. 827, 2017, pp. 394--414.

\bibitem[{Sun and Zhang(2016)}]{sun2016nonlinear}
Sun, X., and Zhang, J.-z., \enquote{Nonlinear vibrations of a flexible membrane
  under periodic load,} \emph{Nonlinear Dynamics}, Vol.~85, No.~4, 2016, pp.
  2467--2486.

\bibitem[{Curet et~al.(2014)Curet, Carrere, Waldman, and
  Breuer}]{curet2014aerodynamic}
Curet, O.~M., Carrere, A., Waldman, R., and Breuer, K.~S., \enquote{Aerodynamic
  characterization of a wing membrane with variable compliance,} \emph{AIAA
  journal}, Vol.~52, No.~8, 2014, pp. 1749--1756.

\bibitem[{Bohnker and Breuer(2019)}]{bohnker2019control}
Bohnker, J.~R., and Breuer, K.~S., \enquote{Control of separated flow using
  actuated compliant membrane wings,} \emph{AIAA Journal}, Vol.~57, No.~9,
  2019, pp. 3801--3811.

\bibitem[{Huang et~al.(2021)Huang, Xia, Dai, Yang, and Wu}]{huang2021fluid}
Huang, G., Xia, Y., Dai, Y., Yang, C., and Wu, Y., \enquote{Fluid--structure
  interaction in piezoelectric energy harvesting of a membrane wing,}
  \emph{Physics of Fluids}, Vol.~33, No.~6, 2021, p. 063610.

\bibitem[{Rojratsirikul et~al.(2010{\natexlab{b}})Rojratsirikul, Wang, and
  Gursul}]{rojratsirikul2010unsteady}
Rojratsirikul, P., Wang, Z., and Gursul, I., \enquote{Unsteady fluid-structure
  interactions of membrane airfoils at low Reynolds numbers,} \emph{Animal
  Locomotion}, 2010{\natexlab{b}}, pp. 297--310.

\end{thebibliography}

\end{document}